\documentclass[apj]{emulateapj}

\usepackage{astro}

\slugcomment{accepted for publication in the Astrophysical Journal}

\shorttitle{Tearing the veil of the Orion Nebula}
\shortauthors{Van der Werf et al.}

\begin{document}

\title{Tearing the veil: interaction of the Orion Nebula with its neutral environment}

\author{Paul P.~van der Werf\altaffilmark{1,2}, W.M.~Goss\altaffilmark{3},
C.R.~O'Dell\altaffilmark{4}}

\altaffiltext{1}{Leiden Observatory, Leiden University, P.O.~Box 9513,
NL - 2300 RA Leiden, The Netherlands}
\altaffiltext{2}{SUPA, Institute for Astronomy, University of Edinburgh,
  Royal Observatory, Blackford Hill, Edinburgh EH9 3HJ, United
  Kingdom}
\altaffiltext{3}{National Radio Astronomy Observatory, P.O.~Box 0, Socorro,
NM 87801, USA}
\altaffiltext{4}{Department of Physics and Astronomy, Vanderbilt University, Box
  1807-B, Nashville, TN 37235, USA}

\begin{abstract}
  We present $\HI$ $21\cm$ observations of the Orion Nebula, obtained
  with the Karl G.~Jansky Very Large Array, at an angular resolution
  of $\secd 7.2 \times \secd 5.7$ and a velocity resolution of
  $0.77\kms$. Our data reveal $\HI$ absorption towards the radio
  continuum of the $\HII$ region, and $\HI$ emission arising from the
  Orion Bar photon-dominated region (PDR) and from the Orion-KL
  outflow. In the Orion Bar PDR, the $\HI$ signal peaks in the same
  layer as the $\Ht$ near-infrared vibrational line emission, in
  agreement with models of the photodissociation of $\Ht$. The gas
  temperature in this region is approximately $540\K$, and the $\HI$
  abundance in the interclump gas in the PDR is 5--10\% of the
  available hydrogen nuclei. Most of the gas in this region therefore
  remains molecular.  Mechanical feedback on the Veil manifests itself
  through the interaction of ionized flow systems in the Orion Nebula,
  in particular the Herbig-Haro object $\HH{202}$, with the
  Veil. These interactions give rise to prominent blueward velocity
  shifts of the gas in the Veil. The unambiguous evidence for
  interaction of this flow system with the Veil shows that the
  distance between the Veil and the Trapezium stars needs to be
  revised downwards to about $0.4\pc$. The depth of the ionized cavity
  is about $0.7\pc$, which is much smaller than the depth and the
  lateral extent of the Veil. Our results reaffirm the blister model
  for the M42 $\HII$ region, while also revealing its relation to the
  neutral environment on a larger scale.
\end{abstract}


\section{Introduction}

The Orion Nebula (M42, $\NGC{1976}$, Orion~A) is the nearest region of
recent massive star formation, containing the densest nearby cluster
of OB stars. Since the optically visible nebula M42 is located in
front of the parent molecular cloud OMC-1, it is accessible for
detailed studies in every region of the electromagnetic spectrum. As a
result, the Orion Nebula has become a cornerstone for our
understanding of massive star formation, as well as its feedback
effects on the star forming environment, which is the subject of the
present paper.

The Orion nebula and OMC-1 are located near
the center of a prominent north-south ridge of dense molecular gas,
shaped approximately like an integral sign
\citep{Bally.etal1987,Castets.etal1990,Heyer.etal1992,Johnstone.Bally1999,Plume.etal2000},
and containing the OMC-1 through OMC-4 molecular
clumps. OMC-1 is the most most prominent of these, with a mass of
approximately $2200\Msun$ \citep{Bally.etal1987}. 
The integral-shaped
ridge is the northern part of the larger Orion~A giant molecular cloud
(GMC), which has a mass of about $10^5\Msun$
\citep{Maddalena.etal1986} and is one of a system of two GMCs (the
Orion~A and Orion~B GMCs, named after the radio sources they contain)
that extends roughly north-south through the belt and sword regions of
the Orion constellation
\citep{Kutner.etal1977,Maddalena.etal1986,Sakamoto.etal1994,Wilson.etal2005}.
These clouds are associated with even larger diffuse $\HI$ clouds
\citep{Chromey.etal1989,Green1991}. An excellent recent review of star
formation and molecular clouds in the greater Orion region has been
presented by \citet{Bally2008}.

The Orion~A molecular cloud hosts several generations of OB star
formation \citep{Blaauw1964}, the youngest of which is the Orion
Nebula Cluster (ONC), ionizing the M42 $\HII$ region \citep[see][for a
  detailed recent review]{Muench.etal2008}. This cluster has a central
density of about $2\times10^4\,$stars$\pun{pc}{-3}$ and a total stellar
mass of about $1800\Msun$ in about 3500 stars
\citep{Hillenbrand.Hartmann1998}, out to a radius of
$\sim2.5\pc$\null. The total mass of the ONC is therefore comparable
to the molecular gas mass of OMC-1, which is $2200\Msun$, within a
similar radius \citep{Bally.etal1987}. Locally, the star formation
efficiency (here quantified as
$\qu{M}{stars}/(\qu{M}{stars}+\qu{M}{gas})$) is therefore quite high
at approximately 50\%. On the scale of the integral-shaped ridge
(linear size about $9\pc$), which has a gas mass of $\sim5000\Msun$
\citep{Bally.etal1987}, this efficiency is somewhat lower,
approximately 25\%.  The ionizing luminosity of the ONC is dominated
by $\theta^1$\,C~Ori.  This star is the most luminous component of the
asterism formed by the Trapezium stars ($\theta^1$\,A--D~Ori).
$\theta^1$\,C~Ori is an oblique magnetic rotator with an effective
temperature $\qu{T}{eff}\approx39000\pm1000\K$ and $\log g=4.1$
\citep{SimonDiaz.etal2006}, implying a spectral type O6Vp.
Observations by \citet{Weigelt.etal1999} revealed that
$\theta^1$\,C~Ori is a close binary, dominated in mass and luminosity
by the star $\theta^1$\,C$_1$~Ori, for which a spectral type O5.5 was
derived by \citet{Kraus.etal2007}. The ionizing photon flux
corresponding to spectral types O6 to O5.5 is
$Q_0=1.0-1.3\cdot10^{49}\ps$ \citep{Martins.etal2005}.

Most visual studies of the Orion Nebula have concentrated on the
$\sim5'$ diameter optically bright region centered on the Trapezium
stars, commonly referred to as the Huygens region, after its first
description by \citet{Huygens1659}. However, lower surface brightness
nebular emission extends significantly towards the southwest.
Including this fainter region the nebula subtends an approximately
circular region on the sky, with a diameter of about half a degree
\citep[e.g., Fig.~1 in][]{Muench.etal2008}. This region is now referred
to as the Extended Orion Nebula \citep[EON,][]{Gudel.etal2008} and
contains the Huygens region at its north-east boundary. The Huygens
region itself is bounded at the northeast side by the Northeast Dark
Lane \citep{ODell.Harris2010}, which separates M42 from the fainter
$\HII$ region M43 towards the northeast. Another prominent dark
feature, already seen by \citet{Huygens1659} is the Dark Bay, which is
a tongue of obscuration, covering part of M42 east of the Trapezium
stars.

The Orion Nebula is a blister-type $\HII$ region, with the ionized gas streaming
away from the high pressure interface with OMC-1
\citep{Zuckerman1973,Balick.etal1974}. Velocities with respect to the Local
Standard of Rest (LSR) in the ionized gas are $\vLSR=7.4\pm1.5\kms$ for the low
ionization lines ([$\OI$], [$\SII$]) arising at the ionization front (IF), but
lower (i.e., more blueshifted) velocities $\vLSR=-0.2\pm1.3\kms$ are found for
higher excitation species ([$\OII$], [$\OIII$], [$\NII$]) and for the bulk
ionized gas traced by hydrogen recombination lines
\citep{Kaler1967,ODell.Wen1992,Doi.etal2004,Henney.etal2007,GarciaDiaz.Henney2007,GarciaDiaz.etal2008}.
The background molecular gas is at
$\vLSR\approx10\kms$\citep{Loren1979}\footnote{In the region under consideration
  in this paper, LSR and heliocentric velocities are related by
  $\vLSR=\vhel-18.1\kms$.}. The ionization front (IF) separating M42 and OMC-1
is located behind the Trapezium stars, at a distance of approximately $0.3\pc$
behind $\theta^1$\,C~Ori \citep{Wen.ODell1995,ODell2001,ODell.etal2008}. A
three-dimensional model of the ionized region has been derived by
\citet{Wen.ODell1995}, who showed that the IF, which is approximately face-on in
the region behind the Trapezium stars, curves to an orientation that is almost
edge-on approximately $100''$ south-east of the Trapezium stars.  In this region
the IF is observed as a prominent, almost linear optical feature commonly
referred to as the Bright Bar. On the molecular side of the IF, a
photon-dominated region (PDR) has formed, which is close to edge-on south-east
of the Bright Bar.  It has been studied in the strong neutral gas cooling lines,
in particular [$\CII$] $158\mum$ \citep{Stacey.etal1993} and [$\OI$] $63\mum$
\citep{Herrmann.etal1997} as well as numerous other species.  Due to its aspect
and proximity, the edge-on Orion Bar PDR has become the most iconic region of
its type.

The OMC-1 molecular cloud behind M42 harbors an obscured region of
young massive star formation, exhibiting luminous infrared emission
with a bolometric luminosity of about $8\cdot10^4\Lsun$
\citep{Gezari.etal1998}, known as the Kleinmann-Low region
\citep[Orion-KL,][]{Kleinmann.Low1967}, and located about $1'$
northwest of the Trapezium stars.  This region contains a complex
system of outflows and masers, various young stellar objects, and the
eponymous Orion Hot Core, a compact region of molecular gas and dust
with high temperature (several $100\K$) and density
($\sim10^6\pcmcub$) driving a complex chemistry.  The high velocity
outflow originating in this region gives rise to the famous
``fingers'', first discussed by \citet{Allen.Burton1993}. All of these
features are the subject of a vast literature, to which we will return
in \secref{sec.KL}. Extensive background can be found in the review by
\citet{Genzel.Stutzki1989}, which contains an overview and synthesis
of earlier results, and the review by \citet{ODell.etal2008}, which
discusses more recent results on this complex region.

A second active star forming region is located about $\mind 1.5$ south
of Orion-KL\null.  This region, referred to as Orion-S, has an
infrared luminosity of about 10\% of that of Orion-KL
\citep{Mezger.etal1990}. Like Orion-KL, Orion-S is a rich source of
molecular line emission, containing several hot cores
\citep{Zapata.etal2007} and multiple bipolar outflows and maser
systems.  However, unlike Orion-KL, Orion-S is an isolated
  molecular core located within the cavity containing the ONC
  \citep{ODell.etal2009}. As a result several of the outflows
originating from Orion-S produce optically detectable features, many
of which are catalogued as Herbig-Haro (HH) objects
\citep{ODell.etal1997,Bally.etal2000,Henney.etal2007,ODell.Henney2008}.

In front of the ionized nebula, several layers of predominantly
neutral atomic gas are found. These were first detected in $21\cm$
$\HI$ absorption towards the nebular radio continuum
\citep{Muller1959,Clark.etal1962,Clark1965,Radhakrishnan.etal1972,Lockhart.Goss1978},
and are collectively referred to as the Veil \citep{ODell2001}.  The
term Veil is appropriate since this feature is largely transparent,
and only becomes opaque (at visual wavelengths) in the Dark Bay and
Northeast Dark Lane regions, where its column density is highest
\citep{ODell.YusefZadeh2000}. The first full $\HI$ aperture synthesis
observations of the Orion Nebula were carried out by
\citet{Lockhart.Goss1978} at an angular resolution of $2'$, using the
Owens Valley Interferometer. These authors first showed the presence
of 3 velocity components in the Veil.  This velocity structure was
confirmed in higher resolution ($16''$) aperture synthesis using the
VLA in C-configuration \citep[][hereafter
  vdWG89]{VanderWerf.Goss1989}, who found LSR velocities of
approximately 6, 4, and $-2\kms$ for the absorbing components A, B and
C (adopting the notation of vdWG89, which we follow in the present
paper).  These observations confirmed the physical association of the
Veil with the Orion Nebula, first suggested by
\citet{Lockhart.Goss1978}, based on the increasing $\HI$ column
density towards the Dark Bay and Northeast Dark Lane in the
velocity components A and B\null. Absorption by components A and B is
also detected towards the smaller $\HII$ region M43 towards the
northeast, confirming that the Veil represents an extended layer
covering the M42/M43 system. The total $\HI$ opacity distribution of
components A and B correlates well with the optical extinction towards
the Huygens region \citep{ODell.etal1992,ODell.YusefZadeh2000}.
Physical conditions in the Veil have been studied further by optical
\citep{ODell.etal1993a} and ultraviolet (UV) absorption lines
\citep{Abel.etal2004,Abel.etal2006,Lykins.etal2010}. Modeling of these
results has resulted in a location for the Veil of a significant, but
not accurately determined, distance of $1-3\pc$ in front of the
Trapezium stars \citep{Abel.etal2004}.

Several additional $\HI$ absorption components have been detected towards the
Huygens region. These cover only small parts of M42 and are not detected towards
M43.  Velocity component C at $\vLSR\approx-2\kms$ was already detected by
\citet{Lockhart.Goss1978}.  The $\HI$ observations described by vdWG89
unexpectedly revealed a remarkable set of small-scale ($0.02-0.06\pc$) $\HI$
absorption components \citep[][hereafter vdWG90]{VanderWerf.Goss1990b}. Most of
these features (D--G in the notation of vdWG90) are blueshifted with respect to
both the molecular and the ionized gas, and have central LSR velocities from
$-7$ to $-17\kms$. Two of the features exhibited several velocity components. In
addition, one feature (H) was detected by vdWG90 in absorption at the velocity
of the background molecular cloud OMC-1. The features are most likely associated
with M42 (vdWG90), but their precise nature remained somewhat unclear.

In the present paper, we return to the Orion Nebula to investigate the
radiative and mechanical feedback of the ONC, the Orion Nebula and the
various outflow systems, on the neutral environment of the nebula. We
use new high resolution $\HI$ radio observations to probe $\HI$
emission from behind the Huygens region and from the Orion Bar PDR, as
well as $\HI$ absorption from the Veil and the small-scale absorption
components. We thus obtain a comprehensive picture of the 
radiative and mechanical feedback effects of massive star formation
in this region on the neutral gas environment. We describe the
observations and data reduction in \secref{sec.ObsRed}. The radio
continuum, $\HI$ emission, and $\HI$ absorption results are presented
in \secsref{sec.Continuum}{sec.AbsResults}. These results are
discussed in detail in
\twosecsref{sec.EmDisc}{sec.HIAbsDisc}. Finally, our conclusions are
summarized in \secref{sec.Conclusions}. Throughout this paper, we
adopt a distance to the Trapezium stars of $436\pm20\pc$ as given by
\citet{ODell.Henney2008}, which is based on a weighted combination of
several parallax measurements
\citep{Genzel.etal1981,Kraus.etal2007,Menten.etal2007,Hirota.etal2007}. At
this distance, $1''=0.0023\pc$ or $436\AU$. Where we use
distance-dependent quantities from earlier publications, we have
tacitly converted these to the distance adopted here.

\section{Observations and reduction}
\label{sec.ObsRed}

\subsection{Observations}

We used the NRAO Karl G.~Jansky Very Large Array (VLA), to obtain
$\HI$ data of the Orion Nebula in two periods in 2006 and 2007
(programs AG738 and AV297). The C and then the B~array of the VLA were
used to extend the angular resolution, sensitivity and velocity
coverage of the old C~array data of vdWG89 and vdWG90, obtained in
1984.  The VLA correlator was used. The total bandwidth was
$781.25\un{kHz}$ with 256~channels and two circular polarizations,
centered at $\vLSR=2.0\kms$. The channel separation was
$3.052\un{kHz}$ ($0.64\kms$ at the $\HI$ line) and the velocity
resolution was $0.77\kms$.

The C~array data was obtained in a series of three 5~hour observations on 2006
September 29, November 9 and November 19. Twenty of the VLA antennas were
used, with no use of the 7 antennas that had been converted to the EVLA at that
time.  The phase calibration was based on frequent observations (once per half
hour) of the quasar $\sou{OG}{050}$ (J0532+0732) with a flux density of
$1.8\Jy$.  The flux density scale was set by observations of $\sou{3C}{48}$
($15.7\Jy$).

The B~array data were obtained during late 2007 in a three times 5~hour
observation on November~15, November~24 and December~3. The flux density scale
was set using observations of $\sou{3C}{138}$ (total flux density $8.3\Jy$).
The observations of $\sou{3C}{138}$ were carried out every 30~minutes for a
period of 4~minutes.

For both the C~array and the B~array data, bandpass responses of each antenna
were determined by observing the strong sources $\sou{3C}{48}$ and
$\sou{3C}{84}$; these observations were shifted by plus and minus $0.7\MHz$
($148\kms$) to avoid the $\HI$ emission near $\vLSR=0\kms$ and absorption lines
of Galactic $\HI$ in the spectra of the calibration sources.

\subsection{Reduction and generation of data cubes}
\label{sec.reduction}

During the 2007 observations, we used EVLA antennas for the first time.  At this
time there were 12 EVLA antennas and 13 VLA antennas. During a test observation
of $\sou{3C}{48}$ obtained on 2007 October~4, an aliasing problem with the old
VLA correlator and the use of EVLA baselines was discovered by a number of NRAO
staff (including M.~Goss). This problem was caused by the hardware used to
convert the digital signals from the EVLA antennas into analogue signals to be
fed in the VLA correlator, which caused power to be aliased into the bottom
$0.5\MHz$ of the baseband. Only EVLA to EVLA antenna correlations were affected.
A number of partial solutions were
found\footnote{http://www.vla.nrao.edu/astro/guides/evlareturn/aliasing/}.  For
the Orion~A $\HI$ observations, the solution adopted was to use observations of
the strong source $\sou{3C}{138}$ every 30~minutes and to apply a time variable
baseline based calibration (as opposed to an antenna based calibration) to
correct for the closure errors due to the mismatched and time variable
bandpasses resulting from the aliasing. This scheme was tested in detail using
the correlator configuration that we used for the Orion~A $\HI$ observations.
We found that tracking the errors over this time interval worked well and the
visibility functions of the calibrator sources for EVLA to EVLA baselines had a
similar behavior as those of the VLA to VLA baselines. Before the phase closure
corrections were made, the amplitude fluctuations were at the level of 15\%
($1\sigma$) of the continuum flux density of $\sou{3C}{138}$; after correction
the fluctuations were reduced to values well below $2\%$. With the advent of the
WIDAR correlator in early 2010, these aliasing problems have disappeared.

The data from the C and B arrays were then combined and the line images were
made after subtracting the continuum in the {\it uv\/} plane (using the AIPS
task UVLSF); 102 of the 255 channels were line free and formed the continuum.
The final images were made using the AIPS task IMAGR with Robust=0 weighting.
The resulting datacube has a synthesized beam of $\secd7.2 \times \secd5.7$ at a
position angle of $29.7\deg$ and an r.m.s.\ noise per channel of $40\K$.  The
conversion factor between brightness temperature and flux density is
$14.9\K\,({\rm mJy}\per{beam})^{-1}$. A $1420.4\MHz$ image was produced using a
multiscale CLEAN algorithm, in order to optimally preserve the large range of
spatial scales present in this image. The
measured r.m.s.\ noise in the continuum image is $3.1\mJy\per{beam}$.

In the spectral line data cube, we discarded channels with elevated
noise at the edges of the band. Our final data cube covers the range
$-62.00\kms<\vLSR<68.56\kms$.

\subsection{Further processing}
\label{sec.processing}

Inspection of the $\HI$ data cube revealed a large and complex set of
features at various velocities, and with various angular
sizes. Remarkably, $\HI$ is detected in both emission and absorption.

Interferometric imaging of extended low-level emission features in the presence
of a strong continuum requires careful processing, because of non-linearities
introduced by deconvolution algorithms such as CLEAN \citep[see
e.g.,][]{VanGorkom.Ekers1989}, which may give rise to spurious features after
continuum subtraction. The best way to avoid these problems consists of first
subtracting the continuum and applying the deconvolution to the continuum-free
line images.  As described above, this is the procedure that was used.  After
continuum subtraction the $\HI$ absorption produces a strong negative signal
carrying the imprint of the subtracted continuum at LSR velocities between $-20$
and $10\kms$; at other velocities any remaining signal results from $\HI$
emission. As a result, $\HI$ emission can only reliably be studied at velocities
outside the range from $-20$ to $10\kms$. In order to increase the S/N ratio of
the $\HI$ emission data, we convolved the channel maps to a $\secd 7.5$ circular
beam, and smoothed the data cube spectrally by a factor 2, i.e., to a velocity
resolution and channel separation of $1.29\kms$. The r.m.s.\ noise in brightness
temperature $\Tb$ in these images is $20\K$.  Since the shortest baselines in
our observations were about $73\un{m}$, our data are insensitive to structures
with scales of about $7'$ or more.

In order to study the $\HI$ absorption, the full spatial and spectral
resolution $\HI$ line data cube was used, with the corresponding
$1420.4\MHz$ continuum image, to derive a data cube of $\HI$ optical
depth $\tau$, following the approach of vdWG89.  Optical depths were
derived by solving the equation of transfer
\begin{equation}
\Tb(v)=[1-e^{-\tau(v)}][\Ts-\Tc-\qu{T}{back}(v)],
\label{eq.Transfer}
\end{equation}
where $\Tb(v)$ is the observed $\HI$ brightness temperature at LSR
velocity $v$, after subtraction of the continuum (which has brightness
temperature $\Tc$).  $\Ts$ is the spin temperature of the absorbing
$\HI$, and $\qu{T}{back}(v)$ is the brightness temperature (at LSR
velocity $v$) of Galactic $\HI$ originating behind the absorbing
$\HI$. The peak brightness temperature of Galactic $\HI$ in the region
of the Orion Nebula is about $60\K$ \citep{Green1991}. It is not
possible to determine what fraction of this signal originates behind
the absorbing $\HI$, and the situation is complicated further by the
fact that this fraction may be a function of $v$. Therefore the
observed Galactic $\HI$ brightness temperature only provides an upper
limit for $\qu{T}{back}(v)$. For $\Ts$ a harmonic mean value can be
determined at positions where $\HI$ $21\cm$ absorption can be combined
with measurements of $\Lya$ absorption. Such measurements are
available at the positions of $\theta^1$~C~Ori
\citep{Shuping.Snow1997} and $\theta^1$~B~Ori \citep{Abel.etal2006},
giving $\Ts\approx90\K$ ($80{-}110\K$) in component~A and
$\Ts\approx135\K$ ($100{-}160\K$) in component~B\null.  Given the
uncertainties in $\Ts$ and $\qu{T}{back}(v)$ we solve
\eqref{eq.Transfer} with the approximation
$\Tc\gg\left|\Ts-\qu{T}{back}(v)\right|$, by only calculating
opacities at positions where $\Tc>450\K$ (corresponding to a surface
brightness level of $30\mJy\per{beam}$), which is almost $10\sigma$ in
the continuum image. While $\HI$ absorption lines can be detected
towards considerably fainter continuum levels, quantitatively reliable
opacities can only be derived where
$\Tc\gg\left|\Ts-\qu{T}{back}(v)\right|$.  With this approach the
precise value of $\Ts$ for determining the opacities (as well as the
implicit assumption of \eqref{eq.Transfer} that $\Ts$ is the same at
every position and for all absorbing velocity components) becomes
irrelevant.  At positions with $\Tc\le 450\K$, this procedure will
give rise to systematic errors in the derived opacities; at these
positions opacities are therefore not calculated.
\footnote{The derived $\HI$ opacity datacube and the corresponding
  $21\cm$ continuum image will be made available in electronic form
  through the Centre de Donn\'ees astronomiques de Strasbourg (CDS) at
  {\tt http://cds.u-strasbg.fr}.}

\begin{figure*}
 \plotone{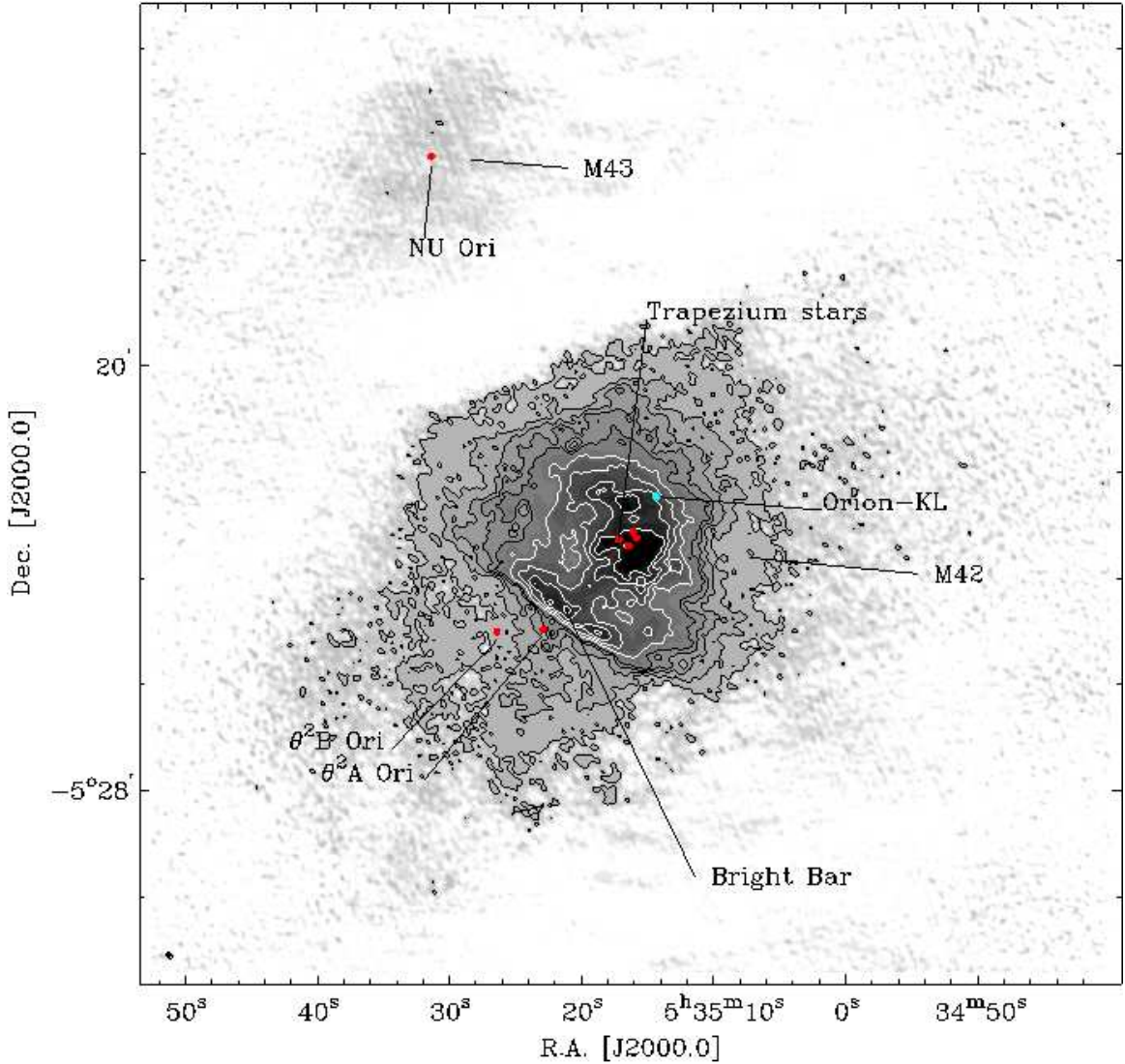}
 \caption{Continuum emission of the Orion Nebula at $1420.4\MHz$,
   constructed using a multi-scale deconvolution (see
   \secref{sec.reduction}).  Contours indicate surface
     brightness levels of 20, 40, 60, 80 and $100\mJy\per{beam}$
   (black contours) and 150, 200, 250, 300 and $350\mJy\per{beam}$
   (white contours). The small ellipse in the lower lefthand corner
   indicates the FWHM size and the orientation of the synthesized beam
   ($\secd 7.2\times \secd 5.7$ at a position angle of
     $29.7\deg$). The image has been corrected for primary beam
   attenuation. The principal massive young stars are indicated by red dots. A
   cyan dot indicates the position of the Orion-KL region.}
\label{fig.continuum}
\end{figure*}

\section{Continuum emission}
\label{sec.Continuum}

The $1420.4\MHz$ continuum image is shown in
\figref{fig.continuum}. This image shows both the main $\HII$ region
M42 and the fainter $\HII$ region M43 in the northeast. The image of
M42 is dominated by the Huygens region and the Bright Bar in the
southeast.  Fainter emission from the EON is seen to extend in all
directions from the southeast counterclockwise to the West, but not in
the other directions. Indications for the presence of low surface
brightness radio emission from the EON were first found by
\citet{Mills.Shaver1968} and \citet{Goss.Shaver1970}.  Emission from
the full EON is detected in the $4.75\GHz$ single-dish image of
\citet{Wilson.etal1997} and in the $330\MHz$ VLA image of
\citet{Subrahmanyan.etal2001}; the $1.5\GHz$ VLA image by
\citet{YusefZadeh1990} and \citet{Subrahmanyan.etal2001} also shows
the extended emission from the EON\null.

The total flux density of M42 in our image is $335\pm15\Jy$. This
value is somewhat lower than the total flux density of $374\Jy$ found
by vdWG89 (which is consistent with the best single-dish value, see
Table~2 in vdWG89).  For M43, we find a total flux density of
$14\pm2\Jy$, in excellent agreement with vdWG89.

The peak continuum flux density of M42 is $376\mJy\per{beam}$, which corresponds
to a peak continuum brightness temperature $\Tc=5600\K$. The electron
temperature in this region is $\Te=8400\pm400\K$, as determined by
\citet{Wilson.etal1997} from measurements of the H64$\alpha$ recombination line.
Since $\Tc=\Te(1-e^{-\qu{\tau}{ff}})$, the peak free-free optical depth is
$\tau_{\rm ff}=1.1$, i.e., the $\HII$ region is significantly optically
thick at $1420.4\MHz$, and will be opaque at lower frequencies, in
agreement with the spectral index distribution between 330 and $1500\MHz$
determined by \citet{Subrahmanyan.etal2001}.
A free-free opacity $\tau_{\rm ff}\approx 1$ is also found at the
brightest peaks of the Bright Bar.

\begin{figure*}
\includegraphics[width=0.1mm]{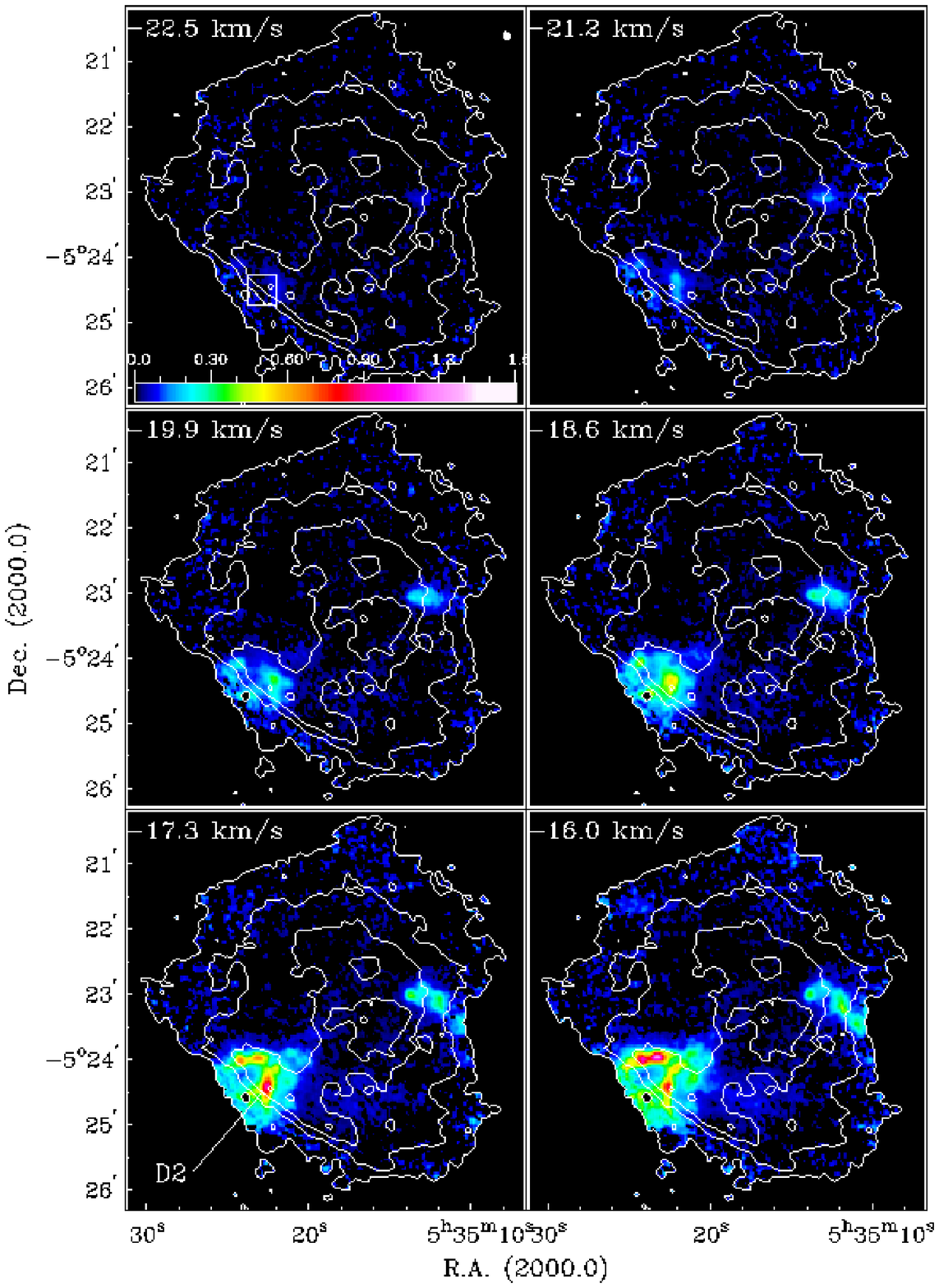}
\end{figure*}

\begin{figure*}
\plotone{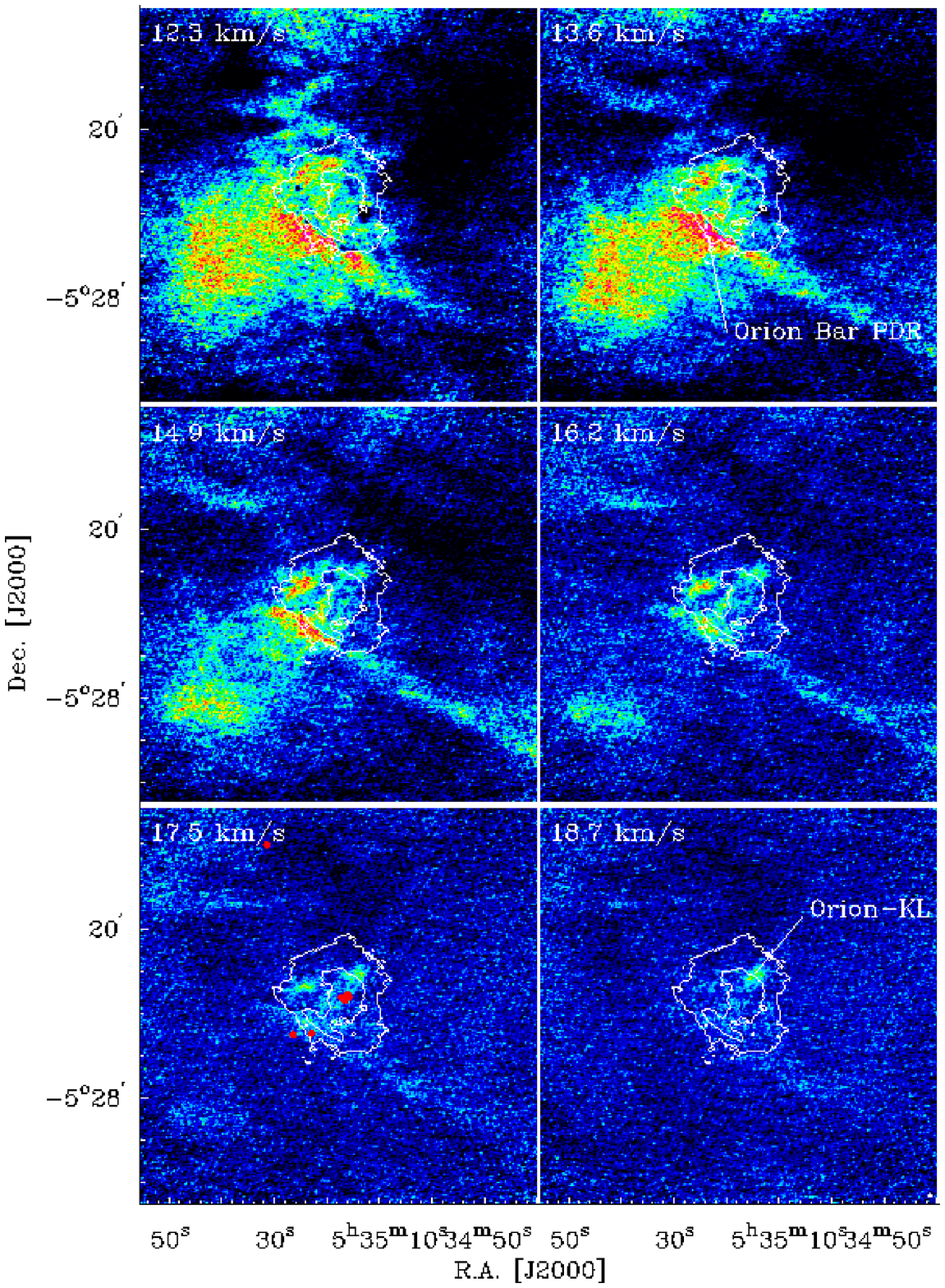}
\caption{Images of $\HI$ emission at the LSR velocities indicated in
  the upper left-hand corner of each frame. The images cover a
    velocity interval of $1.28\kms$ each.  Colors ranging from dark
  blue to red indicate a brightness temperature range from 0 to
  $250\K$. White contours indicate the $21\cm$ continuum at levels of
  50 and $200\mJy\per{beam}$. These images have been corrected for
  primary beam attenuation. The $\secd 7.5$ circular beam of these
  images is indicated in the lower righthand corner. The red dots in the lower
    left panel indicate the principal young massive stars (see
    \figref{fig.continuum} for legend).}
\label{fig.EmissionMaps}
\end{figure*}

\section{$\HI$ emission features}
\label{sec.HIEmission}

\subsection{$\HI$ emission images}
\label{sec.HIEmIm}

As noted in \secref{sec.processing}, $\HI$ emission can be studied at
velocities avoiding strong absorption, i.e., outside the velocity
interval from $-20$ to $10\kms$. Since the prominent molecular cloud
associated with the Orion Nebula is at $\vLSR\sim10\kms$, $\HI$
emission from the neutral environment of the $\HII$ region can be
probed only in its red line wing. Inspection of the $\HI$ emission
data cube revealed $\HI$ emission at LSR velocities from 10 to
$19\kms$, and six images covering this velocity range are presented in
\figref{fig.EmissionMaps}.

Inspection of \figref{fig.EmissionMaps} reveals that the strongest
$\HI$ emission is found in the region directly southeast of the Bright
Bar. $\HI$ brightness temperatures in this region are approximately
$250\K$ (coded red in \figref{fig.EmissionMaps}), with peaks reaching
$300\K$, indicating that the gas is quite warm.  While the brightest
$\HI$ emission is found closest to the Bar, the emission extends
towards the southeast over a distance of about $6'$ ($0.8\pc$), at
brightness temperature levels of about $120\K$.  Other features worthy
of note are an elongated $\HI$ feature extending from the Bar region
towards the southwest at velocities of $13-16\kms$, and compact $\HI$
emission features with higher velocities ($\vLSR$ up to $19\kms$).
$\HI$ emission features northeast of M42 trace the direct environment
of M43.

\begin{figure*}[ht]
\plotone{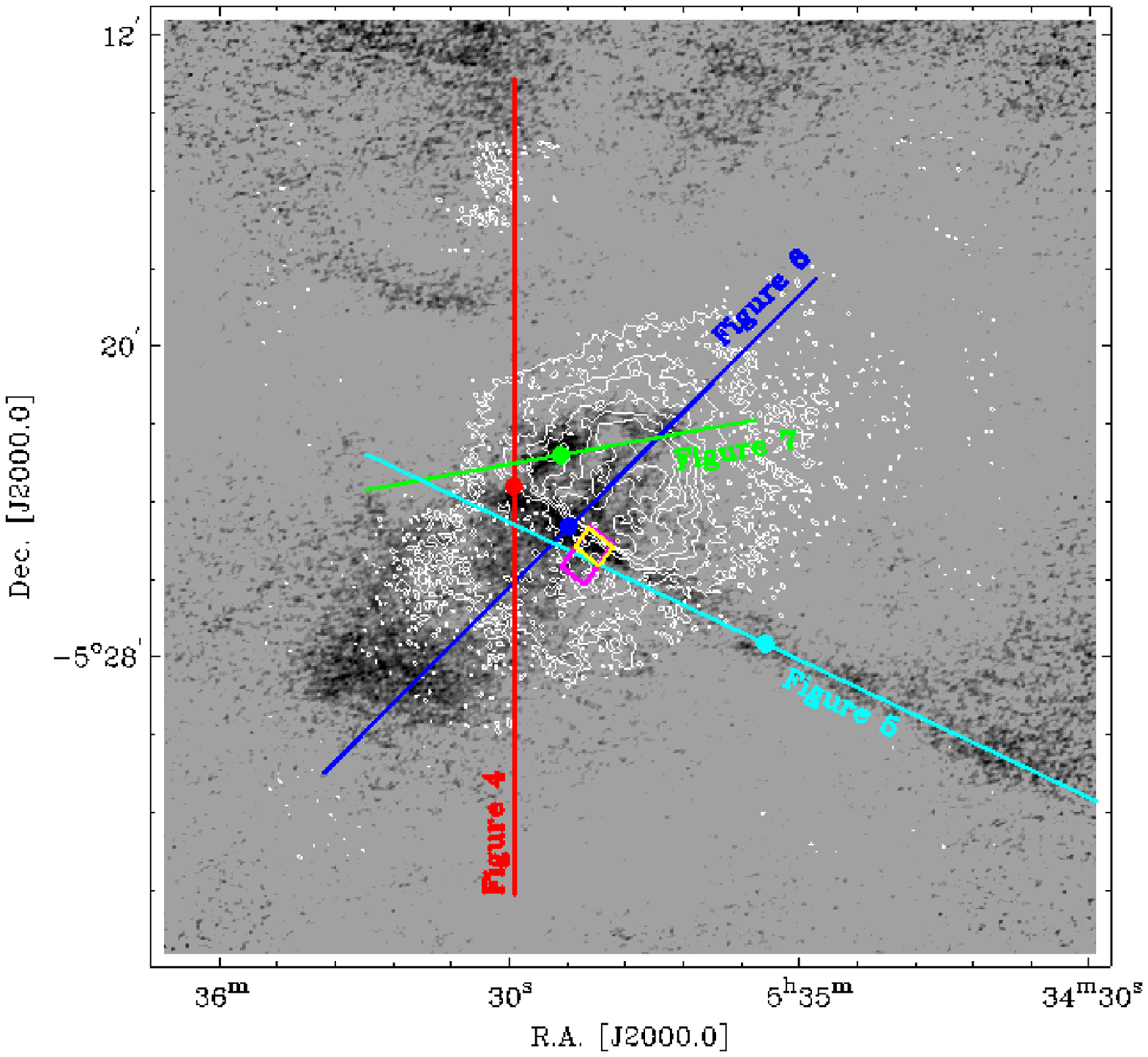}
\caption{Location of the position-velocity (PV) diagrams of $\HI$
  emission shown in \figsref{fig.XVEm1}{fig.XVEm4}, superposed on an $\HI$
  emission image at $\vLSR=14.9\kms$ (greyscale). The precise lengths,
  orientations and positions of these PV diagrams are indicated by the colored
  bars. In each bar, a thick dot of the same color indicates the zero position
  of the spatial coordinate in the corresponding PV diagram. The relevant figure
  numbers are indicated. White contours indicate $1420.4\MHz$ continuum surface
  brightnesses of 15, 40, 70, 100, 150, 200 and $300\mJy\per{beam}$. The purple
  rectangle in the Bright Bar region indicates the region used for constructing
  the strip shown in \figref{fig.BarStrips}. The yellow rectangle (which is
  contained in the purple one) shows the region over which the spectrum shown in
  \figref{fig.EmSpectrum} was averaged.}
\label{fig.EmFinder}
\end{figure*}

\begin{figure}
\includegraphics[width=\linewidth]{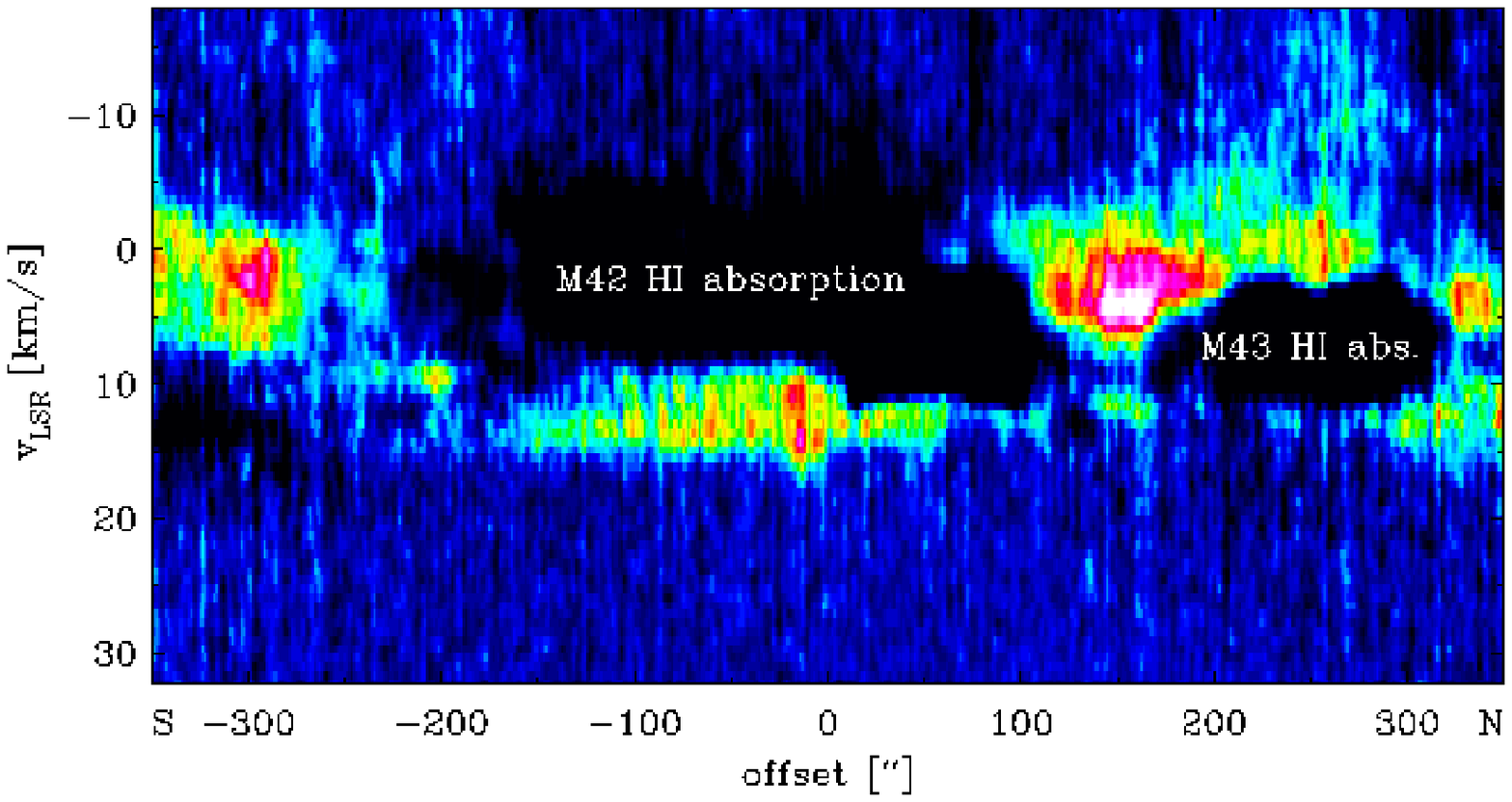}
\caption{Position-velocity diagram of $\HI$ emission along the red
  north-south bar shown in \figref{fig.EmFinder}.  The spatial axis
  has its zero position at $\RA=\hmsd 5h35m29.5s$, $\Dec=\dms
  -5d23m37s$ and a position angle of $0\deg$. Spatial offsets are
  positive towards the north and negative towards the south. Features
  discussed in the text are indicated.}
\label{fig.XVEm1}
\end{figure}

\begin{figure}
\includegraphics[width=\linewidth]{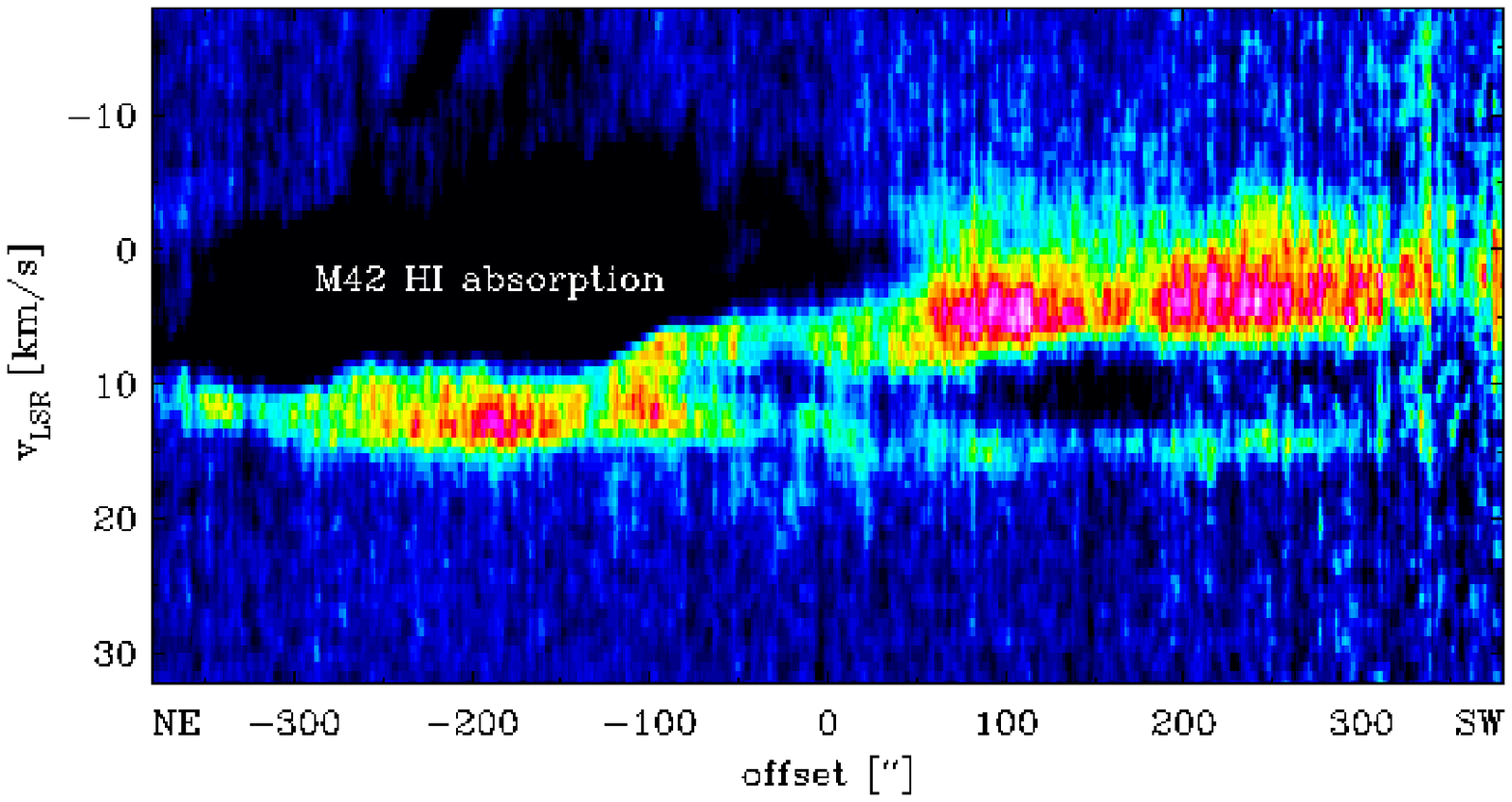}
\caption{Position-velocity diagram of $\HI$ emission along the light
  blue northeast-southwest bar shown in \figref{fig.EmFinder}.  The
  spatial axis has its zero position at $\RA=\hmsd 5h35m03.5s$,
  $\Dec=\dms -5d27m40s$ and a position angle of $64.66\deg$. Spatial
  offsets are negative towards the northest and positive towards the
  southwest. Features discussed in the text are indicated.}
\label{fig.XVEm2}
\end{figure}

\begin{figure}
\includegraphics[width=\linewidth]{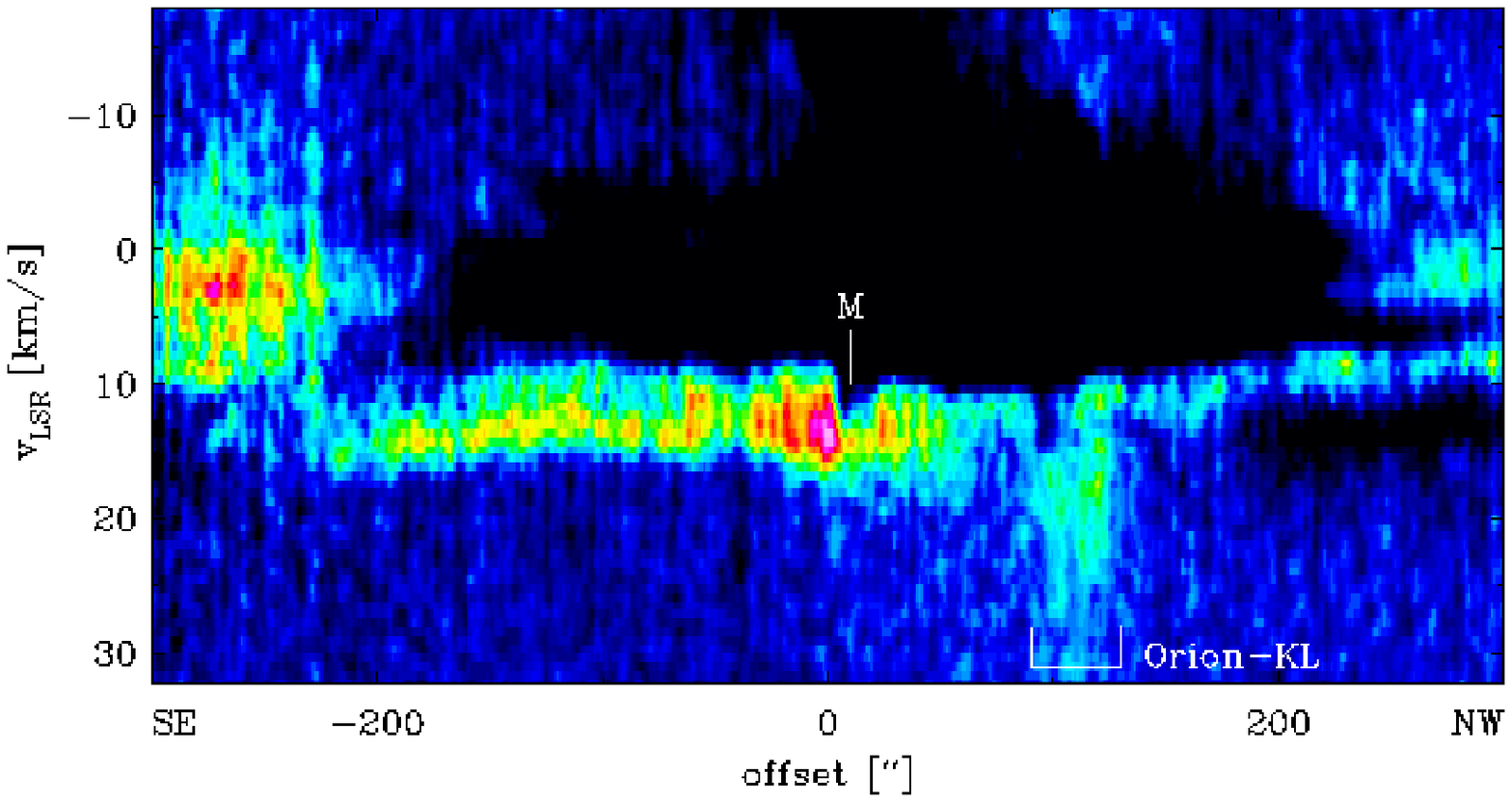}
\caption{Position-velocity diagram of $\HI$ emission along the dark
  blue northwest-southeast bar shown in \figref{fig.EmFinder}.  The
  spatial axis has its zero position at $\RA=\hmsd 5h35m23.8s$,
  $\Dec=\dms -5d24m38s$ and a position angle of $-45\deg$. Spatial
  offsets are negative towards the southeast and positive towards the
  northwest. Features discussed in the text are indicated.}
\label{fig.XVEm3}
\end{figure}

\begin{figure}
\includegraphics[width=\linewidth]{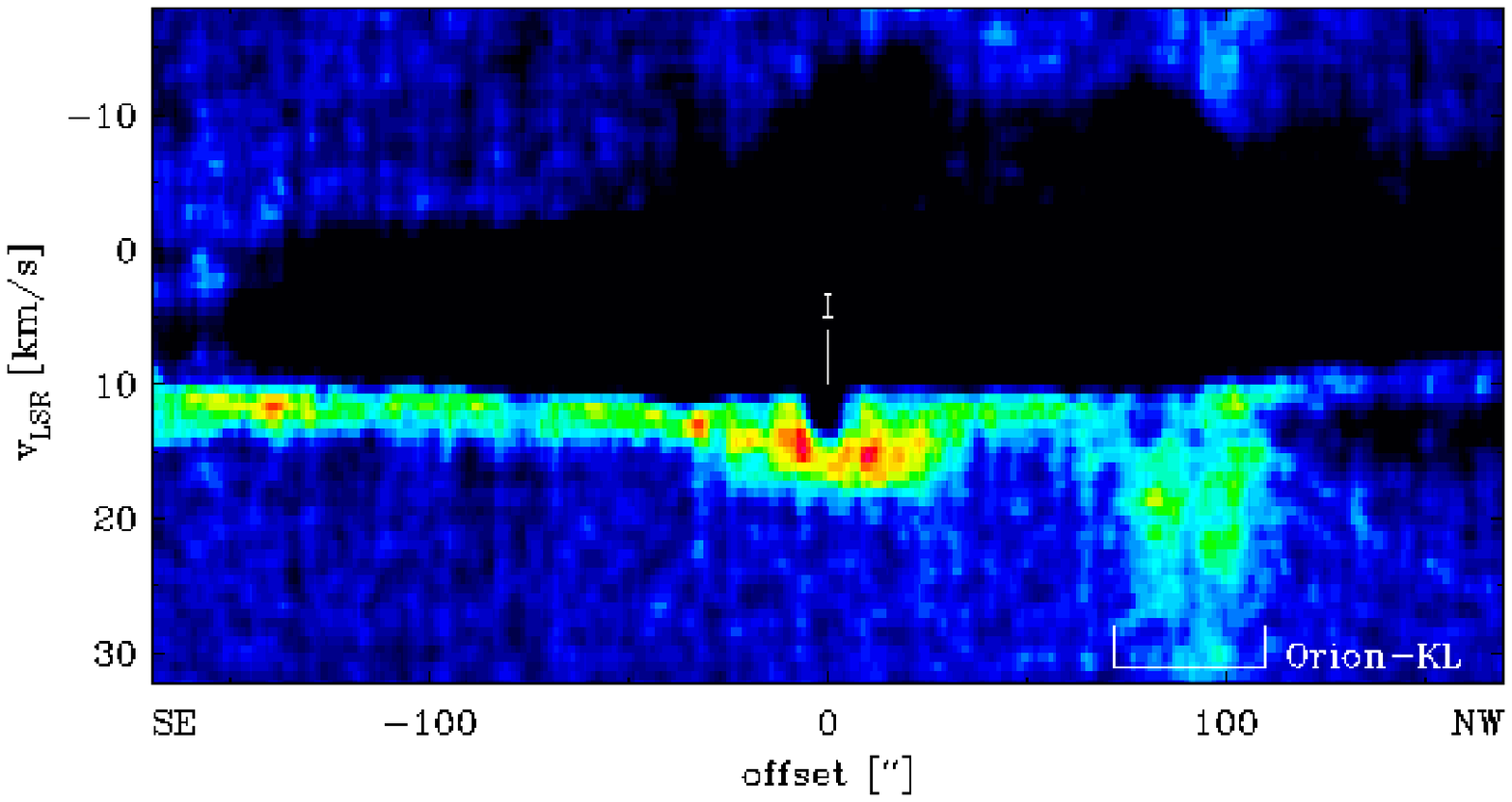}
\caption{Position-velocity diagram of $\HI$ emission along the green
  northwest-southeast bar shown in \figref{fig.EmFinder}.  The spatial
  axis has its zero position at $\RA=\hmsd 5h35m25.4s$, $\Dec=\dms
  -5d22m48s$ and a position angle of $-80\deg$. Spatial offsets are
  negative towards the southeast and positive towards the
  northwest. Features discussed in the text are indicated.}
\label{fig.XVEm4}
\end{figure}

\subsection{Position-velocity diagrams of $\HI$ emission}
\label{sec.HIEmPV}

In order to study the velocity structure of the $\HI$ emission and its
relation to the $\HI$ absorption (to be discussed below), we have
constructed a number of position-velocity (PV) diagrams of the $\HI$
emission. The orientations of the spatial axes of these diagrams are
shown in \figref{fig.EmFinder}.

\Figref{fig.XVEm1} shows the velocity structure of the extended $\HI$
layers in the region of the Orion Nebula. A prominent $\HI$ layer can
be seen in emission in the south (at an offset of approximately
$-300''$ in \figref{fig.XVEm1}). Following this layer northwards, it
produces strong $\HI$ absorption in front of the strong radio
continuum of M42. Between M42 and M43, in the Northeast Dark Lane,
strong $\HI$ emission is found. These emission features have central
velocities $\vLSR\sim2\kms$.  The absorbing $\HI$ in front of M43 is
at $\vLSR\sim7\kms$.

In the region southwest of the Bright Bar (offsets $-160$ to $0''$ in
\figref{fig.XVEm1}) $\HI$ emission is found with a peak velocity of about
$12-13\kms$, i.e., displaced in velocity from the $\HI$ absorption by about
$10-11\kms$. This $\HI$ emission feature is also detected at offsets $-300$ to
$-100''$ in \figref{fig.XVEm2}, which presents a PV diagram through the
elongated $\HI$ feature detected at $14\kms$ in \figref{fig.EmissionMaps}. This
PV diagram clearly reveals the elongated $\HI$ feature as a kinematically
separate entity, detected at offsets from $280$ to $-80''$.  At the latter
position, it connects to the $\HI$ at $12\kms$ southeast of the Bright Bar.

\Figref{fig.XVEm3} shows a PV diagram crossing the Orion Bar PDR
orthogonally, with the IF at offset $0''$, and also crossing the
compact high-velocity $\HI$ emission feature detected in
\figref{fig.EmissionMaps} at $\vLSR\sim18\kms$. The $\HI$ emission at
$\vLSR\sim14\kms$ is located in the region of the Orion Bar PDR
(offsets $-220$ to $0''$), but also extends slightly northwest of the
IF (offsets 0 to $50''$).  The feature M is an absorption feature
associated with the IF that will be discussed in
\twosecsref{sec.HIabsspectrum}{sec.HIAbsBackgr}. The high velocity
$\HI$ feature at offsets $100-120''$ reaches velocities up to
$31\kms$, and contains two distinct components separated by a local
emission minimum. This structure is well detected in
\figref{fig.XVEm4}. In this diagram, faint high velocity $\HI$
emission from the northwest component (offset $100''$) is also
detected at negative velocities. The PV diagram in \figref{fig.XVEm4}
also crosses a region of $\HI$ emission at $15\kms$ (at offsets from
$-30$ to $30''$) located at the position of the optical Dark Bay. This
feature is remarkable since it contains at its center an absorption
feature (marked I in \figref{fig.XVEm4} and discussed further in
\secref{sec.HIAbsBackgr}). \Figref{fig.XVEm4} also shows $\HI$
emission at $\vLSR\sim12\kms$ at offsets from $-160''$ to $120''$,
associated with OMC-1, but strong foreground $\HI$ absorption
precludes further study of this feature.

\begin{figure}
  \plotone{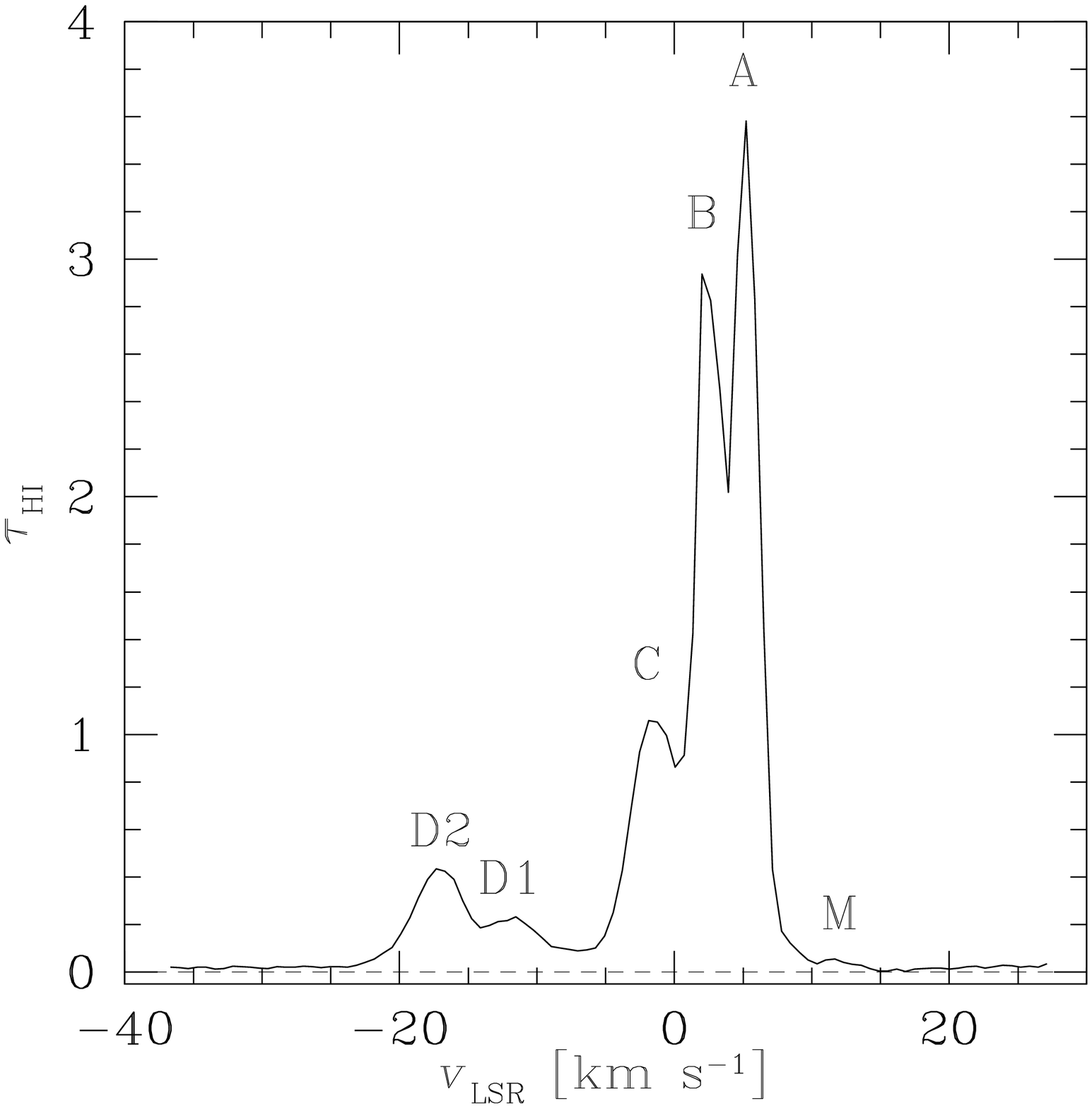}
\caption{Spectrum of $\HI$ opacity, averaged over a $27''\times 27''$
  area centered on RA=$\hmsd 5h35m22.88s$, Dec=$\dmsd -5d24m30.7s$, as indicated
  in \figref{fig.channelmaps}, in the panel at $\vLSR=-22.5\kms$.}
\label{fig.spectrum}
\end{figure}

\section{$\HI$ absorption features}
\label{sec.AbsResults}

\subsection{Overall velocity structure of the absorbing $\HI$}
\label{sec.HIabsspectrum}

The velocity structure of the $\HI$ absorption is illustrated in
\figref{fig.spectrum}, which shows the $\HI$ opacity spectrum averaged
over a region in the southeast part of the Huygens region, close to
the Bright Bar. This spectrum matches that shown in Fig.~1a of
vdWG90, corresponding to approximately the same region.  Only
unsaturated points were included in the calculation of the average
spectrum. Therefore, the spectral shape in the vicinity of the peaks
of components A and B in this figure should be treated with
caution.

Five prominent $\HI$ velocity components are seen in this figure,
corresponding to the components A, B, C, D1 and D2 of vdWG89 and
vdWG90. The $\HI$ components A and B cover the entire nebula. The
difference in the central velocities of these two components decreases
towards the northeast. As a result, and due to the increasing opacity
towards this region (resulting in saturation of the line peaks),
components A and B become difficult to separate in the region towards
the Northeast Dark Lane. Over most of the nebula, these components can
however be traced as two kinematically distinct features.  Component~C
is prominent in the southwest part of the Huygens region. It is
however not detected in the region of the Trapezium stars and towards
the Northeast Dark Lane. Towards M43 (even farther to the northeast)
only the components~A and B are detected.

\Figref{fig.spectrum} also shows two velocity components, D1 and D2,
at significantly negative velocities. These are examples of features with small
spatial scales compared to components A and B, as discussed by
vdWG90. Two of these features (D and F) revealed several velocity
components.  Because of the limited frequency coverage of the
observations by vdWG90, components D2 and F2 were not observed over
their full velocity extent.  \Figref{fig.spectrum} shows that D2 is
completely within the spectral band of the observations presented
here; this is also the case for F2 (not shown in
\figref{fig.spectrum}). In addition, our observations enable the
detection of further $\HI$ components at velocities not covered by
vdWG90. One new velocity component is indicated in
\figref{fig.spectrum} as component~M\null. This feature, which is also
indicated in \figref{fig.XVEm3}) will be discussed in
\secref{sec.HIAbsBar}. The detection of this very faint feature
illustrates the high sensitivity and dynamic range of this dataset.

\begin{figure*}
\includegraphics[width=1pt]{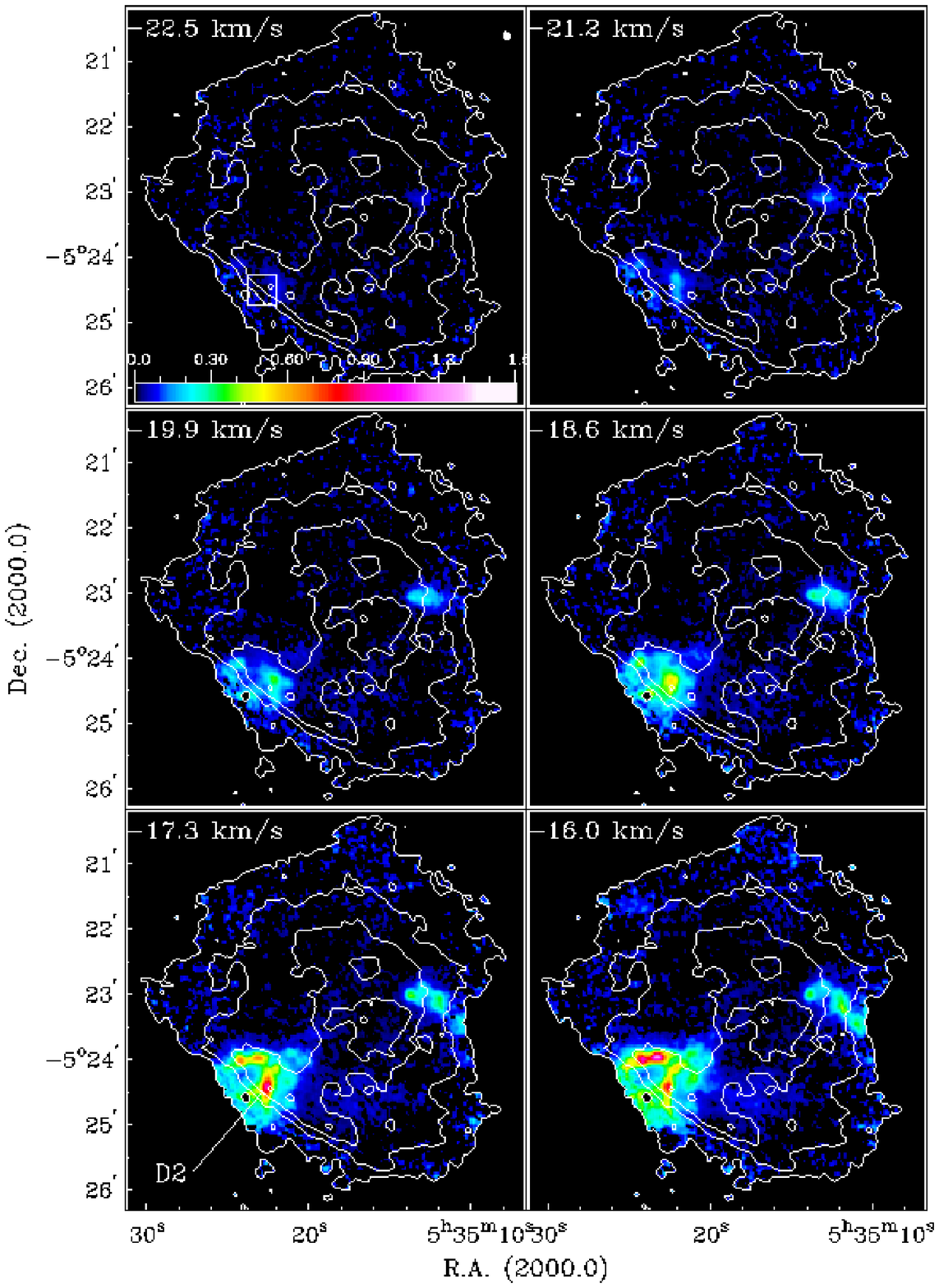}
\end{figure*}

\begin{figure*}
\plotone{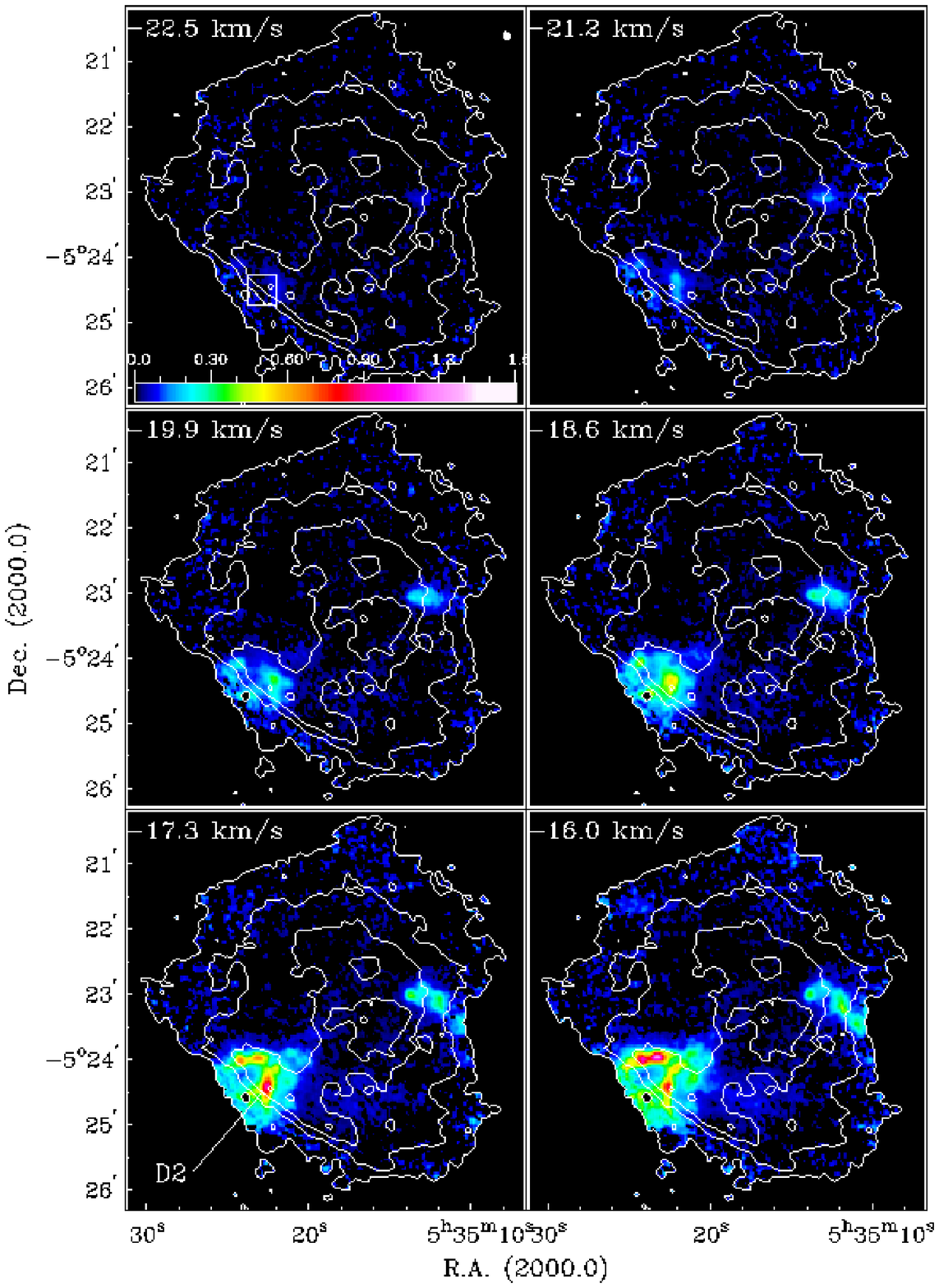}
\caption{Images of $\HI$ opacity (color images), at the LSR velocities
  indicated in the upper left-hand corner of each frame. White contours indicate
  the background $21\cm$ continuum at levels of 50, 100, 200, 300 and
  $400\mJy\per{beam}$.  Black pixels within the faintest continuum contour
  denote positions affected by saturation. The opacity scale is indicated by the
  color bar in the first frame. In this frame also the region over which the
  opacity spectrum shown in \figref{fig.spectrum} has been averaged, is shown.
  The size and orientation of the synthesized beam ($\secd 7.2\times \secd 5.7$
  at a position angle of $29.7\deg$) is indicated by the ellipse in the top
  righthand corner of the first frame. Letters C--M indicate various absorption
  features discussed in the text.}
\label{fig.channelmaps}
\end{figure*}

\begin{figure*}
\figurenum{9}
\plotone{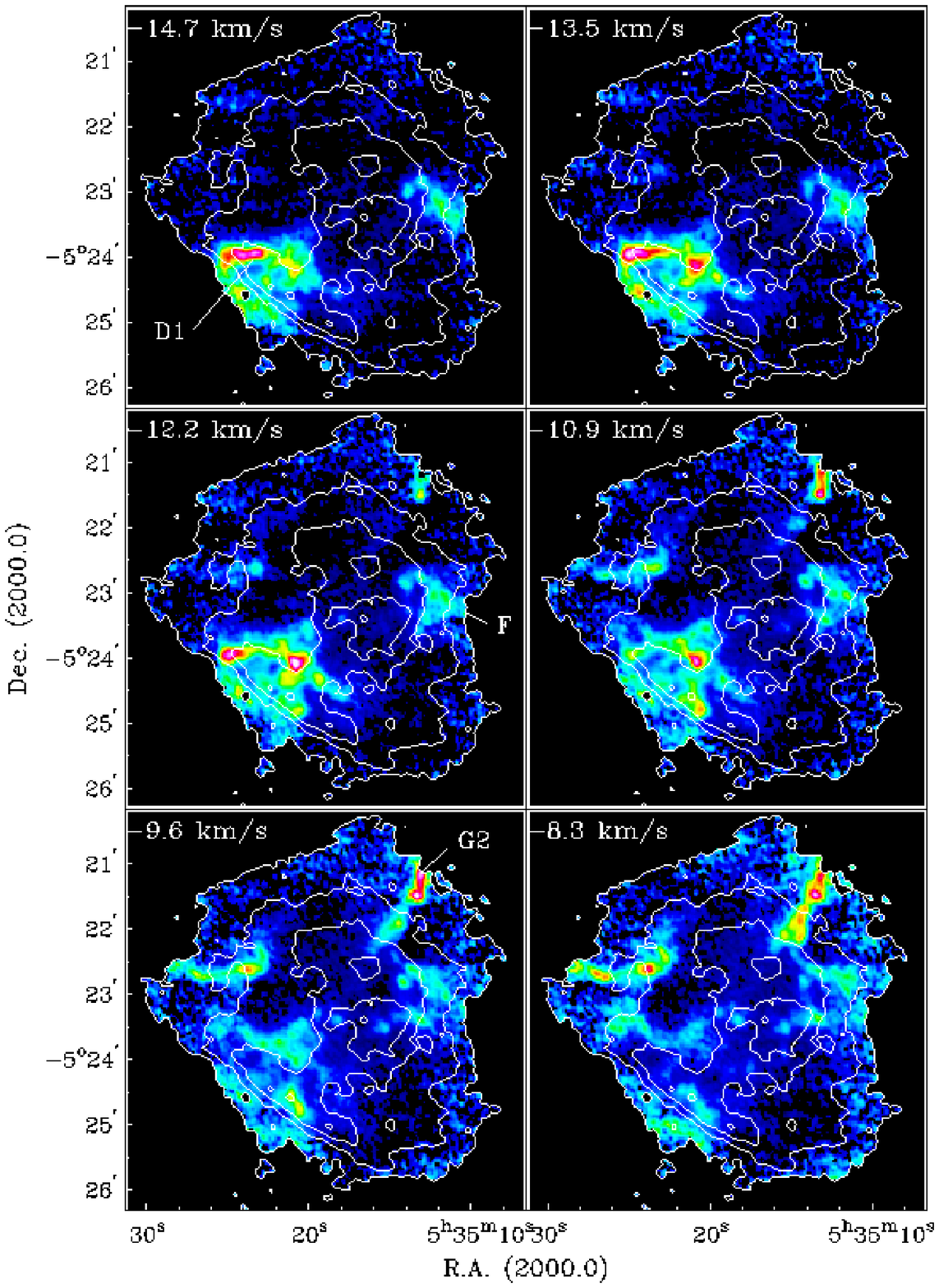}
\caption{{\it (continued)}
}
\end{figure*}

\begin{figure*}
\figurenum{9}
\plotone{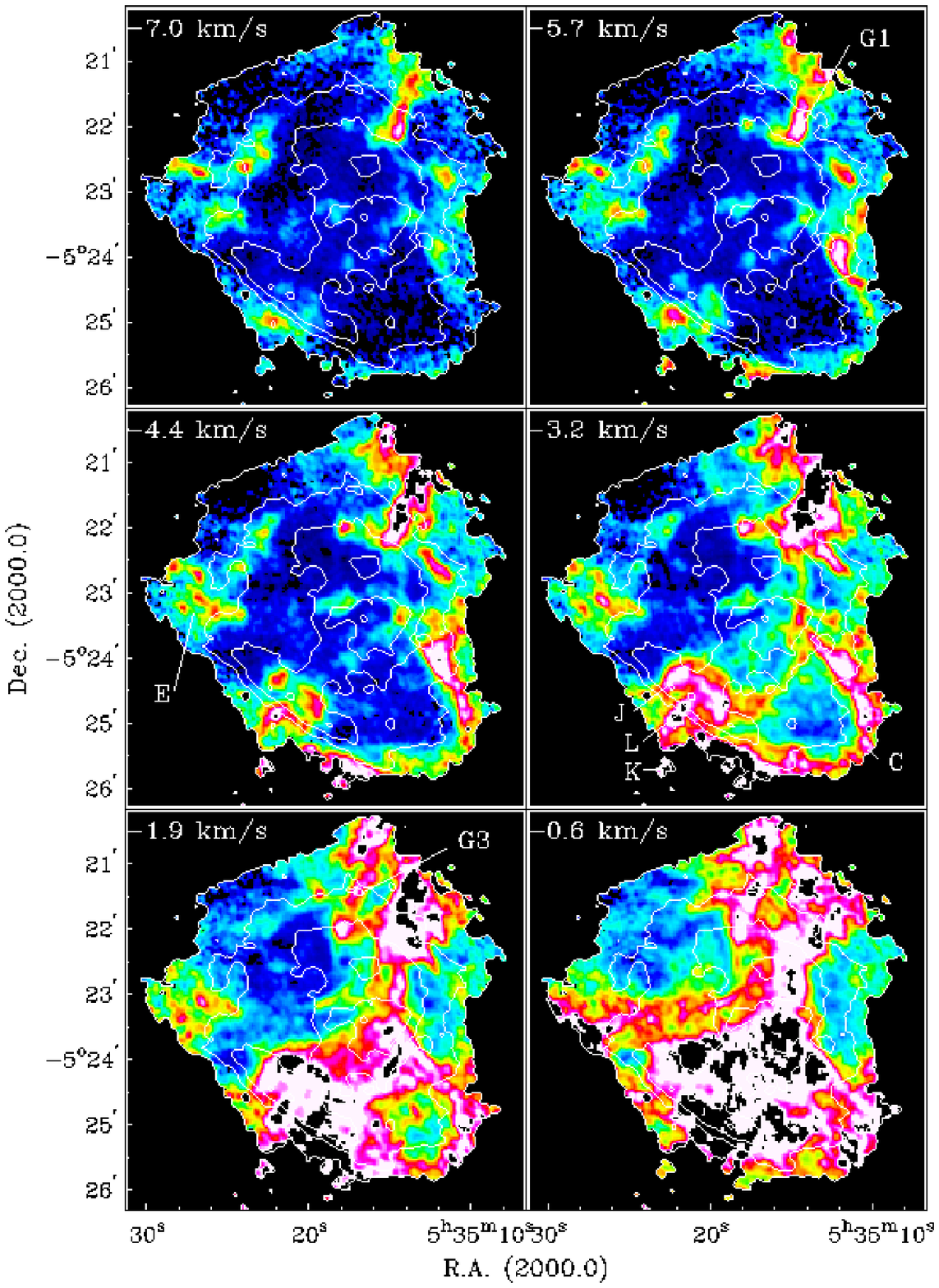}
\caption{{\it (continued)}
}
\end{figure*}

\begin{figure*}
\figurenum{9}
\plotone{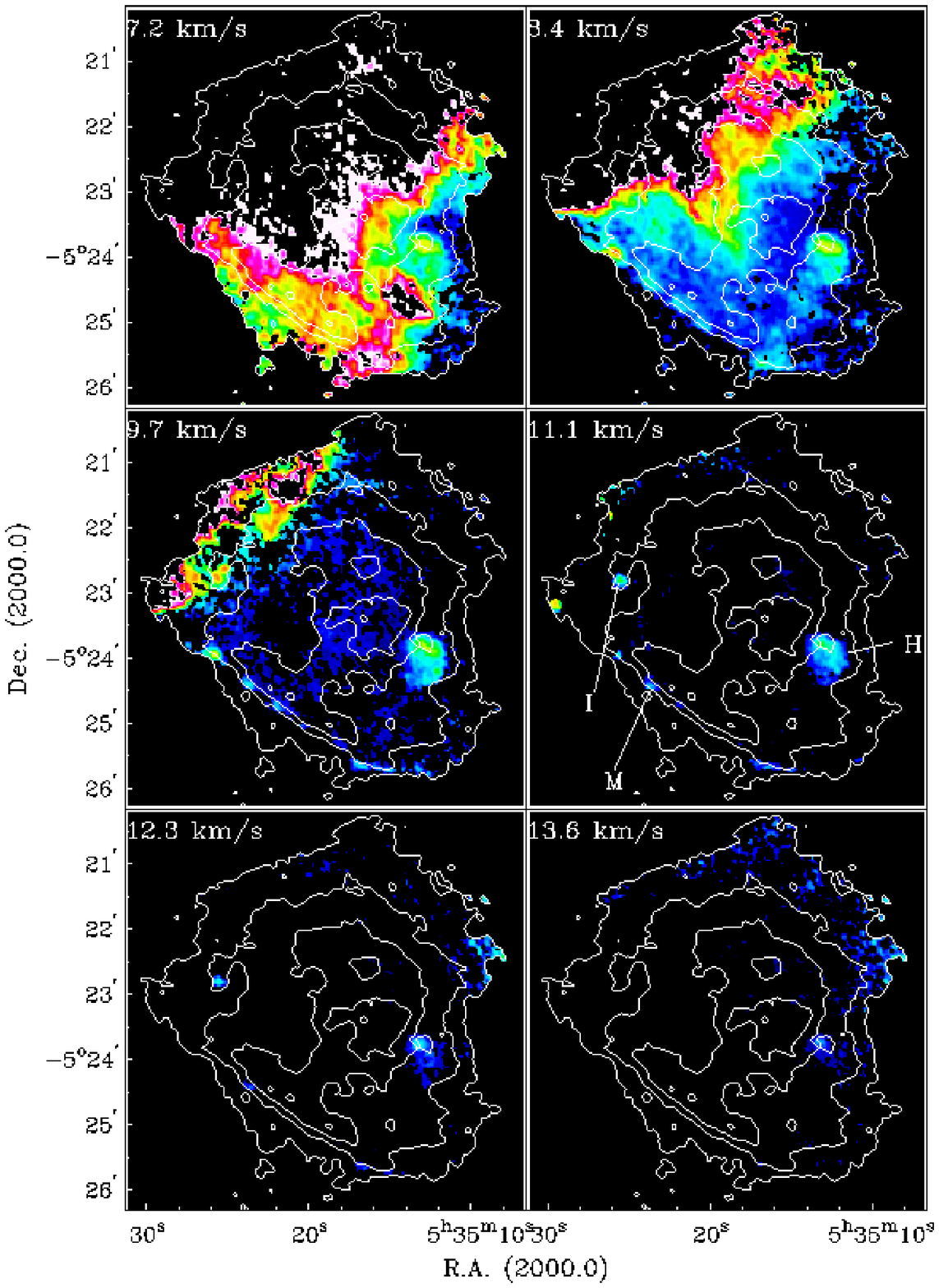}
\caption{{\it (continued)}
}
\end{figure*}
       
\subsection{$\HI$ opacity images}
\label{sec.channelmaps}

In order to study the morphology of the $\HI$ opacity as a function of
$\vLSR$, we present a sequence of opacity images. For presentation
purposes, the data cube was spectrally smoothed (ignoring saturated
pixels) to a channel resolution and separation of $1.28\kms$. A
sequence of these images is shown in \figref{fig.channelmaps}.  We
note that the smoothing in velocity was only done for the purpose of
creating these figures; all analysis was done on the full velocity
resolution data cube. The range $0.6\kms < \vLSR < 5.9\kms$ is heavily
affected by saturation.
Therefore, this region, covering part of component~A and most of
component~B, was omitted from \figref{fig.channelmaps}.

\subsubsection{Large-scale $\HI$ absorption features}

We focus first on the components~A, B and C identified by
vdWG89.  Referring to \figref{fig.spectrum}, the red line wing
of component~A is traced in the opacity image at
$\vLSR=8.4\kms$. A strong increase in opacity is observed towards the
Northeast Dark Lane, in agreement with the results of vdWG89.

Component~C of vdWG89, located at the southwestern edge of the Huygens
region, can be traced in the opacity images between $-4.4$ and
$-0.6\kms$. The morphology of this component is obvious at
$\vLSR=-3.2\kms$, where it displays a striking arc of high $\HI$
opacity tracing the extreme southwest edge of the Huygens region, and
broken into a number of $\HI$ peaks. Several other $\HI$ features are
detected in the same velocity range, most notably near the Bright Bar
and towards the northern part of Huygens region. However, as will be
discussed below (\twosecsref{sec.G}{sec.L}), those features are not
physically related to the prominent $\HI$ opacity arc at the southwest
edge of the Huygens region. Therefore in our nomenclature component ~C
will only denote this $\HI$ arc.

\begin{deluxetable*}{l c c c c c r c c l}
\tablecaption{Global parameters of small-scale $\HI$ components
\label{tab.smallscale}}
\tablehead{Name & R.A.\tablenotemark{a} & Dec.\tablenotemark{a} & $\vLSR$ range
  & \multicolumn{2}{c}{Diameter} & \multicolumn{2}{c}{$N(\HI)\left({\Ts\over
  100\K}\right)$} & $M_{\HI}\left({\Ts\over 100\K}\right)$ & Associated with\\
  & \multicolumn{2}{c}{[J2000]} & [km\,s$^{-1}$] & & [pc] &
  \multicolumn{2}{c}{[$10^{19}\pcmsqu$]} & [$M_\odot$]\\
 & & & & & & peak\tablenotemark{b} & mean}
\startdata
D\tablenotemark{c,d} & $\hmsd 5h35m23.0s$ & $\dms -5d24m30s$ & $-21.2\to-10.9$ & $\mind 1.5$ & 0.21
  & 100 & 6.6 & 0.07 & $\theta^2$B~Ori\\
E\tablenotemark{e} & $\hmsd 5h35m25.0s$ & $\dms -5d23m00s$ & $-10.9\to-1.9$ & $\mind 1.5$ & 0.21
  & 130 & 6.2 & 0.07 & Dark Bay\\
F\tablenotemark{c,d} & $\hmsd 5h35m14.0s$ & $\dms -5d23m00s$ & $-21.2\to-10.9$ & $40''$ & 0.09 & 24
  & 9.2 & 0.02 & $\HH{202}$\\
G1\tablenotemark{f} & $\hmsd 5h35m14.0s$ & $\dms -5d21m40s$ & $-9.6\to-1.9$ & $\mind 1.5$ & 0.21
  & $>350$ & 3.6 & 0.04 &\\
G2\tablenotemark{f} & $\hmsd 5h35m13.0s$ & $\dms -5d21m20s$ & $-12.2\to-8.3$ & $30''$ & 0.07 &
  $>350$ & 12 & 0.02 & $\Ht$ finger\\
G3\tablenotemark{f} & $\hmsd 5h35m17.5s$ & $\dms -5d21m40s$ & $-3.2\to-0.6$ & $\mind 1.5$ & 0.21
  & 27 & 2.7 & 0.03 &\\
H & $\hmsd 5h35m13.0s$ & $\dms -5d24m00s$ & $+7.2\to+12.3$ & $40''$ & 0.09 & 34 
& 9.9 & 0.02 & Orion-S\\
I & $\hmsd 5h35m25.6s$ & $\dms -5d22m49s$ & $+9.7\to+13.0$ & $8''$ & 0.02 & 23 &
  20 & 0.002 & extinction Knot~2\\
J\tablenotemark{f} & $\hmsd 5h35m22.0s$ & $\dms -5d24m52s$ & $-4.4\to-1.2$ & $30''$ & 0.07 &
  $>340$ & 3.3 & 0.04 & $\HH{203{/}204}$\\
K\tablenotemark{g} & $\hmsd 5h35m23.0s$ & $\dms -5d25m45s$ & $-5.7\to-0.6$ & $10''$ & 0.02 & & & & shock south of $\HH{204}$\\
L\tablenotemark{d} & $\hmsd 5h35m20.4s$ & $\dms -5d24m25s$ & $-5.7\to-0.6$ & $1'$ & 0.14 & $>330$
  & 10 & 0.05 & $\theta^2$A~Ori\\
M\tablenotemark{h} & $\hmsd 5h35m25.8s$ & $\dms -5d23m57s$ & $+7.8\to+11.7$ & $12''$ & 0.03 & 46 
& 4.5 & 0.001 & Bright Bar \\
\enddata
\tablenotetext{a}{Positions for extended features are approximate
  center positions.}
\tablenotetext{b}{Lower limits are affected by saturation.}
\tablenotetext{c}{Two velocity components}
\tablenotetext{d}{Arc}
\tablenotetext{e}{Ten compact clumps}
\tablenotetext{f}{Elongated}
\tablenotetext{g}{Column density and mass cannot be calculated due to saturation.}
\tablenotetext{h}{One of several compact clumps at the edge of the Bright Bar}
\end{deluxetable*}

\subsubsection{Small-scale $\HI$ absorption features}

To trace the small-scale $\HI$ absorption components, we begin at the
most negative velocities and follow the opacity images towards more
positive velocities.  Central positions and velocity ranges over which
these features are detected are summarized in
\tabref{tab.smallscale}. The components are also indicated in
\twofigsref{fig.channelmaps}{fig.HST}.  \Tabref{tab.smallscale} also
summarizes $\HI$ masses, and peak and average column densities, as
well as approximate total sizes of the various features. The $\HI$
masses and column densities scale with $\Ts$, here assumed to be
$100\K$ (see \secref{sec.DiscVeil}). For convenience, the formulae
relating the $\HI$ emission and absorption measurements for $\HI$
column densities, and the underlying assumptions, are summarized in
the Appendix.

The most negative velocity components in our data are components D (in
the region of the Bright Bar) and F (in the western part of the
Huygens region). Both are detected over a considerable velocity range
($-21.2<\vLSR<-10.9\kms$).  The more opaque component~D
displays obvious extended structure with embedded higher opacity
clumps.  Both components contain several velocity components, in
agreement with vdWG90.

Continuing to more positive velocities, component~E of vdWG90 is detected in the
region of the Dark Bay. The data shown by vdWG90 already indicated that this
component consists of several clumps, which are more prominent in the
present higher resolution data. Inspection of \figref{fig.channelmaps} in the
region of component~E reveals the presence of 10 compact $\HI$
opacity clumps.  At the resolution of $\sim6''$, the diameters of the unresolved
clumps are less than $0.014\pc$ or $2800\AU$.

The remaining blueshifted small-scale feature identified by vdWG90,
component~G, is detected as a conspicuous elongated feature towards the northern
part of M42 at velocities from $-4$ to $-11\kms$.

One compact feature was identified by vdWG90 at positive $\vLSR$
(their component~H). This feature is at the velocity of OMC-1 and it
is detected at $\vLSR=7-12\kms$ in \figref{fig.channelmaps} in the western
part of the Huygens region.

The data cube was carefully searched for additional velocity
components. No additional $\HI$ components were found at velocities
more negative than that of component~D, even though $\ion{Na}{i}$ and
$\ion{Ca}{ii}$ absorption line measurements towards the Trapezium
stars and $\theta^2$\,A~Ori reveal several of components at these
velocities, e.g., at $\vLSR\approx-32.8$, $-20.6$, $-17.0$, and
$-12.1\kms$ \citep{ODell.etal1993a}.  Likewise, no absorption was
found at velocities more positive than that of
component~H\null. However, our search revealed several new small-scale
features within the velocity range shown in \figref{fig.channelmaps},
spatially and kinematically distinct from the features described
above. Some of these features are quite compact, or are located close
to other features, which accounts for their non-detection in the lower
resolution data of vdWG90. Several features are quite close in
velocity to component~C, but spectra and position-velocity diagrams
reveal that these features are kinematically distinct. Their global
properties are summarized in \tabref{tab.smallscale}. These features
are best seen in PV diagrams discussed below.

\begin{figure*}
\includegraphics[width=\linewidth]{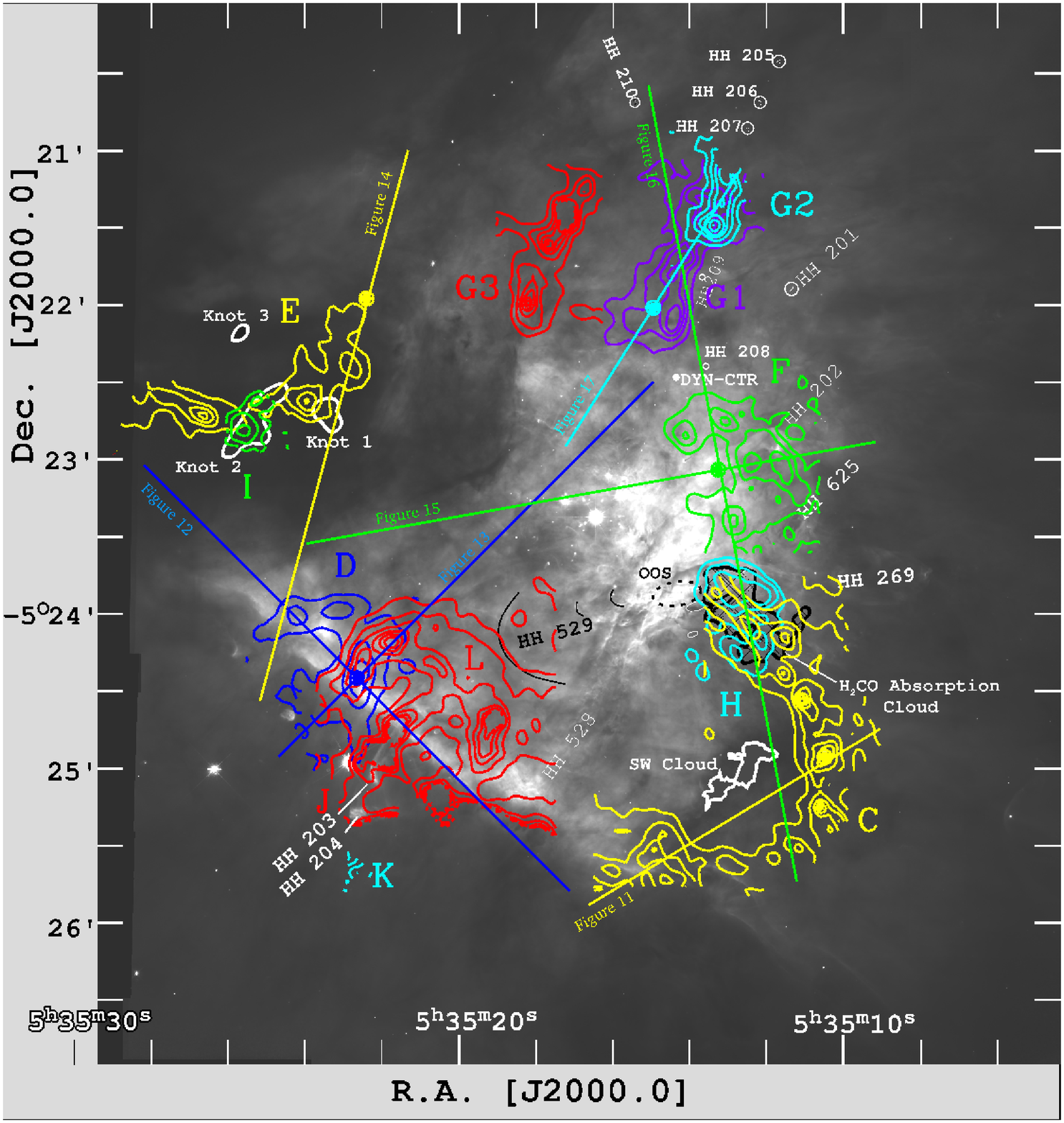}
\caption{Location of the position-velocity (PV) diagrams of $\HI$
  optical depth shown in \figsref{fig.PVC}{fig.PVG}, superposed on an
  HST/WFPC2 image (greyscale). The lengths, orientations and positions
  of these PV diagrams are indicated by the colored bars. In each bar,
  a thick dot of the same color indicates the zero position of the
  spatial coordinate in the corresponding PV diagram. Finally, contour
  sets indicate the $\HI$ opacity of the various $\HI$ features shown
  in \figsref{fig.PVC}{fig.PVG}. These contours are taken from the
  opacity images at $-2.5\kms$ (yellow contours in southwest,
  indicating component~C, and red contours near the Bright Bar,
  indicating components~J and L); $-17.3\kms$ (dark blue contours near
  the Bright Bar, indicating component D, and green contours
  indicating component~F); $-7.0\kms$ (yellow, component~E);
  $-7.7\kms$ (purple contours in northwest, indicating component G1);
  $-9.6\kms$ (light blue, component~G2); $-1.3\kms$ (red contours in
  the north, indicating component~G3); $8.4\kms$ (light blue contours
  southwest of the Trapezium stars, indicating component~H);
  $10.3\kms$ (green, component~I); and $1.9\kms$ (light blue contours
  in the south, indicating component~K).  The optical HST image is a
  combination of the WFPC2 F502N and F658N filter images from
  \citet{ODell.Wong1996}, which contain respectively the [$\OIII$]
  $5007\Ang$ and the [$\NII$] $6583\Ang$ and $\Ha$ lines, thus the
  extremes of both high and low excitation levels. The features
  indicated by ``SW Cloud'' and ``Knot 1$-$3'' are features of
  enhanced extinction identified in the extinction study of
  \citet{ODell.YusefZadeh2000}.  ``Dyn-ctr'' indicates the dyanmical
  center of the Orion-KL molecular outflow, discussed further in
  \secref{sec.KL}. ``$\HtCO$ absorption cloud'' denotes the outline of
  the $\HtCO$ absorption signal of Orion-S\null. The optical outflow
  source associated with Orion-S is indicated by ``OOS''. These features
  are discussed in \secref{sec.OrionS}.}
\label{fig.HST}
\end{figure*}

\begin{figure}
\includegraphics[width=\linewidth]{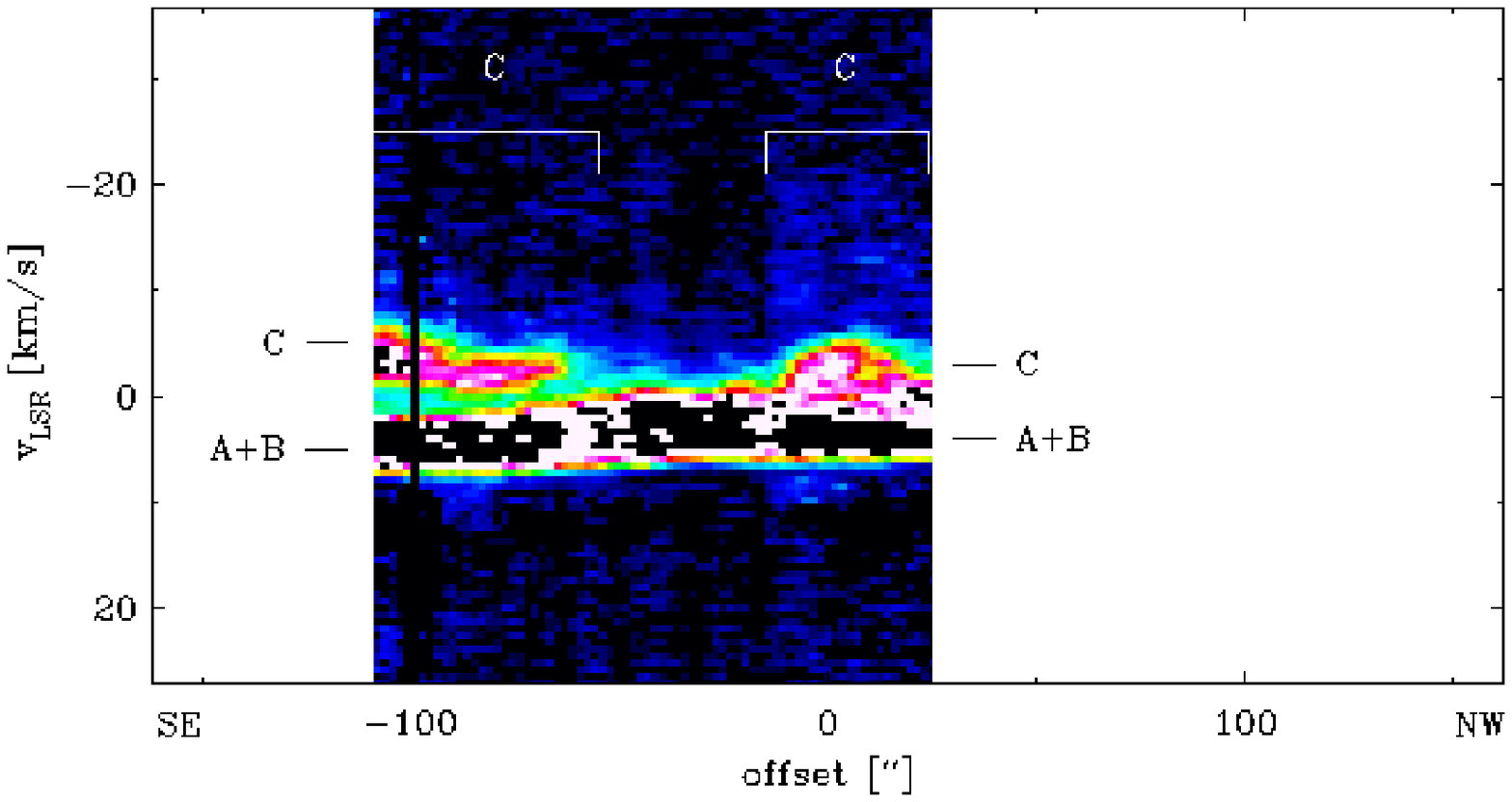}
\caption{Position-velocity diagram of $\HI$ optical depth along the
  yellow southeast-northwest bar shown in \figref{fig.HST} towards the
  southwest of the Huygens region.  The spatial axis has its zero
  position at $\RA=\hmsd 5h35m10.5s$, $\Dec=\dms -5d24m56s$ and a
  position angle of $120\deg$.  Spatial offsets are negative towards
  the southeast and positive towards the northwest.  Velocity
  components discussed in the text are indicated.}
\label{fig.PVC}
\end{figure}

\begin{figure}
\includegraphics[width=\linewidth]{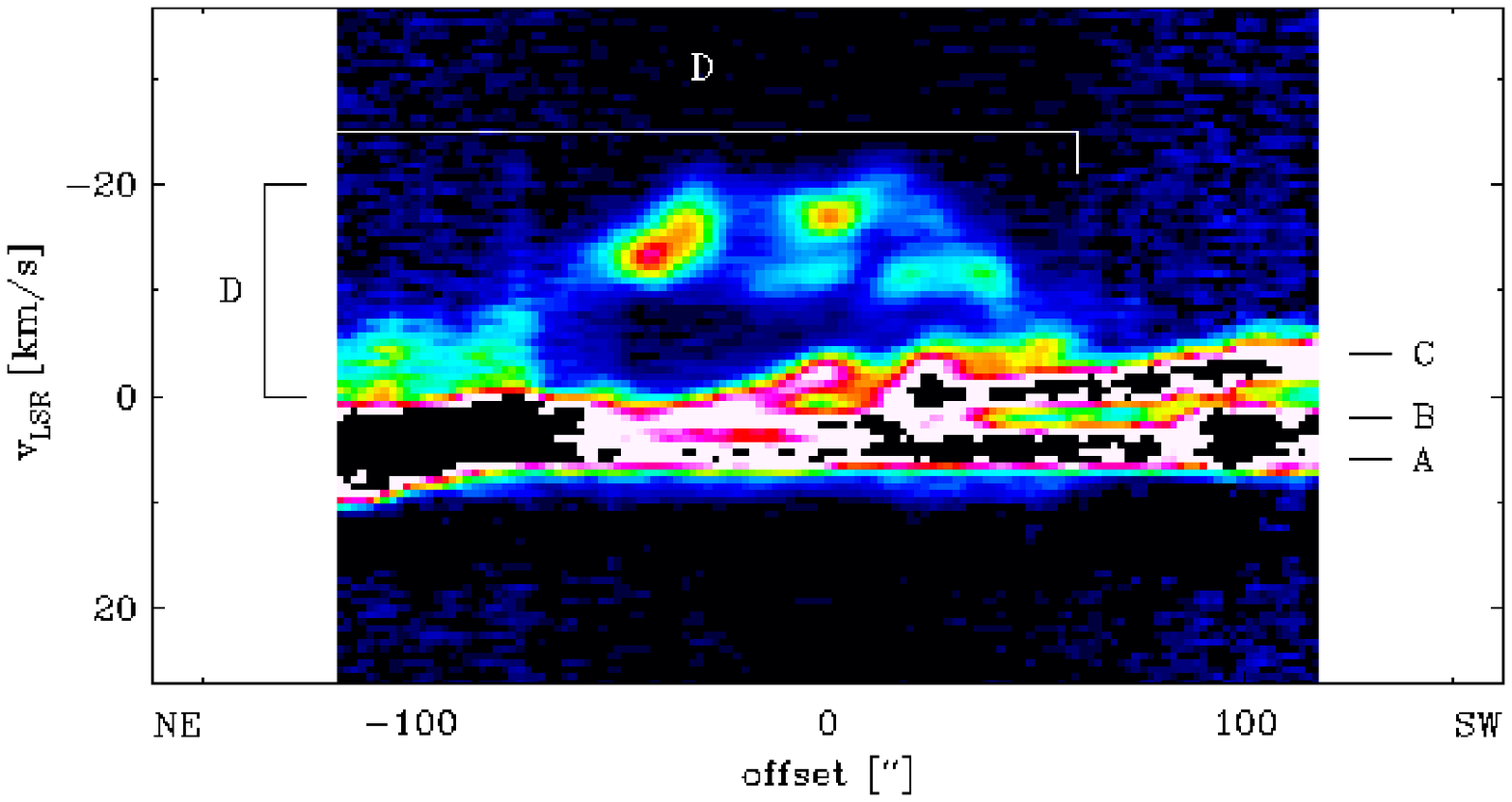}
\caption{Position-velocity diagram of $\HI$ optical depth along the
  dark blue northeast-southwest bar shown in \figref{fig.HST}.  The
  spatial axis has its zero position at $\RA=\hmsd 5h35m22.6s$,
  $\Dec=\dms -5d24m25s$ and a position angle of $45\deg$. Spatial
  offsets are negative towards the northeast and positive towards the
  southwest.  Velocity components discussed in the text are indicated.
}
\label{fig.PVD1}
\end{figure}

\begin{figure}
\includegraphics[width=\linewidth]{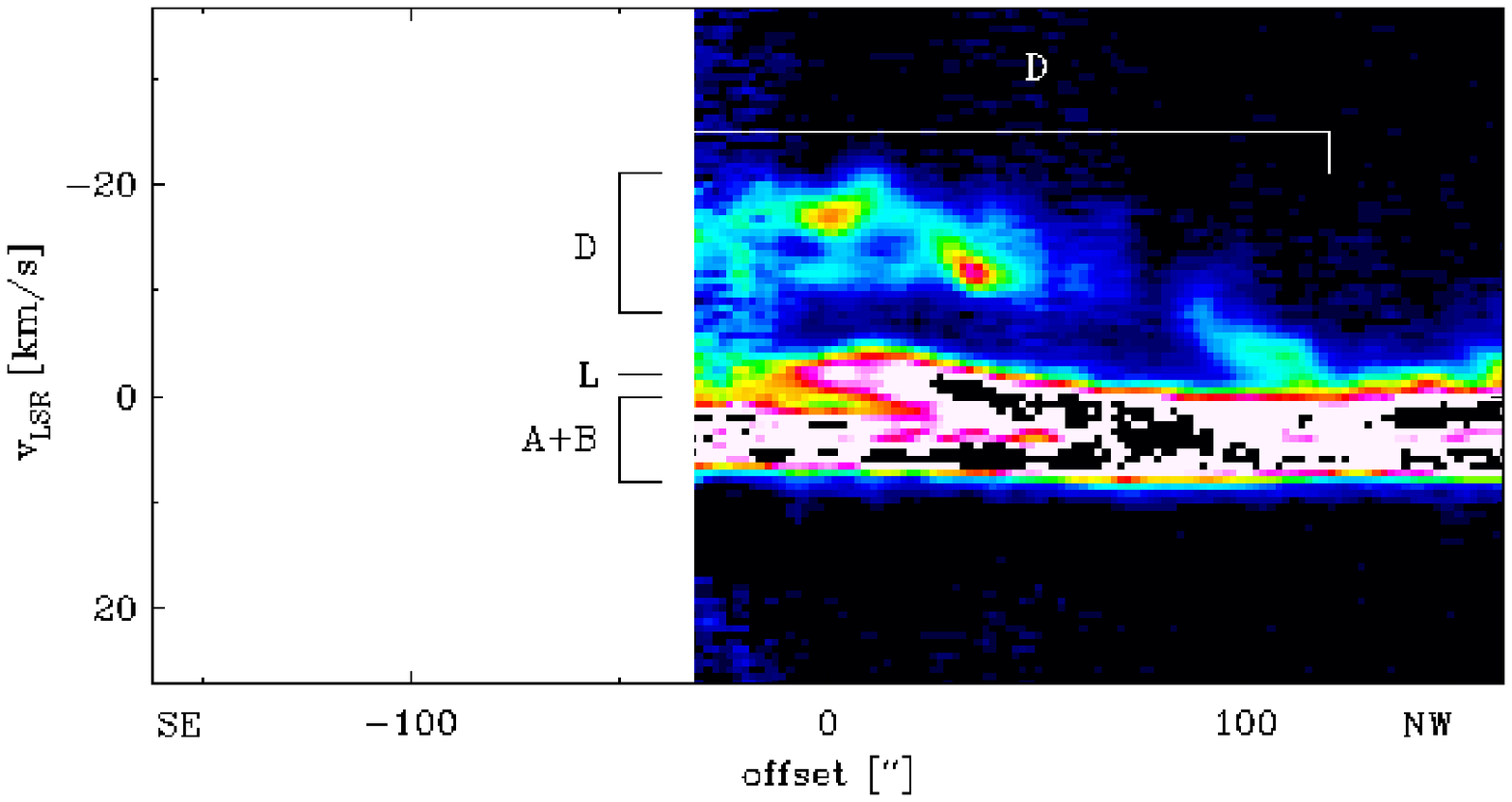}
\caption{Position-velocity diagram of $\HI$ optical depth along the
  dark blue southeast-northwest bar shown in \figref{fig.HST}.  The
  spatial axis has the same zero position as for \figref{fig.PVD1},
  but has a position angle of $135\deg$, perpendicular to the PV
  diagram shown in \figref{fig.PVD1}.  Spatial offsets are negative
  towards the southeast and positive towards the northwest.  Velocity
  components discussed in the text are indicated.}
\label{fig.PVD2}
\end{figure}

\begin{figure}
\includegraphics[width=\linewidth]{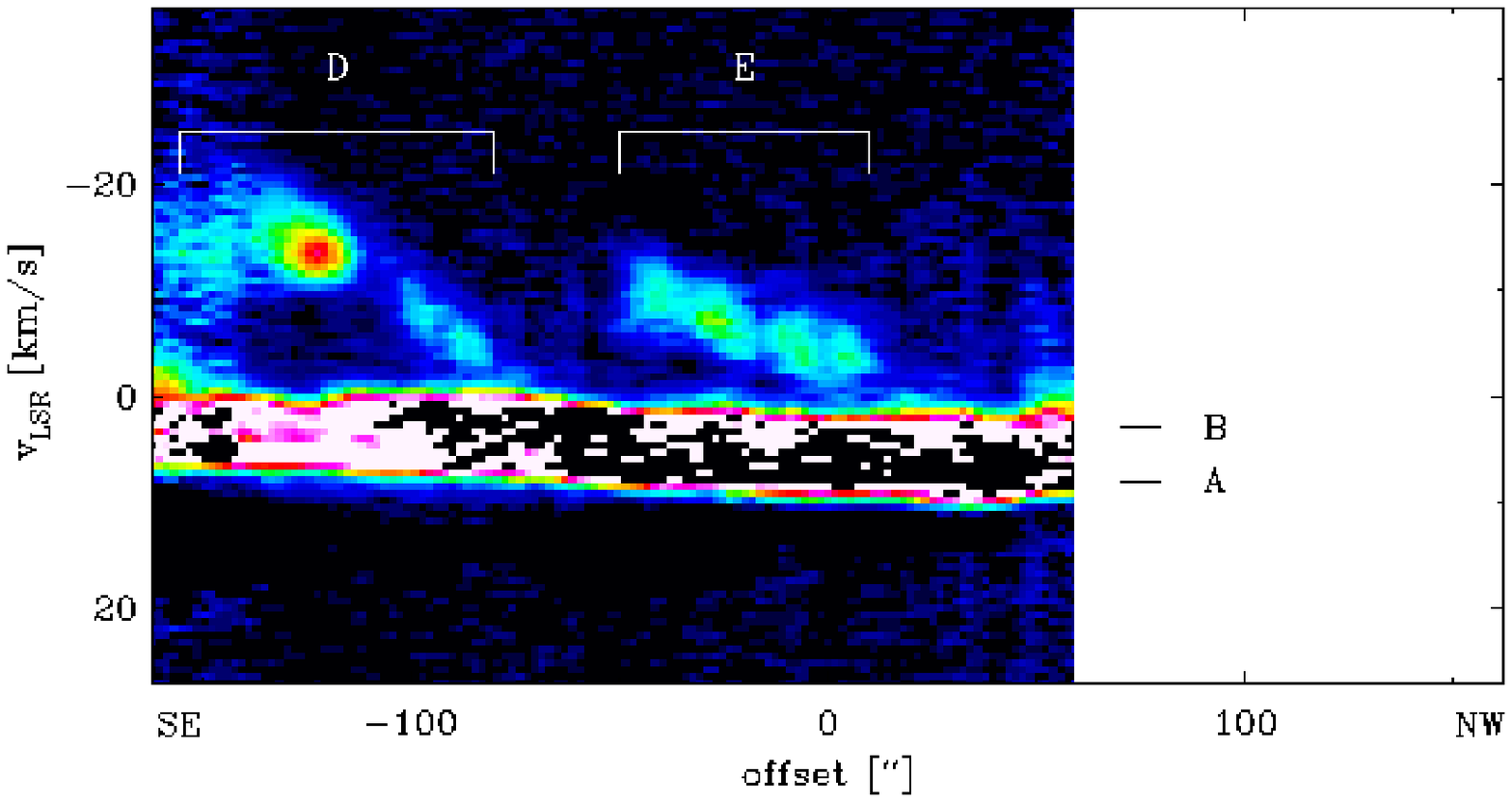}
\caption{Position-velocity diagram of $\HI$ optical depth along the
  yellow southeast-northwest bar west of the Trapezium stars, shown in
  \figref{fig.HST}.  The spatial axis has its zero position at
  $\RA=\hmsd 5h35m22.4s$, $\Dec=\dms -5d21m58s$ and a position angle
  of $345\deg$.  Spatial offsets are negative towards the southeast
  and positive towards the northwest.  Velocity components discussed
  in the text are indicated.}
\label{fig.PVE}
\end{figure}

\begin{figure}
\includegraphics[width=\linewidth]{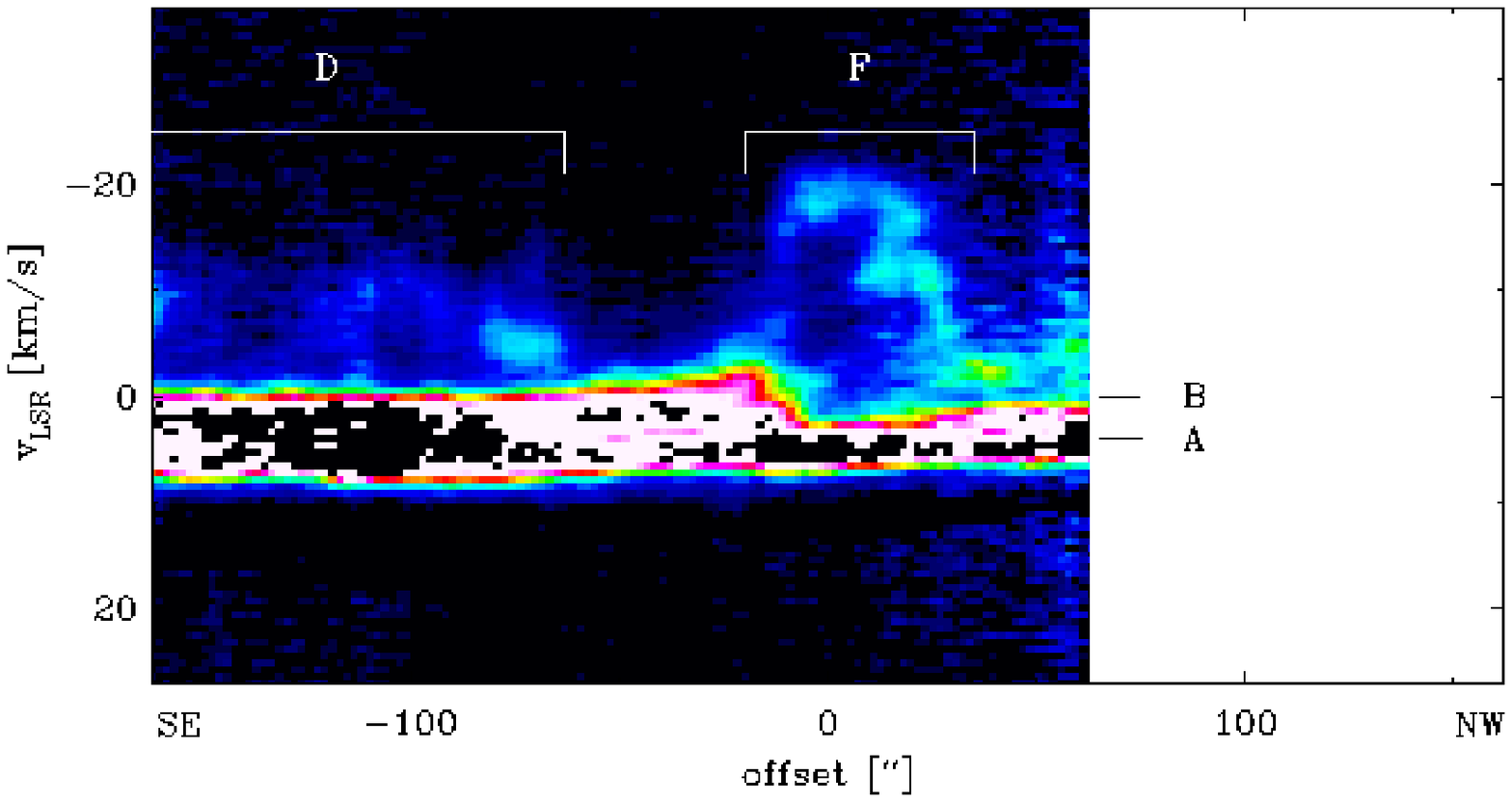}
\caption{Position-velocity diagram of $\HI$ optical depth along the
  green southeast-northwest bar shown in \figref{fig.HST}.  The
  spatial axis has its zero position at $\RA=\hmsd 5h35m13.2s$,
  $\Dec=\dms -5d23m04s$ and a position angle of $100\deg$. Spatial
  offsets are negative towards the southeast and positive towards the
  northwest.  Velocity components discussed in the text are indicated.
}
\label{fig.PVF}
\end{figure}

\begin{figure}
\includegraphics[width=\linewidth]{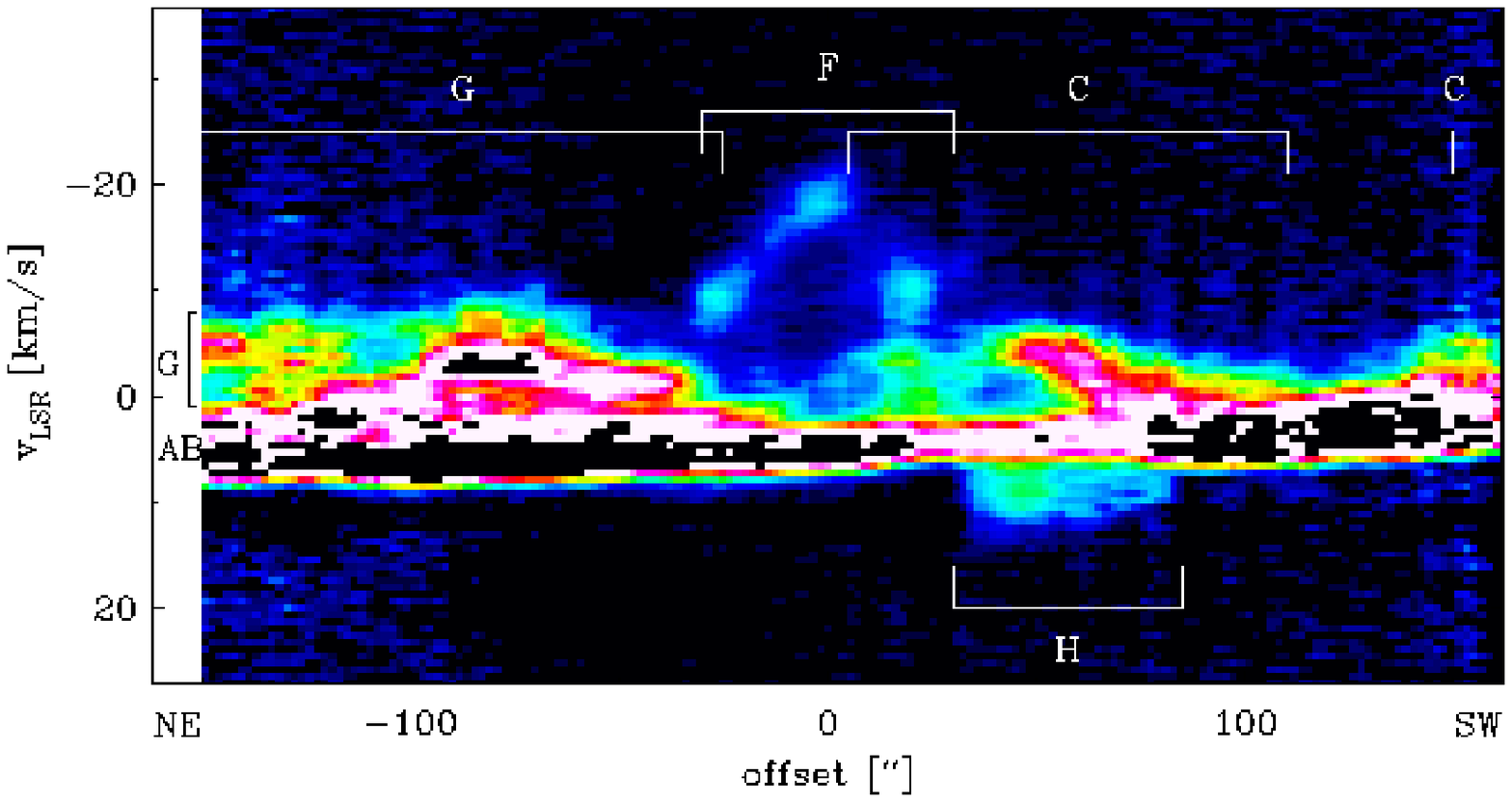}
\caption{Position-velocity diagram of $\HI$ optical depth along the
  green northeast-southwest bar shown in \figref{fig.HST}.  The
  spatial axis has the same zero position as for \figref{fig.PVF}, but
  has a position angle of $10\deg$, perpendicular to the PV diagram
  shown in \figref{fig.PVF}. Spatial offsets are negative towards the
  northeast and positive towards the southwest.  Velocity components
  discussed in the text are indicated.}
\label{fig.PVH}
\end{figure}

\begin{figure}
\includegraphics[width=\linewidth]{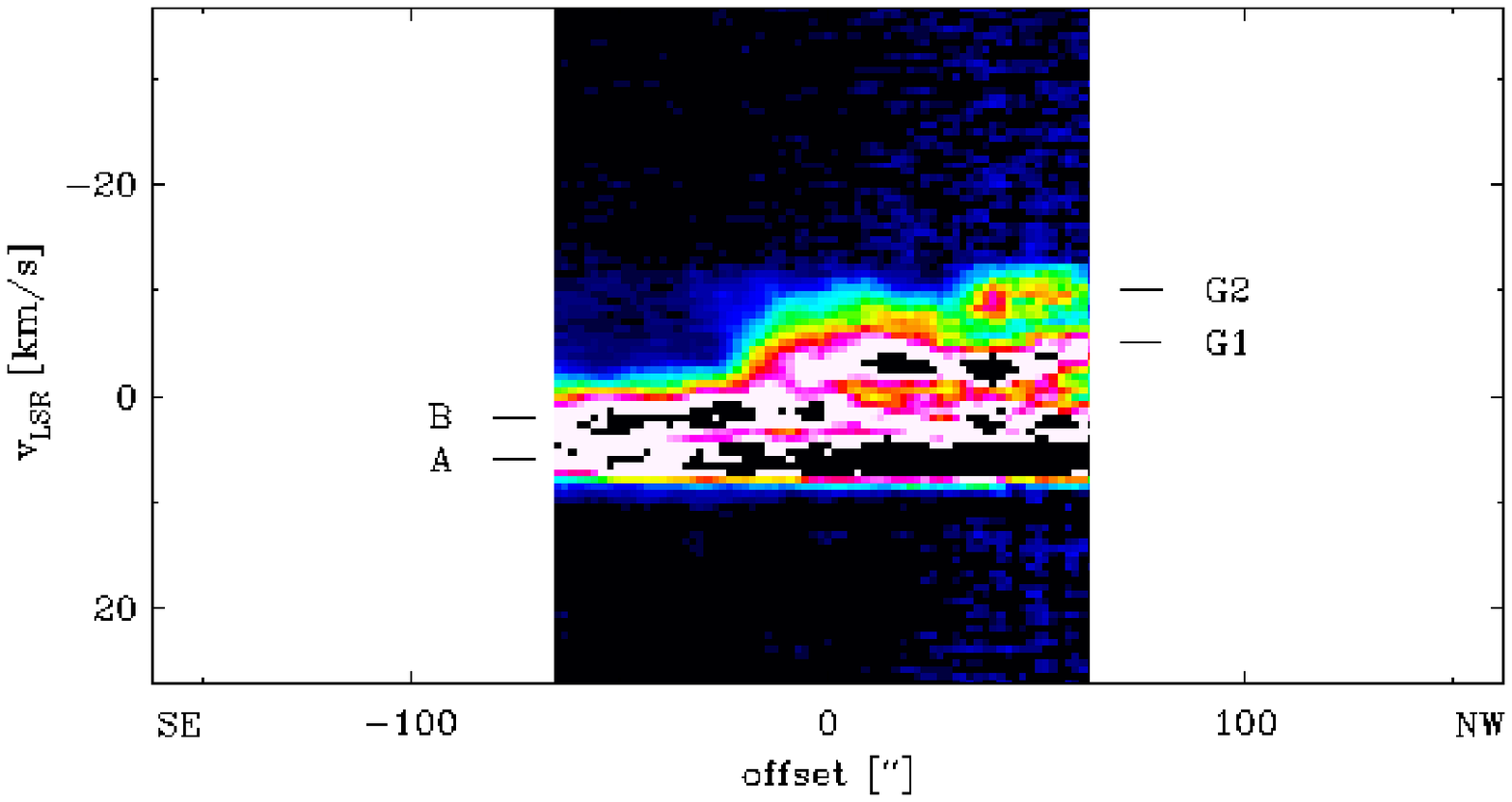}
\caption{Position-velocity diagram of $\HI$ optical depth along the
  light blue southeast-northwest bar shown in \figref{fig.HST}.
  The spatial axis has its zero position at $\RA=\hmsd 5h35m14.9s$,
  $\Dec=\dms -5d22m01s$ and a position angle of $328\deg$. Spatial
  offsets are negative towards the southeast and positive towards the
  northwest.  Velocity components discussed in the text are indicated.}
\label{fig.PVG}
\end{figure}

\subsection{Position-velocity diagrams of $\HI$ opacity}
\label{sec.PV}

The location of a set of illustrative PV diagrams is
indicated in \figref{fig.HST}. These diagrams are presented in
\figsref{fig.PVC}{fig.PVG}. The various $\HI$ components are indicated
in these figures.

\subsubsection{Components A, B and C}

\Figref{fig.PVC} shows the saturated signal of the large-scale
components~A and B at positive velocities. In addition, component~C is
detected at slightly negative velocities ($-4$ to $-2\kms$). As
already indicated by the channel map at $-3.2\kms$, \figref{fig.PVC}
shows that this component does not cover the entire nebula (in
contrast to components A and B), but is found only at the edge of the
Huygens region, where it forms a large arc in the shape of an
incomplete semicircle.

\subsubsection{Structure and extent of component~D}
\label{sec.D}

The velocity structure and extent of the most extended small-scale
component~D is shown in \twofigsref{fig.PVD1}{fig.PVD2}, which show
crosscuts through this feature in orthogonal directions.
\Figref{fig.PVD1}, which shows a PV diagram approximately along the
Bright Bar, shows that this component is extended over about
$160''$. The prominent high opacity features seen at the most negative
velocities appear to be connected to the main $\HI$ components through
gas at intermediate velocities. Inspection of \figref{fig.PVD2}, which
presents a PV diagram roughly perpendicular to the Bright Bar, shows a
velocity gradient, in the sense that the most negative velocities
occur to the southeast.  The overall velocity structure therefore
resembles that of an expanding shell.  This impression is reinforced
by the morphology of the absorbing $\HI$, in particular at velocities
between $-17.3$ and $-10.9\kms$, and also illustrated by the dark blue
contours in \figref{fig.HST}. \Figref{fig.PVD2} also shows that the
blueshifted gas of component~D connects to the larger scale $\HI$
layers through gas at intermediate velocities, located approximately
$20''$ east of the Trapezium stars.

\subsubsection{Component~E: $\HI$ in the Dark Bay}
\label{sec.E}

A PV diagram through component~E, which is located in the area of the
Dark Bay, is shown in \figref{fig.PVE}. While this component is
resolved into a number of compact opacity peaks, these are obviously
part of one coherent velocity structure. This diagram also
demonstrates that although in projection components~D and E are almost
contiguous, component~E is actually a kinematically separate feature.

\subsubsection{Stucture of the high velocity component~F}
\label{sec.F}

The PV diagrams of the high velocity component~F in
\twofigsref{fig.PVF}{fig.PVH} reveal clearly that the high (negative)
velocity gas is connected with the main $\HI$ components through
features at intermediate velocities. This is particularly clear in
\figref{fig.PVF}, where the high velocity gas (which shows several
velocity components) is connected to the lower velocity gas through a
prominent $\HI$ feature at its northwestern side. Careful inspection
of \figref{fig.PVF} and of the $\HI$ opacity data cube reveals that a
fainter connection between the high velocity gas and the large-scale
components is also present at the southeast side of the feature.

The fact that the location of component~F spatially coincides with a
gap in the extended gas layer at less negative velocities is
striking. This behavior argues that component~F is physically part of
the larger scale $\HI$ layer, but that it has been accelerated to
negative velocities. The situation is further illustrated by
\figref{fig.PVH}, which shows a PV diagram along a position angle
perpendicular to \figref{fig.PVF}. The connection with the lower
velocity gas is clearly observed on both sides of the feature, as well
as the gap in less negative velocity gas at the position
component~F\null.

Like component~D, the velocity structure of component~F resembles that
of an expanding bubble or shell, although this interpretation ignores
the fact that at some positions several velocity components are
present. However, component~F is much more confined than component~D,
with a diameter of about $60''$ ($0.14\pc$).

\subsubsection{The elongated component~G}
\label{sec.G}

A PV diagram over the long axis of the elongated component~G is shown
in \figref{fig.PVG}. Inspection of the line images in
\figref{fig.channelmaps} shows that multiple elongated $\HI$
absorption features are present towards the northern part of M42 at
the full velocity range from $-0.6$ to $-12.2\kms$, which we
collectively denote component~G\null. The PV diagram in
\figref{fig.PVG} shows a prominent feature at velocities of $-3$ to
$-4\kms$. However, at more negative velocities (approximately
$-10\kms$), an additional feature is detected towards the northern
part of the nebula. We denote the features at $\sim-3\kms$ G1 and the
feature at $\sim-10\kms$ G2, as indicated in \figref{fig.PVG}. The
latter feature matches component~G of vdWG90.  Finally, we note that a
parallel and similarly elongated $\HI$ absorption feature appears
$\sim45''$ east of component~G1 at velocities of $-1.9$ and $-0.6\kms$
in \figref{fig.channelmaps}; we denote the eastern feature
component~G3.

\subsubsection{The arclike component~L}
\label{sec.L}

At $\vLSR=-3.2\kms$ in \figref{fig.channelmaps}, a prominent
absorption feature is detected at approximately the location of the
Bright Bar. This feature, indicated by red contours in
\figref{fig.HST}, was included by vdWG89 in component~C, based on the
agreement in velocity. The higher resolution provided by the present
data however shows that this feature is distinct from component~C, as
will be discussed in \secref{sec.arclike}.  This feature, which we
denote by L, has an arc-like structure, open towards the southeast,
similar to the shape of component~D (indicated by dark blue contours
in \figref{fig.HST}). It reveals a velocity gradient with more
negative velocities towards the southeast (e.g., \figref{fig.PVD2}),
similar to component~D\null.

\subsubsection{The elongated component~J}

Southeast of component~L, and at approximately the same velocity
($\sim-3\kms$), an elongated $\HI$ feature is found in the opacity
images, crossing the Bright Bar orthogonally. A PV diagram through
this feature, which we label J, shows that it is kinematically
distinct from component~L, and at slightly more negative
velocities. Component~J is similar to component~G in appearing
elongated, and has a velocity gradient oriented along its long axis.

\subsubsection{Component~K}

A compact absorption feature at $\sim-3.2\kms$, which we label component~K, is
detected in \figref{fig.channelmaps} south of component~J\null.

\subsubsection{Features at the velocity of OMC-1}
\label{sec.HIAbsBackgr}

Several features at the velocity of the background molecular cloud
OMC-1 are found in the present data. Component~H was already
discovered by vdWG90 and is detected as an extended feature in
\figref{fig.channelmaps} at $7.2-11.1\kms$, and in the PV diagram in
\figref{fig.PVH}. In addition, a system of compact features is found
directly southeast of the Bright Bar and most likely associated with
it. These features can be seen in \figref{fig.channelmaps}, at a
velocity of $9.7\kms$, and we denote them collectively as
component~M\null. The most prominent of these is located at the
extreme eastern edge of the Bright Bar at $\RA=\hmsd 5h35m25.8s$,
$\Dec=\dms -5d23m57s$. Following the Bright Bar towards the southwest,
several similar features are detected. One of these components can be
seen in the sample spectrum shown in \figref{fig.spectrum} and the
$\HI$ emission PV diagram shown in \figref{fig.XVEm3}.

In the Dark Bay region, a single compact $\HI$ absoption feature is
found at the velocity of OMC-1. This component (component~I) can
clearly be seen in the opacity images at $11.1$ and $12.3\kms$. It was
also shown in the PV diagram of $\HI$ emission in \figref{fig.XVEm4}.

\section{$\HI$ emission associated with the Orion Nebula}
\label{sec.EmDisc}

The $\HI$ emission images and PV diagrams presented in
\secref{sec.HIEmission} show a number of separate features revealing
the neutral environment of the Orion Nebula and the effects of $\HII$
region and the ONC on this environment.
\begin{enumerate}
\item $\HI$ emission at approximately the velocity of the main
  absorbing components~A and B is detected in regions where strong
  absorption is absent. This layer, which is shown in the PV diagrams
  in \figsref{fig.XVEm1}{fig.XVEm3}, contains a bright $\HI$ emission
  feature, detected in \figref{fig.XVEm1} at a spatial offset of
  appoximately $150''$ This feature is therefore located between M42
  and M43 and may represent photodissociated gas outside an IF
  bounding M42 on the side of the Northeast Dark Lane. This region
  lies immediately northeast of sample 5-east in
  \citet{ODell.Harris2010}. Their Figure~1 shows that this region
  corresponds to an overlap of a northern protusion from the Dark Bay
  and the Northeast Dark Lane. Since $\HI$ emission could arise from
  both features, it is impossible to unambiguously assign the observed
  emission.
\item A prominent $\HI$ emission feature is detected in
  \figref{fig.EmissionMaps} southeast of the Bright Bar, i.e., the
  Orion Bar PDR\null. As shown in \twofigsref{fig.EmissionMaps}{fig.XVEm2},
  an elongated $\HI$ emission system extends from this feature towards
  the southwest at $\vLSR\sim14\kms$ along.
\item High-velocity $\HI$ emission arising from a small region is seen at
  $\vLSR\sim18\kms$ in \figref{fig.EmissionMaps} and in the PV diagrams in
  \twofigsref{fig.XVEm3}{fig.XVEm4}.
\item $\HI$ emission is detected from the Dark Bay region, and
  this feature is centered on the $\HI$ absorption component~I, as
  shown in \figref{fig.XVEm4}.
\end{enumerate}

\begin{figure}
\plotone{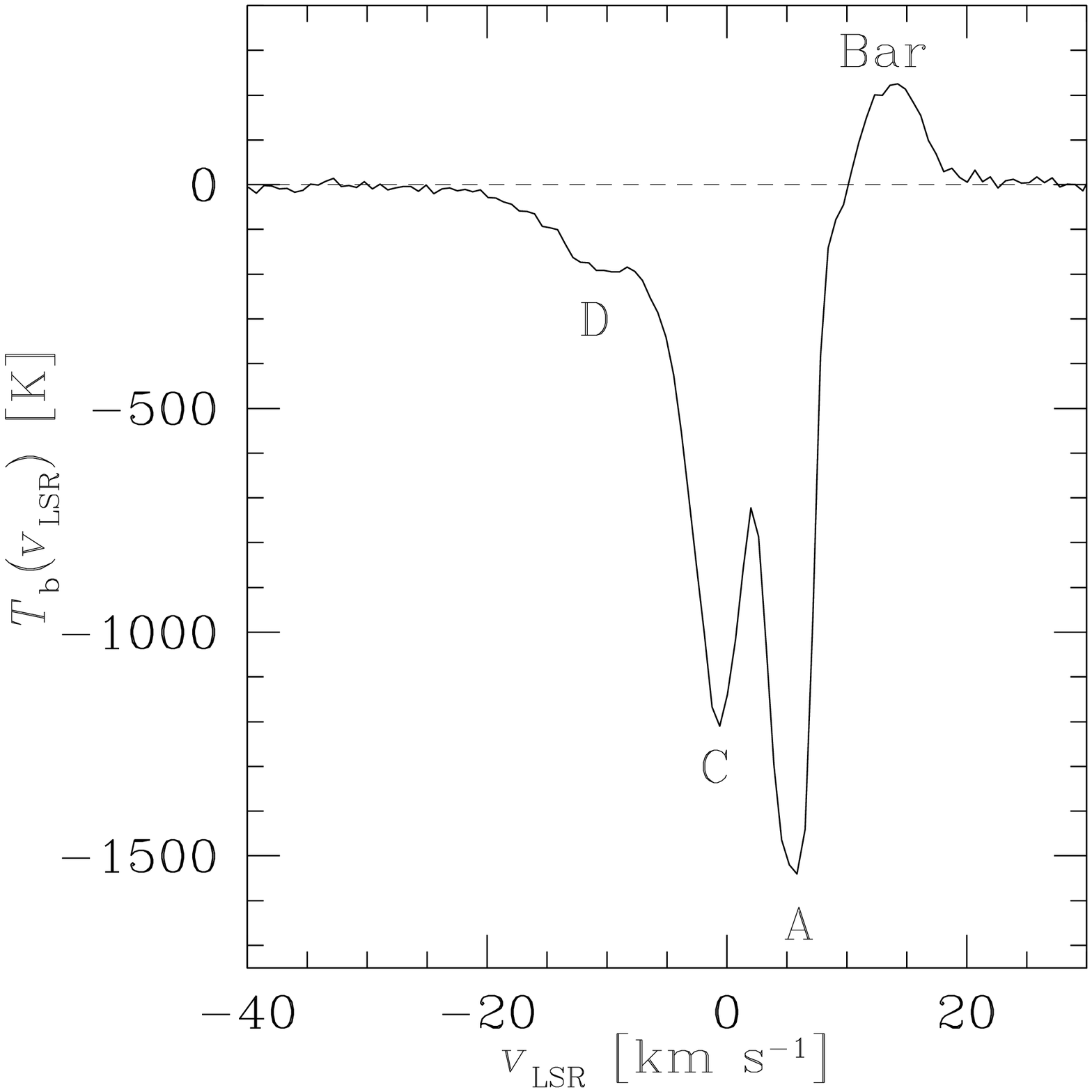}
\caption{Spectrum of $\HI$ brightness temperature $\Tb$, averaged over
  a $45''\times 36''$ rectangular region parallel to the Bright Bar
  (position angle $55\deg$), centered on RA=$\hmsd 5h35m21.3s$,
  Dec=$\dms -5d25m08s$. This region is indicated by a yellow
    rectangle in \figref{fig.EmFinder}.}
\label{fig.EmSpectrum}
\end{figure}

\begin{figure}
\plotone{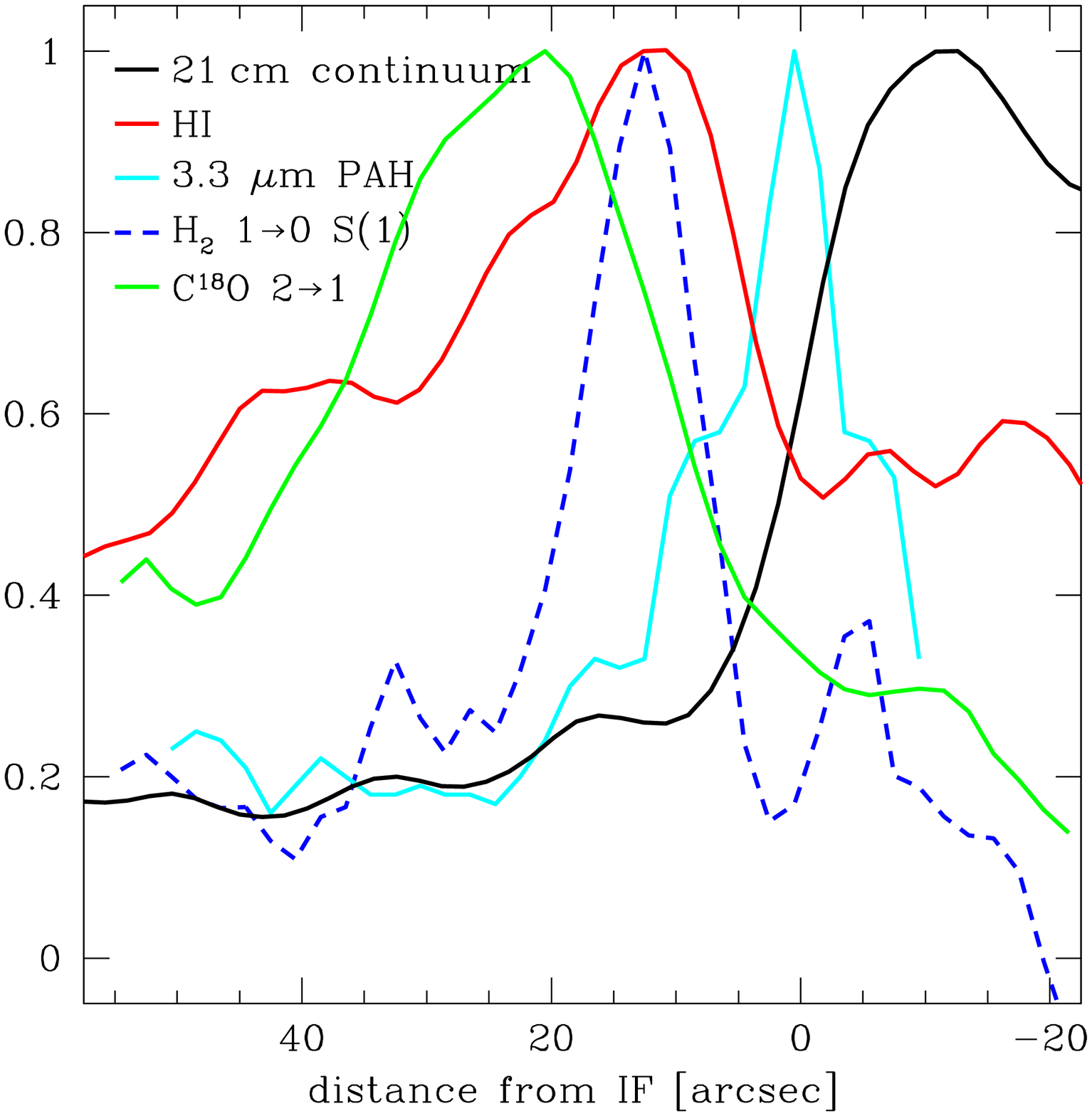}
\caption{Normalized crosscuts across the Bright Bar and the Orion
    Bar PDR in various tracers.  The crosscuts cover a length of
  $75''$ perpendicular to the Bar at a position angle of $145\deg$,
  and have been averaged parallel to the bar over a strip with a
  total width of $45''$. The center of this strip passes through
  coordinates $\RA=\hmsd 5h35m21.3s$, $\Dec=\dms -5d25m08s$, i.e., the
  central position for the spectrum shown in \figref{fig.EmSpectrum}.
  The location of this strip is shown by a purple rectangle in
  \figref{fig.EmFinder}, and it is identical to that used by
  \citet{VanderWerf.etal1996} to construct their Fig.~9. The
  abscissa shows distance (positive towards the southeast) from the
  IF, where the 0 position has been chosen as the point of maximum
  $3.3\mum$ PAH emission, which marks the location of the IF\null. The
  $21\cm$ continuum and $\HI$ brightness temperature data are
  from the present paper, while the $3.3\mum$ PAH data are from
  \citet{Tielens.etal1993} and \citet{Bregman.etal1994}. The $\Ht$
  $v=1{-}0$ S(1) and $\CO{18}$ $2{-}1$ data are from
  \citet{VanderWerf.etal1996}.}
\label{fig.BarStrips}
\end{figure}

\subsection{$\HI$ emission from the Orion Bar PDR}
\label{sec.BarPDR}

The brightest $\HI$ emission in \figref{fig.EmissionMaps} is found directly
southeast of the Bright Bar, thus arising in the prominent edge-on PDR\null. An
$\HI$ emission spectrum, averaged over the region shown by the yellow box in
\figref{fig.EmFinder}, is shown in \figref{fig.EmSpectrum}.  This spectrum is
very similar to the $\HI$ spectrum of this region obtained almost 40 years
earlier with the Parkes $64\un{m}$ telescope shown in Fig.~6 of
\citet{Radhakrishnan.etal1972}.

Our observation of $\HI$ emission from the Orion Bar probes a significantly
younger PDR, with much higher gas density than previous studies of PDRs
associated with evolved $\HII$ regions such as $\NGC{281}$
\citep{Roger.Pedlar1981}, $\NGC{1579}$
\citep{Dewdney.Roger1982,Dewdney.Roger1986}, $\IC{5146}$
\citep{Roger.Irwin1982}, S187 \citep{Joncas.etal1992}, S185 \citep[which
contains the well-studied PDR $\IC{63}$][]{Blouin.etal1997} and S270
\citep{Roger.etal2004}.  Due to the high resolution and the edge-on orientation
of the Orion Bar, our observations offer for the first time the opportunity to
use $\HI$ to study the stratified structure of such a PDR\null. We therefore
construct a crosscut perpendicular to the bar in various tracers, as shown in
\figref{fig.BarStrips}).  This figure dramatically confirms the layered
structure of the PDR, and for the first time observationally pinpoints the
location of atomic hydrogen in a resolved edge-on PDR\null. The edge-on IF is
marked by the peak in $3.3\mum$ PAH emission, which is strongly excited in the
neutral UV-exposed layer directly outside the IF
\citep{Tielens.etal1993,Kassis.etal2006}.

The most important parameter in determining the thickness of a
homogeneous PDR and therefore the separation of the various tracers in
\figref{fig.BarStrips} is the {\it dissociation parameter\/}
$\chi/\nH$
\citep[e.g.,][]{Tielens.Hollenbach1985a,Sternberg1988,Burton.etal1990b,Draine.Bertoldi1996,Hollenbach.Tielens1999},
where $\chi$ represents the intensity of the incident UV radiation
field, and $\nH$ is the number density of hydrogen nuclei. The
structure can be modified if the gas is clumpy
\citep{Burton.etal1990b,Meixner.Tielens1993,Tauber.etal1994,YoungOwl.etal2000},
since in this case UV photons penetrate deeper into the cloud through
the interclump medium \citep[e.g.,][]{Boisse1990}.
For the Orion Bar
PDR, the incident UV radiation field at the IF is
$\chi=2.6\times10^4$, where $\chi$ is expressed in units of the
\cite{Draine1978} interstellar radiation field of
$2.7\times10^{-3}\esc$. This value of $\chi$ was derived directly from
the properties of the Trapezium stars and their distance from the Bright Bar
\citep{Tielens.Hollenbach1985b}.  Analysis of the stratified structure
then yields a gas density $\nH=1-2\times10^5\pcmcub$,
\citep{Tielens.etal1993,Simon.etal1997,VanderWiel.etal2009}, with
higher density (up to $\nHt=10^7\pcmcub$) embedded clumps
\citep{Tauber.etal1994,VanderWerf.etal1996,YoungOwl.etal2000,Lis.Schilke2003}.
We now discuss our $\HI$ observations of the Orion Bar in the context
of this model.

As shown in \figref{fig.EmSpectrum}, the peak $\HI$ brightness
temperature in the Orion Bar PDR is approximately $230\K$, and the
$\HI$ spin temperature and kinetic temperature are therefore at least
this high. Such temperatures are easily reached in PDRs exposed to an
intense UV radiation field.  For the values of $\chi$ and $\nHt$
applicable to the Orion Bar, PDR models predict gas kinetic
temperatures of approximately $1000\K$ at the UV-exposed surface
\citep{LePetit.etal2006,Meijerink.etal2007,Kaufman.etal1999},
decreasing to values of a few hundred K at larger distances
($>0.01\pc$) from the IF\null. An upper limit for the kinetic
temperature follows from the observed linewidth, using
\begin{equation}
\qu{\Delta v}{FWHM}=2\sqrt{2\ln 2{k\qu{T}{kin}\over \mHI}},
\end{equation}
where $\mHI$ is the mass of the $\HI$ atom.
While the low
velocity side of the emission feature in \figref{fig.EmSpectrum} may
be affected by absorption, the high velocity side appears to be
unaffected, and at that side the HWHM of the line is $2.6\kms$,
corresponding to a FWHM of $5.2\kms$. This line width implies tha
$\Ts<590\K$. The turbulent linewidth in this region, as measured from
the optically thin $\CO{17}$ $3{-}2$ line \citep{Johnstone.etal2003}
is $1.3\kms$. Allowing for this turbulent velocity gives a thermal
FWHM of $5.0\kms$, implying $\Tkin=540\kms$.  Since the peak
brightness temperature is $230\K$, the implied peak optical depth is
$\tau_0=0.6$, so the $\HI$ line emission is only marginally optically
thin.  The implied $\HI$ column density is
$N(\HI)=3\times10^{21}\pcmsqu$, averaged over the region used to
construct \figref{fig.EmSpectrum}. The line of sight molecular
hydrogen column density in the bar is
$N(\Ht)\approx6.8\times10^{22}\pcmsqu$
\citep{Hogerheijde.etal1995,Jansen.etal1995,VanderWerf.etal1996,VanderWiel.etal2009},
so that only about 2\% of all hydrogen nuclei are in atomic form.
Since the dense clumps in the PDR will likely remain molecular, it is
more appropriate to consider only the interclump medium with
$\nH=1-2\times10^5\pcmcub$ for calculating the local $\HI$
abundance. The line-of-sight depth of the IF in the Bar region is
approximately $0.12\pc$, as derived from modeling by
\citet{Pellegrini.etal2009}. Using these numbers, the $\HI$ abundance
in the Bar is $5-10\%$ at the position of the $\HI$ peak.

The derived atomic fraction indicates that the observed $\HI$ emission
originates in the region where the transition towards molecular gas occurs. This
conclusion is supported by the $\Ht$ temperature in this region from rotational
lines, which is $400-700\K$ \citep{Allers.etal2005,Shaw.etal2009}, in excellent
agreement with our estimate of $540\K$ for the $\HI$.

Inspection of \figref{fig.BarStrips} reveals strong $\HI$ emission
from the region where $\Ht$ vibrational line emission shows a
maximum. This agreement is physically significant. Direct $\Ht$
photodissociation from the ground state is strongly forbidden, since
the molecule is homonuclear.  The actual
photodissociation of $\Ht$ is a two-step process, initiated by the
absorption of UV photons in the Lyman or Werner bands, as first
proposed in 1965 by Solomon \citep[private communication
  in][]{Field.etal1966}.  The UV absorption is followed by a radiative
cascade, in which there is an 11\% chance of dissociation
\citep{Stecher.Williams1967}. In the remaining 89\% of cases the
molecule cascades down to the ground state, and the resulting
fluorescent photon emission produces the $\Ht$ vibrational lines
\citep[e.g.,]{Black.VanDishoeck1987,Sternberg1988,Sternberg.Dalgarno1989}. The
strong $\HI$ emission from the region of maximum $\Ht$ vibrational
line emission therefore directly supports the photodissociation
mechanism, and our data reveal this agreement for the first time.

At larger distances from the IF, the $\HI$ brightness temperature decreases
slowly. The optically thin $\CO{18}$ $J=2{-}1$ emission peaks at a larger
distance from the IF than the $\HI$ emission ($21''$ or $0.05\pc$). The
brightest $\HI$ emission is thus located in the region where CO is
photodissociated.  In this region the gas-phase carbon is singly ionized, and
detected through the [$\CII$] $158\mum$ fine structure line
\citep{Stacey.etal1993,Herrmann.etal1997} and recombination lines in the radio
\citep{Jaffe.Pankonin1978,Natta.etal1994,Wyrowski.etal1997} and in the
near-infrared \citep{Walmsley.etal2000}.  The slower decrease of the $\HI$
brightness temperature towards larger distances from the IF compared to $\Ht$
$v=1{-}0$ S(1) results from the fact that collisional excitation contributes to
the flux of this $\Ht$ line \citep{VanderWerf.etal1996}. Observations of the
fainter $\Ht$ $v=2{-}1$ S(1) line, which is dominated by UV-pumped fluorescence
reveal a slower decline from the IF
\citep{Hayashi.etal1985,VanderWerf.etal1996,Luhman.etal1998,Marconi.etal1998,Walmsley.etal2000},
matching the decreasing $\HI$ brightness temperature in the same region.

The Orion Bar arises from an escarpment protruding from the OMC-1
molecular cloud towards the observer. Southeast of the Bright Bar the
IF curves back to a more face-on aspect \citep{Wen.ODell1995}. This
geometry is illustrated schematically for instance in Fig.~3 of
\citet{Pellegrini.etal2009} and Fig.~13 of
\citet{ODell.Harris2010}. The present $\HI$ emission results show that
the northeast-southwest extent of this escarpment is much larger than
the length of the bright IF in the Huygens region. As the highly
elongated $\HI$ emission feature at $\vLSR=14.9\kms$
(\figref{fig.EmissionMaps}) shows, the escarpment extends
significantly beyond the Bright Bar towards the southwest. This $\HI$
emission feature can be seen as a separate velocity component at
$\vLSR\sim15\kms$ in \figref{fig.XVEm2}, extending over about $11'$
($1.5\pc$). This $\HI$ feature is located in the EON, which is much
less well studied than the bright Huygens region.  The extended $\HI$
feature has a counterpart in a similarly extended feature observed in
$8\mum$ emission using IRAC on the {\it Spitzer Space Telescope\/},
representing emission from PAHs at 7.6, 7.8 and
$8.6\mum$\footnote{http://www.spitzer.caltech.edu/images/1648-ssc2006-16b-The-Sword-of-Orion}.
An IF bounding the elongated $\HI$ feature on its northern side is
detected in the wide-field optical images of the Orion Nebula obtained
with ACS/HST \citep{Henney.etal2007}.  This geometry confirms that the
Orion Bar PDR, and thus the escarpment from OMC-1, extends
significantly beyond the bright section in the Huygens Region.

Southeast of the brightest section of the Orion Bar PDR the IF curves
back to a more face-on aspect
\citep{Wen.ODell1995,Hogerheijde.etal1995,Jansen.etal1995,Pellegrini.etal2009,Ascasibar.etal2011}.
A recent Spitzer IRS study has revealed [$\ion{Ne}{iii}$] $15.56\mum$
and other ionized gas lines out to $12'$ southeast of the Trapezium
stars, i.e., far beyond the Bright Bar \citep{Rubin.etal2011}.
Our data reveal an extended $\HI$ cloud in this region with a central
velocity $\vLSR\sim13\kms$. This feature may represent atomic gas in
the extended PDR with a more face-on aspect here than in the
bright Bar region.  All details of the geometry of the Orion Bar
region are represented in the diagram shown in the upper panel of
Fig.~13 of \citet{ODell.Harris2010}.

\subsection{$\HI$ emission from the Dark Bay region}
\label{sec.DarkBay}

The $\HI$ emission images in \figref{fig.EmissionMaps} also reveal a
$\sim45''$ diameter $\HI$ emission feature at velocities up to
$17\kms$ located in the Dark Bay area. The peak of this feature is at
$\RA=\hmsd 5h35m24.0s$, $\Dec=\dms -5d22m30s$ at $14.9\kms$; the
feature extends towards the southeast where it crosses the compact
$\HI$ absorption feature I at $\RA=\hmsd 5h35m25.4s$, $\Dec=\dms
-5d22m38s$. Given the excellent agreement both in velocity and
position, as shown in \figref{fig.XVEm4}, the emission and absorption
feature are almost certainly physically related.

Given that the continuum brightness temperature of the $\HII$ region
in this area is $\Tc\sim1500\K$, the emitting $\HI$ must be located
behind the ionized gas, and is therefore not associated with the Dark
Bay, which represents a tongue of absorption in front of the $\HII$
region, with a velocity $\vLSR\sim 6\kms$ as measured from radio
recombination lines of partly ionized gas
\citep{Jaffe.Pankonin1978}.
  
The $\HI$ emission in the Dark Bay area is located east of the region
where the IF curves to a more edge-on orientation, as shown by
\citet{Wen.ODell1995}.  This location suggests a geometry analogous to
that of the Bright Bar, discussed in \secref{sec.BarPDR}. This idea is
supported by the presence of a $\CO{13}$ $J=3{-}2$ emission feature in
this region, approximately coinciding in orientation and extent with
the $\HI$ emission \citep[see Fig.~12 of][]{Buckle.etal2012}. Dust
emission from this region has been detected with SCUBA at $850$ and
$450\mum$ \citep{Johnstone.Bally1999}. Thus the $\HI$ emission in the
Dark Bay may originate in an approximately edge-on PDR located behind
the $\HII$ region.  The $\HI$ absorption feature~I, which is
associated with the emission, most likely represents a dense region of
this PDR\null. The expanding IF will propagate more slowly where it
encounters dense neutral gas, creating a concave region in the
approximately edge-on IF\null. The $\HI$ in this concave region then
has a background radio continuum, giving rise to a detection in
absorption. The dust in component~I will obscure the section of
the IF lying behind it (as seen from Earth). Therefore this model also
predicts enhanced extinction at the position of the absorbing
clump. As shown in \figref{fig.HST}, this enhanced extinction is in
fact observed, since $\HI$ component~I corresponds accurately in
position, extent and orientation with extinction ``Knot 2'' of
\citet{ODell.YusefZadeh2000}.

\begin{figure*}
  \includegraphics[angle=270,width=\textwidth]{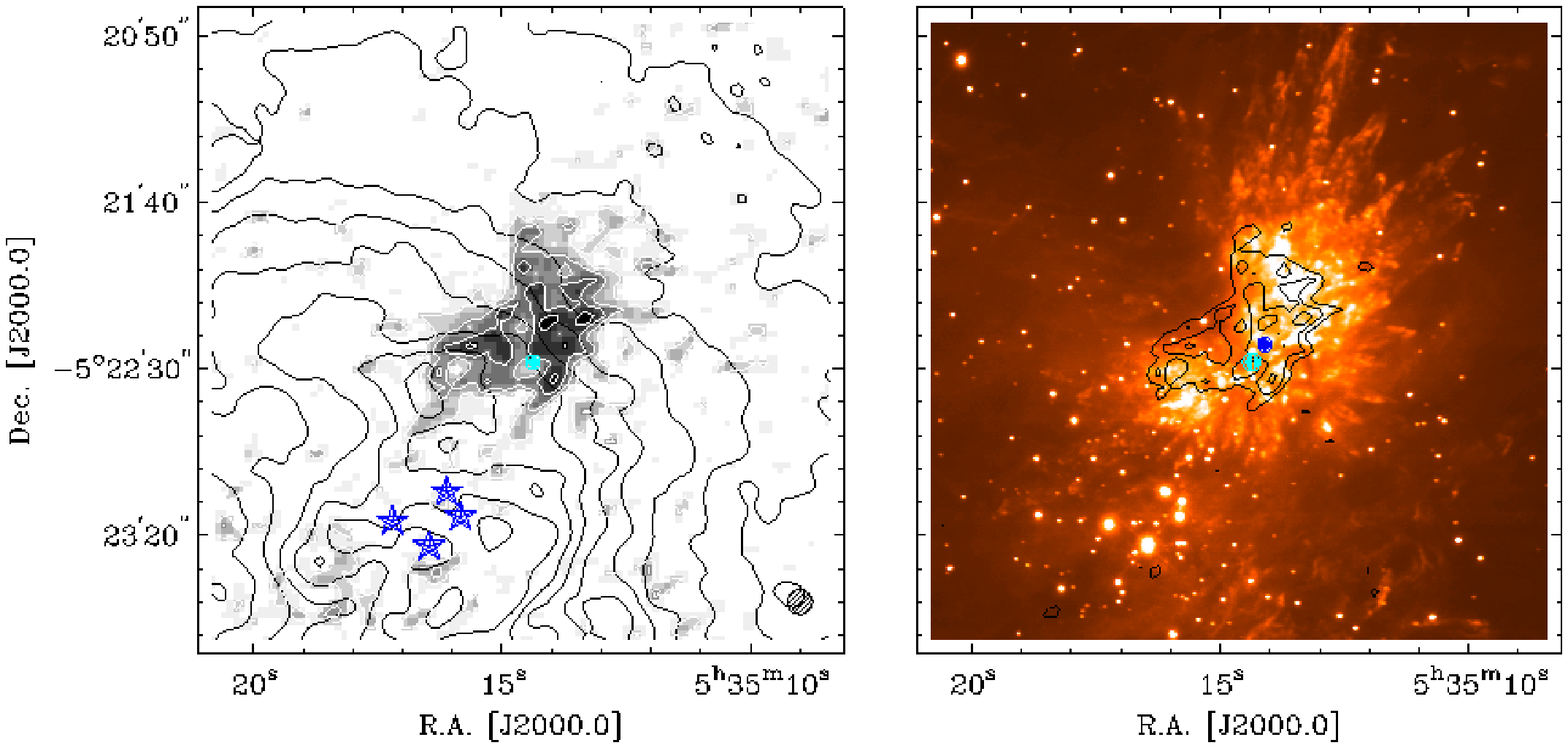}
\caption{Left panel: image of $\HI$ emission associated with Orion-KL,
  integrated over $18.7<\vLSR<31.\kms$ (grayscale with white
  contours). Black contours indicate $1420\MHz$ continuum surface
  brightness levels of 0 to $0.5\mJy\per{beam}$ in steps of
  $0.04\mJy\per{beam}$.  The cyan circle indicates the origin of the
  high velocity outflow as derived from the CO observations
  \protect\citep{Zapata.etal2009}.  The dark blue stars represent the
  Trapezium stars.  The resolution of the $\HI$ data is indicated by
  the $\secd 7.5$ circular beam in the lower righthand corner of the
  diagram. Right panel: the same $\HI$ emission image represented by
  black contours, superposed on the $\Ht$ $v=1{-}0$ S(1) image of
  \protect\citet{Bally.etal2011}. The cyan dot is as in the left
  panel, and the blue dot represent the position, taken from
  \protect\citet{Goddi.etal2011}, of the Becklin-Neugebauer object.}
\label{fig.BN}
\end{figure*}

\subsection{High velocity $\HI$ emission from the Kleinmann-Low region}
\label{sec.KL}

High positive velocity $\HI$ from a region approximately $45''$ in
diameter (e.g., at $\vLSR=18.7\kms$ in \figref{fig.EmissionMaps})
extends to LSR velocities of approximately $31\kms$, as shown in
\twofigsref{fig.XVEm3}{fig.XVEm4}. An image of this $\HI$ emission,
integrated over the LSR velocity range from 18.7 to $31.0\kms$ is
shown in \figref{fig.BN}.  The high velocity $\HI$ emission coincides
in position with the high velocity outflow in the Orion-KL region,
embedded in the OMC-1 cloud behind the optical Huygens region.

The high velocity outflow in the Orion-KL region exhibits a wide opening angle
and line-of-sight velocities of up to $100\kms$ in CO lines
\citep{Zuckerman.etal1976,Kwan.Scoville1976}.  The shocked gas in this system
produces luminous $\Ht$ vibrational line emission
\citep{Gautier.etal1976,Nadeau.Geballe1979}, with a spectacular morphology
displaying multiple ``bullets'' (bow-shock tips) and ``fingers'' (bow-shock
wakes) as shown by \citet{Allen.Burton1993}, \citet{Stolovy.etal1998},
\citet{Schultz.etal1999}, and illustrated in the $\Ht$ $v=1{-}0$ S(1) image
shown in \figref{fig.BN}.  High resolution CO observations reveal similar high
velocity ``bullets'' and ``fingers''
\citep{Chernin.Wright1996,RodriguezFranco.etal1999a,RodriguezFranco.etal1999b,Zapata.etal2009},
emerging from a common position.  A number of optically identified HH objects
originate from the same position
\citep{Doi.etal2002,ODell.Henney2008,Zapata.etal2009}.

The lack of correspondence of the high velocity $\HI$ emission with
prominent $\Ht$ and CO ``fingers'', and the absence of $\HI$
emission at velocities $\vLSR>31\kms$, indicates that the $\HI$ is not
associated with the highest velocity outflowing gas. Nevertheless, the
velocity of the observed $\HI$ with respect to OMC-1 shows that the
material does participate in the outflow.  The two high velocity
features separated by a local minimum
(\twofigsref{fig.XVEm3}{fig.XVEm4}) suggest that both outflow lobes
are detected, although the northwest lobe is much more prominent in
$\HI$ emission than the southeast lobe (\figref{fig.BN}).  The
northwest lobe displays both redshifted and blueshifted $\HI$
emission, and this situation matches that observed in CO
\citep[e.g.,][]{Chernin.Wright1996,Zapata.etal2009}. The southeast
lobe is much less well-defined (both in $\HI$ and in CO), and it
has been suggested that the outflow towards the southeast is blocked
by the Orion-KL Hot Core \citep{Chernin.Wright1996,Zapata.etal2011b}.

\Figref{fig.BN} shows a positional match between the high velocity $\HI$
emission and the shocked $\Ht$ vibrational line emission, but no detailed
correspondence. In fact a detailed correspondence is not expected since
\figref{fig.BN} shows only the redshifted $\HI$, which is moving into OMC-1,
away from the observer. The $\Ht$ emission on the other hand shows all shocked
$\Ht$, regardless of velocity, but may be biased in favor of blueshifted gas
with lower extinction. The $\HI$ is clearly concentrated in the region of the
flow closest to the principal heating sources of the Orion-KL region.
The lack of more extended $\HI$ emission may
result from a decreasing column density as the flow expands away from its
source, but may also represent photodissociation of the molecular material by UV
radiation close to the heating sources.

The high velocity $\HI$ matches well, both in size and velocity, with the
expanding CO shell recently discovered in Orion-KL by \citet{Zapata.etal2011a},
and centered approximately on the outflow origin.  This bubble has a diameter of
about $25''$, an expansion velocity of about $10\kms$, and a dynamical age of
$500-1000\un{years}$. The left panel of \figref{fig.BN} indeed suggests
that the origin of the outflow is located at a local column density minimum,
bounded north, east and west by an incomplete and clumpy shell.  Inspection of
the CO PV diagram shown by \citet{Zapata.etal2011a} shows that the northwest
region of the CO shell is prominent in both redshifted and blueshifted emission,
while in the southeast part the redshifted emission dominates, in agreement with
the situation observed in $\HI$. The precise relation of the expending CO shell
to the rest of the outflow system is unknown. However, our high velocity $\HI$
data show features of both the CO shell and the inner region of the $\Ht$
emission, suggesting that these are directly related, and plausibly trace the
same gas.

The mass in redshifted $\HI$ represented in \figref{fig.BN} is only
$2\times10^{-5}\Msun$. Based on the limited velocity range sampled,
the total mass of $\HI$ associated with the outflow could well be a
factor of $3{-}5$ higher, i.e., $(0.6-1.0)\times10^{-4}\Msun$
($0.06-0.1$ Jupiter masses).  This mass is insignificant compared to
the total mass of molecular gas in the expanding bubble ($\sim5\Msun$)
as derived from CO \citep{Zapata.etal2011a}.

\section{$\HI$ absorption associated with the Orion Nebula}
\label{sec.HIAbsDisc}

The $\HI$ absorption dataset presented in \secref{sec.AbsResults} 
shows a striking number of structures. Nevertheless, the various
velocity features can be subdivided into a small number of groups,
according to common properties:
\begin{enumerate}
\item {\it large-scale features\/} covering M42 as well as M43. These are the
components~A and B, discussed by vdWG89.
\item {\it arclike features}. These are components D and F of vdWG90
  and L identified in the present dataset. in addition, component~C
  of vdWG89 can be recognized as an extended arc;
\item {\it elongated features\/} displaying unique kinematic
  signatures.  In the present data, these are components G and J\null.
\item {\it features at the velocity of the background molecular
  cloud.} These components are H, I and M\null.
\end{enumerate}

Based on the agreement in velocity with the ionized gas, which is
streaming away from OMC-1 towards us, vdWG89 interpreted component~C
as neutral material closely associated with the ionized gas, and
suggested that this material was entrained in the ionized flow. For
the blueshifted small-scale components, vdWG90 proposed acceleration
by the rocket effect \citep{Oort.Spitzer1955} as the origin of the
blueshift with respect to the bulk of the neutral gas. For these
explanations to be valid, ionization fronts would have to be present
on the surfaces of the $\HI$ features.  However, HST and ground-based
observations have failed to reveal optical rims associated with these
components \citep{Henney.etal2007,GarciaDiaz.Henney2007}.  Therefore,
new interpretations for these features will be developed in the
following subsections.

\subsection{The large-scale atomic Veil}
\label{sec.DiscVeil}

The large-scale structure of the atomic Veil covering M42 and M43 is
dominated by $\HI$ velocity components A and B, and the red wing
of the absorption lines is detected at
$7.2\kms<\vLSR<9.7\kms$ in \figref{fig.channelmaps}. In agreement with
vdWG89, a strong opacity gradient found towards the
Northeast Dark Lane; a positive velocity gradient of
about $2.4\kms\per{pc}$ is found in the same direction. This
velocity gradient is detected in both components,
indicating that they
are related (vdWG89). vdWG89 suggested that component~A
represents $\HI$ in the envelope of OMC-1, extending in
front of M42 and M43. A small velocity difference ($\HI$ component~A
is blueshifted with respect to the molecular gas by $\sim2\kms$)
represents a slow expansion of the cloud envelope.


\subsection{Arc-like $\HI$ features}
\label{sec.arclike}

\subsubsection{The southwest arc: component~C}
\label{sec.DiscC}

The $\HI$ absorption feature at $\vLSR\sim-3.2\kms$, forming an almost
semicircular arc in the southwest part of M42, is $\HI$ component~C of
vdWG89. The opacity images at $\vLSR=-4.4$, $-3.2$ and $-1.9\kms$ show
that this feature contains a string of clumps.

The system of optical flows in the southern part of M42 originates
mostly from a region on the east side of the Orion-S region;
\citet{Henney.etal2007} referred to this location as the Optical
Outflow Source (OOS; indicated as such in \figref{fig.HST}).
Molecular outflows originate from the Orion-S cloud \citep[summarized
  in Fig.~1 of][]{ODell.etal2009}. Flows originating in this area
branch out in multiple directions, driving several well-known HH
objects \citep{Bally.etal2000,Henney.etal2007}. The arc formed by
$\HI$ component~C opens towards Orion-S\null. This feature may thus
result from a flow originating from the Orion-S region.

A flow system located at the position angle from Orion-S towards
component~C was detected as an extended arc of optical emission, denoted as the
Southwest Shock by \citet[][see their Fig.~6]{Henney.etal2007}. This feature is
located approximately twice as far from the OOS as the arc formed by
component~C, but its shape is very similar to the $\HI$. These features may
therefore result from two separate ejections from the same object in the
OOS\null. However, further data would be needed to verify this hypothesis.

The PV diagram shown in \figref{fig.PVD1} shows the relation between
component~C and components A and B, that define the large-scale
structure of the Veil. At small negative spatial offsets (i.e.,
towards the northeast), both components A and B are detected, but
component~C is not observed.  For instance, at offset $\sim-30''$,
where the lines are not saturated, both components~A and B are clearly
detected as kinematically separate components. This situation is
different at positive offsets.  Where component~C appears (at offset
$\sim15''$), component~B disappears.  This result clearly suggests
that the arc formed by component~C consists of material swept up from
component~B, displaced towards a more negative velocity. The fact that
component~B is not detected in the region inside the curved arc
(\figref{fig.PVC}) is consistent with this interpretation.

\subsubsection{The arc-like component~D}
\label{sec.DiscD}

Component~D was first identified by vdWG90 and is the largest of the
small-scale components, extending over a significant part of the
Huygens region between the Trapezium stars and the Bright Bar.  As
noted by vdWG90, this component coincides in position and velocity
with the region where velocity splitting is observed in several
optical emission lines from M42.  This velocity splitting was first
studied in detail by \citet{Deharveng1973}, whose [$\NII$] $6583\Ang$
line-splitting region~A corresponds to our $\HI$ component~D\null. A
full kinematic atlas of several optical lines has been presented by
\citet{GarciaDiaz.Henney2007} and \citet{GarciaDiaz.etal2008}; this
dataset shows the line splitting in detail. The ionized gas component
matching our $\HI$ component~D is referred to as the ``Southeast
Diffuse Blue Layer'' by \citet{GarciaDiaz.Henney2007}, who note that
this component is detected in [$\SII$] 6716 and $6731\Ang$ but not in
[$\SIII$] $6312\Ang$. This result is important since it implies that
the ionizing spectrum is rather soft, and not provided by
$\theta^1$\,C~Ori. \citet{GarciaDiaz.Henney2007} suggest that the
blueshifted velocity component is ionized by the star
$\theta^2$\,A~Ori, located southeast of the Bright Bar.

The present results shed new light on this issue.  The arc-like
morphology of component~D was already pointed out in \secref{sec.D}.
The kinematic structure of this component suggests an expanding shell
with center of expansion in the direction towards which the arc opens,
i.e., southeast of the Bright Bar.  \Figref{fig.HST} shows that the
arc opens towards the star $\theta^2\,$B~Ori, located approximately on
its axis of symmetry; this star has spectral type B0.7V
\citep{SimonDiaz2010}. The observed geometry suggests that this star
provides the ionization for the Southeast Diffuse Blue Layer. In this
model $\HI$ component~D represents the neutral material outside of
this $\HII$ region. This explanation implies that an IF should be
present between the Southeast Diffuse Blue Layer and $\HI$
component~D; the presence of an IF is confirmed by the detection of
weak [$\OI$] $6300\Ang$ emission associated with this layer
\citep{GarciaDiaz.etal2008}.

\subsubsection{The arc-like component~L}
\label{sec.DiscL}

The $\HI$ component~L at $\vLSR=-3.2\kms$ (shown by the red contours
northwest of the Bright Bar in \figref{fig.HST}) displays an arc-like
structure similar to that of component~D, also opening towards the
southeast.  The characteristics of expansion are seen in
\figref{fig.PVD2}, where it exhibits a velocity gradient with more
negative velocities towards the southeast.  For this component, the
star $\theta^2\,$A~Ori is located close to the symmetry axis of the
arc. The spectral type of this star is O9V
\citep{SimonDiaz.etal2006}. This configuration indicates a model
identical to that described above (\secref{sec.DiscD}) for
component~D, except that here $\theta^2$\,A~Ori is the exciting star.

\subsubsection{The expanding shell component~F}
\label{sec.DiscF}

The third component to display an expanding shell-like morphology is
component~F. This component is located close to a large complex of
extended bow shocks in the ionized gas related to the Herbig-haro
object $\HH{202}$. As can be seen in \figref{fig.HST}, $\HH{202}$ is
located (in projection) at the western edge of the arc defined by
$\HI$ component~F\null. At this position, component~F splits into two
separate velocity components (already recognized by vdWG90 and
indicated F1 and F2 in that paper). The velocity splitting is evident in
\figref{fig.PVF}, where component~F is observed to split into two
subcomponents with different spatial and kinematic offsets. The
subcomponent at the most positive spatial offset matches well in
position with $\HH{202}$.

The positions of high proper motion features in the ionized gas,
shaped liked parabolic arcs, have been established by
\citet{ODell.Henney2008}, using HST images from several epochs.  The
driving source for $\HH{202}$ is located in the OOS region associated
with Orion-S \citep{Henney.etal2007}. Indeed, a jet connecting
$\HH{202}$ to this region has been detected in the shock-tracing
[$\FeII$] $1.257\mum$ line \citep{Takami.etal2002}.

Two explanations can be considered for the presence of atomic hydrogen
associated with, but not precisely superposed on $\HH{202}$.
\begin{enumerate}
\item It is possible that $\HH{202}$ and $\HI$ component~F are formed
  by a jet creating a bow shock in both the ionized gas (observed as
  $\HH{202}$) and the neutral Veil (observed as $\HI$ component~F), as
  proposed earlier by \citet{ODell.etal1997}. 
\item Alternatively, it is possible that $\HI$ component~F results
  from rapid recombination in the dense post-shock gas associated with
  $\HH{202}$. This model has been proposed by
  \citet{MesaDelgado.etal2009}, who show that in $\HH{202{-}S}$ (the
  brightest knot in $\HH{202}$) the post-shock density is sufficiently
  high to trap the IF\null. As a result, the gas behind this dense
  post-shock region is shielded from the ionizing radition of the
  Trapezium stars, and rapidly recombines and cools.
\item Finally, it is possible for this component that the acceleration is
  provided by the rockets effect described by \citet{Oort.Spitzer1955}.
\end{enumerate}
In the second and third models, the neutral $\HI$ should be located
behind the IF in the direction away from the Trapezium stars, i.e.,
west of the bright rims of $\HH{202}$. However, none of the absorbing
$\HI$ is located in this region, and most of it is in fact located
east of $\HH{202}$. This geometry argues against the last two models.
In addition, a strong argument in favour of the first of these three
models comes from the fact that the blueshifted $\HI$ gas forming
component~F corresponds to a gap in the Veil, as observed in
\figref{fig.PVF}. This is exactly the geometry that is expected when
part of the Veil gas is accelerated to a more negative velocity by the
impact of a jet. In summary, the location of the $\HI$ with respect to
$\HH{202}$, and its detailed kinematic structure argue in favour of an
interaction of the jet driving $\HH{202}$ with the neutral Veil.

\begin{figure}
\plotone{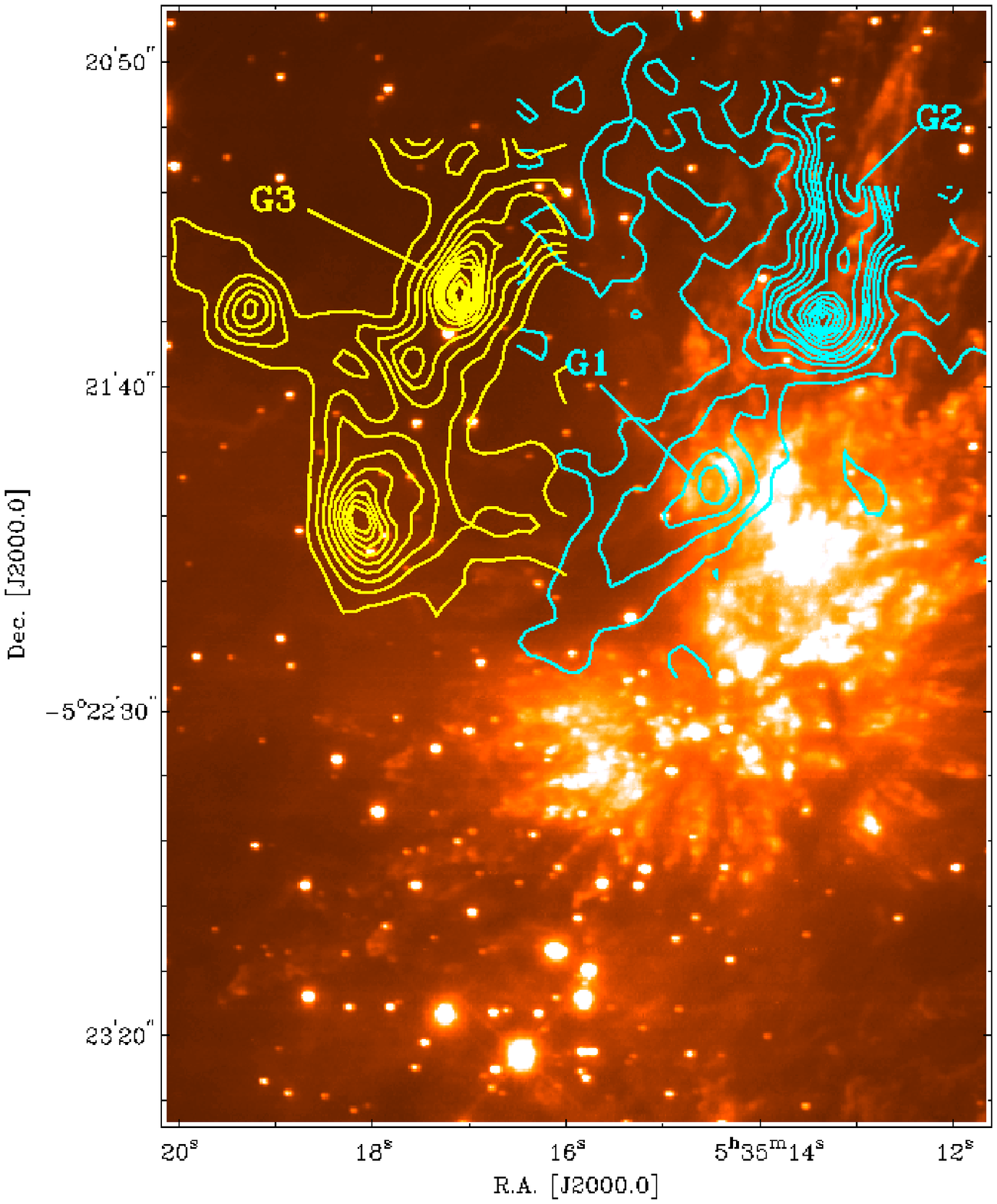}
\caption{$\HI$ opacity contours overlayed on the $\Ht$ $v=1{-}0$ S(1)
  image of \protect\citet{Bally.etal2011}. Light blue contours show
  $\HI$ opacity at $\vLSR=-9.6\kms$ from $\HI$ components~G1 and G2,
  with contour levels from 0.1 to 0.6 in steps of 0.1, then from 0.8
  to 1.8 in steps of 0.2. Yellow contours show $\HI$ opacity at
  $\vLSR=-0.6\kms$ from component~G3, with contour levels from 0.50 to
  2.25 in steps of 0.25, then from 2.5 to 4.0 in steps of 0.5.}
\label{fig.H2G}
\end{figure}

\subsection{Elongated $\HI$ features}
\label{sec.elongated}

Among the kinematic $\HI$ features identified, components~G and J have
an obvious elongated appearance, quite distinct from the arc-like
features discussed above.

\subsubsection{Component G}
\label{sec.absG}

The Northern part of M42, where component~G is located, harbors a
system of optical HH objects (e.g., $\HH{201}$, $\HH{205{-}210}$, all
indicated in \figref{fig.HST}), which are associated with the complex
and spectacular system of ``fingers'' observed in $\Ht$ vibrational
line emission \citep{Allen.Burton1993,Salas.etal1999,Bally.etal2011}
and CO \citep{Zapata.etal2009}.  Emission in the shock tracing
[$\FeII$] $1.64\mum$ line is found at the tips of the fingers.  The
fact that these ``fingers'' display optical line emission shows that
in this region they have broken out of the background molecular cloud
towards us \citep{ODell.etal1997}.

In the present $\HI$ data, several velocity systems are found in this
region, as can be seen in \figref{fig.PVG}. The highest (negative)
velocities are found for component~G2 (light blue contours in
\figref{fig.HST}), and this component is, in location as well as
orientation, closely aligned with the line connecting the ``dynamical
center'' where this outflow system originates (as discussed in
\secref{sec.KL}) to the Herbig-Haro objects $\HH{205{-}207}$.  This
line also corresponds to one of the most prominent molecular
``fingers'': the 1:00 system in the (hour dial) notation of
\citet{ODell.etal1997}. This agreement is shown in \figref{fig.H2G},
where component~G2 is shown in the blue contours at Declination north
of $\dms -5d21m40s$. South of this Declination, the blue contours
trace component~G1, and \figref{fig.H2G} shows that this feature does
not correspond to any of the shocked $\Ht$ features and, given its
orientation, does not originate in the Orion-KL region. At less
negative velocities, component~G3, which is approximately aligned with
G1 but displaced eastwards, is prominent, as shown in \figref{fig.H2G}
by the yellow contours. If components~G1 and G3 originate from a
common source, this source must be located approximately $30''$ east
of the Trapezium stars. However, this region does not harbour any
known driving source for these flows.

The asssociation of component~G2 with the $\Ht$ ``finger'' system
related to the Orion-KL outflow enables a new estimate of the
line-of-sight location of the dynamical center of the outflow behind
the IF\null. The Herbig-Haro object $\HH{205}$ is associated with this
$\Ht$ ``finger'', and closely aligned with $\HI$ component G2
(\figref{fig.HST}).  The radial velocity of $\HH{205}$ is
$\vLSR=-200\kms$ \citep{ODell.etal1997}, while its tangential velocity
has been measured to be $320\kms$ \citep{Bally.etal2000}. This system
thus follows a trajectory with an angle of $\sim30\deg$\ relative to
the plane of the sky. The position where the $\Ht$ ``finger'' breaks
through the IF is displaced from the dynamical center of the outflow
by a projected distance of
$0.14\pc$. Combining these results yields a distance of the
dynamical center of the outflow of $\sim0.1\pc$ behind the
IF\null. This distance is somewhat smaller than the $\sim0.2\pc$
derived by \citet{Doi.etal2004} using a similar calculation based on
the velocity vector and location of the Herbig-Haro object $\HH{201}$
\citep{Graham.etal2003}.

\subsubsection{Components J and K}
\label{sec.DiscJ}

The other elongated $\HI$ feature in our dataset is component~J, which
crosses the Bright Bar orthogonally. This feature is best seen at
$\vLSR=-3.2\kms$ in \figref{fig.channelmaps}. Both in position and
orientation, this feature closely coincides with the large, bright
pair $\HH{203{/}204}$.  Inspection of the $\HI$ opacity images in
\figref{fig.channelmaps} shows a peak in $\HI$ opacity located at the
position of $\HH{203}$ at LSR velocities from $-7.0$ to
$-4.3\kms$. However, at $-3.2\kms$ an extension towards $\HH{204}$ is
apparent, so that $\HI$ is associated with both Herbig-Haro objects.
\citet{ODell.etal1997} suggested that these objects mark the points
where collimated flows impact the extended $\HI$ Veil.  The detection
of $\HI$ associated with these features provides support for this
model.

$\HI$ component~K, located south of J, coincides with a region
resembling a bow shock \citep[see Fig.~6 of][]{Henney.etal2007}. This
region displays high ionization emission ([$\OIII$]) as well as
prominent lower ionization ([$\NII$] and [$\SII$]) ridges. This region
may represent an outer shock related to the $\HH{203{/}204}$ system
and would again be the result of a collimated flow striking the Veil.

\subsection{The blueshifted $\HI$ feature E}

The remaining blueshifted $\HI$ feature, component~E, appears neither
arc-like nor elongated and thus does not fit into any of the above
categories.

Component~E is located in the region of the Dark Bay. This feature
consists of 10 unresolved or barely resolved clumps, forming a
coherent velocity structure (\figref{fig.PVE}).  Due to the lack of
detailed optical information in the Dark Bay region, no associations
with ionized gas flows can be established. However, the different
morphology of component~E as well as the larger linewidth compared to
the other $\HI$ components suggests that component~E does not arise
from the same process. As noted in \secref{sec.DarkBay}, the Dark Bay
gas is at $\vLSR\approx 6\kms$. Component~E is therefore signficantly
blueshifted with respect to the bulk of the Dark Bay gas. An
attractive hypothesis is that $\HI$ component~E arises in a PDR
related to the Dark Bay, at the side facing the ONC\null. The thermal
expansion of the $\HII$ region may then account for the blueshift of
$\HI$ absorption with respect to rest of the Dark Bay gas.

\subsection{$\HI$ features at the velocity of the background molecular cloud}

$\HI$ components~H, I and M have central velocities $\vLSR=9-10\kms$,
corresponding to the velocity of OMC-1.  Since these features are
observed in absorption, a significant radio continuum must arise
behind them. Component~I has already been discussed above
(\secref{sec.DarkBay}), in connection with the $\HI$ emission
associated with that component.

\subsubsection{Component~H: $\HI$ absorption from Orion-S}
\label{sec.OrionS}

The $\HI$ absorption component~H (which was already noted by vdWG90)
corresponds closely in location, velocity, extent and morphology with
an $\HtCO$ absorption feature first identified by
\citet{Johnston.etal1983} and studied at higher resolution by
\citet{Mangum.etal1993}, as indicated in \figref{fig.HST}. This
feature is related to the Orion-S molecular core. The detection of
Orion-S in absorption indicates that the IF separating M42 from OMC-1
must be located {\it behind\/} Orion-S\null. The implication is that
the Orion-S molecular core must be located {\it within\/} the ionized
nebula. \citet{ODell.etal2009} were the first to argue for this
geometry, based on the detection of $\HtCO$ absorption and anomalies
in the derived extinction; the present $\HI$ data confirm this
picture. The detailed 3-dimensional structure and
the physical association of the various features in this region are
discussed in detail by \citet{ODell.etal2009}. These authors also
construct a physical model of the region, showing that the density and
column density are sufficient for Orion-S to survive in the harsh
radiation environment of M42.

\subsubsection{$\HI$ absorption associated with the Bright Bar}
\label{sec.HIAbsBar}

$\HI$ produced in the Orion Bar PDR will be at $\vLSR=10\kms$, as derived from
molecular lines in this region \citep[e.g., Fig.~8 of][]{VanderWerf.etal1996}.
The collection of clumps which we have collectively labeled component~M (as
indicated in the opacity image at $11.1\kms$ in \figref{fig.channelmaps}),
closely trace the outside of the Bright Bar and most likely represent clumps of
photodissociated $\HI$ in the Orion Bar PDR\null.  The small but non-zero angle
of the IF with respect to edge-on causes most of the photodissociated gas to be
located behind the radio continuum of the Bar; this $\HI$ cannot be dectected in
absorption.  However, since the IF will not be flat, a background radio
continuum will be available over a small distance towards the southeast in
regions where small amounts of $\HI$ are located in concave sections of the
IF\null.  Towards this continuum the $\HI$ can be observed in absorption.  At
larger distances from the IF, photodissociated $\HI$ in the Orion Bar PDR is
observed in emission, as discussed in \secref{sec.BarPDR}.

\subsection{Non-detection of $\HI$ from proplyds}

No $\HI$ absorption was found associated with known proplyds in the
Orion Nebula, with upper limits of about a Jupiter mass. This result
indicates that evaporation and ablation of molecular gas associated
with the proplyds
\citep{Chen.etal1998,Henney.ODell1999,Henney.etal2002} is rapidly
followed by photoionization.
Alternatively, the external IF confining
the proplyds may be opaque in the $21\cm$ continuum, prohibiting the
detection of neutral hydrogen.

\section{The Veil of Orion}
\label{sec.Veil}

\subsection{Location of the Veil}
\label{sec.VeilLoc}

Several of the blueshifted velocity components show clear evidence of
interaction with the Veil. This behavior is most obvious for
component~F, where the Veil shows a gap corresponding closely in
position with the blueshifted gas of component~F
(\secref{sec.F}). Similar evidence is found for component~C
(\secref{sec.DiscC}).  For component~G the evidence for interaction
with the Veil is weaker, but it clearly reveals interaction with a
layer of neutral gas.

Component~F therefore plays a crucial role in analysis of the
interaction of the ionized flows with the Veil. This component is
driven by the prominent HH object $\HH{202}$, for which the space
velocity vector is known, since both radial and tangential
velocities have been determined
\citep{Doi.etal2002,Doi.etal2004,Henney.etal2007,ODell.Henney2008}.
This flow system originates in the OOS associated with Orion-S;
therefore its radial distance with respect to Orion-S can be
calculated.  This distance is found to be $0.17\pc$
\citep{Doi.etal2004}. The lateral extent of the $\HH{202}$-component~F
system is about $50''$ or $0.11\pc$, and this size can be used as the
maximum distance between $\HH{202}$ and $\HI$ component~F\null. We
conclude that $\HI$ component~F, and therefore the neutral Veil, is
located at most $0.3\pc$ in front of Orion~S\null.

This distance is to be compared to the $\sim1\pc$ distance between the Trapezium
stars and the Veil, estimated by \citet{Abel.etal2004}. The relative
line-of-sight distances of Orion-S and the Trapezium stars are unknown.
\citet{ODell.etal2009} argue that Orion-S cannot be much behind the Trapezium
stars, or else it would not be possible to distinguish it from main IF
separating M42 from OMC-1. On the other hand, if Orion-S were significantly in
front of the Trapezium stars, it would be in front of almost all optical line
emission, and create a conspicuous extinction feature. While an extinction
feature at this position has indeed been detected, it is not very prominent
\citep{ODell.YusefZadeh2000}, and significantly less pronounced than for
instance the Dark Bay, which is indeed located in front of the ionized nebula.
Considering the large column density of Orion-S, its relatively modest derived
extinction must be due to the fact that a significant fraction of the nebular
emission arises in front of it. Therefore Orion-S is unlikely to be much closer
to the Veil than the Trapezium stars.

An additional argument can be found in the morphology of $\HI$ component~H,
which corresponds to Orion-S\null. The absorbing $\HI$ displays an opacity peak
at its north edge, i.e., approximately facing the Trapezium stars. At the same
position a small local maximum in radio continuum is found. This geometry
suggests that the ionizing and dissociating radiation has created an IF and PDR
at the side of Orion-S facing the Trapezium. The implication is that this
radiation must come mostly from the side, and that the line of sight connecting
Orion-S and the Trapezium stars is unlikely to be inclined more than $45\deg$
from the plane of the sky. This argument suggests that Orion-S is unlikely to
lie more than $0.1\pc$ in front of the Trapezium stars. The surface of the Veil
facing the ionized nebula will then be at most $0.4\pc$ in front of the
Trapezium stars.

Observations of UV absorption lines reveal local densities in the Veil, also
providing limits on the distance of the Veil from the Trapezium stars
\citep{Abel.etal2004,Abel.etal2006}. An upper distance limit of $3\pc$ was
determined by the absence of strong $\Ht$ UV absorption lines, which implies a
significant UV field to keep the molecular hydrogen fraction at the observed
value. A lower limit to the distance of the Veil to the Trapezium stars was
provided by \citet{Abel.etal2004}. These authors show the detailed dependence of
a large number of observables on the Veil-Trapezium distance. In addition they
used limits on $\Hb$ and [$\NII$] $6583\Ang$ emission from the surface of the
Veil facing the Trapezium to derive a minimum Veil-Trapezium distance of
$0.33\pc$ (for $\Hb$) and $1\pc$ (for [$\NII$]).  The [$\NII$] emission was
however considered in more detail by \citet{Abel.etal2006}, who demonstrated the
presence of a density-bounded layer of ionized gas located between the Trapezium
and the Veil and moving towards the Veil.  These authors conclude that this
layer absorbs most of the ionizing radiation from the Trapezium stars, which
accounts for the absence of a pronounced IF at the surface of the Veil facing
the Trapezium. This result has the important implication that the
emission line surface brightness of the illuminated side of the Veil cannot be
used to constrain its distance to the Trapezium. In summary, we conclude that
the distance between the Veil and the Trapezium of $0.4\pc$ derived in this
paper is consistent with the analysis by \citet{Abel.etal2004} and
\citet{Abel.etal2006}.

\begin{figure*}
\includegraphics[width=\textwidth]{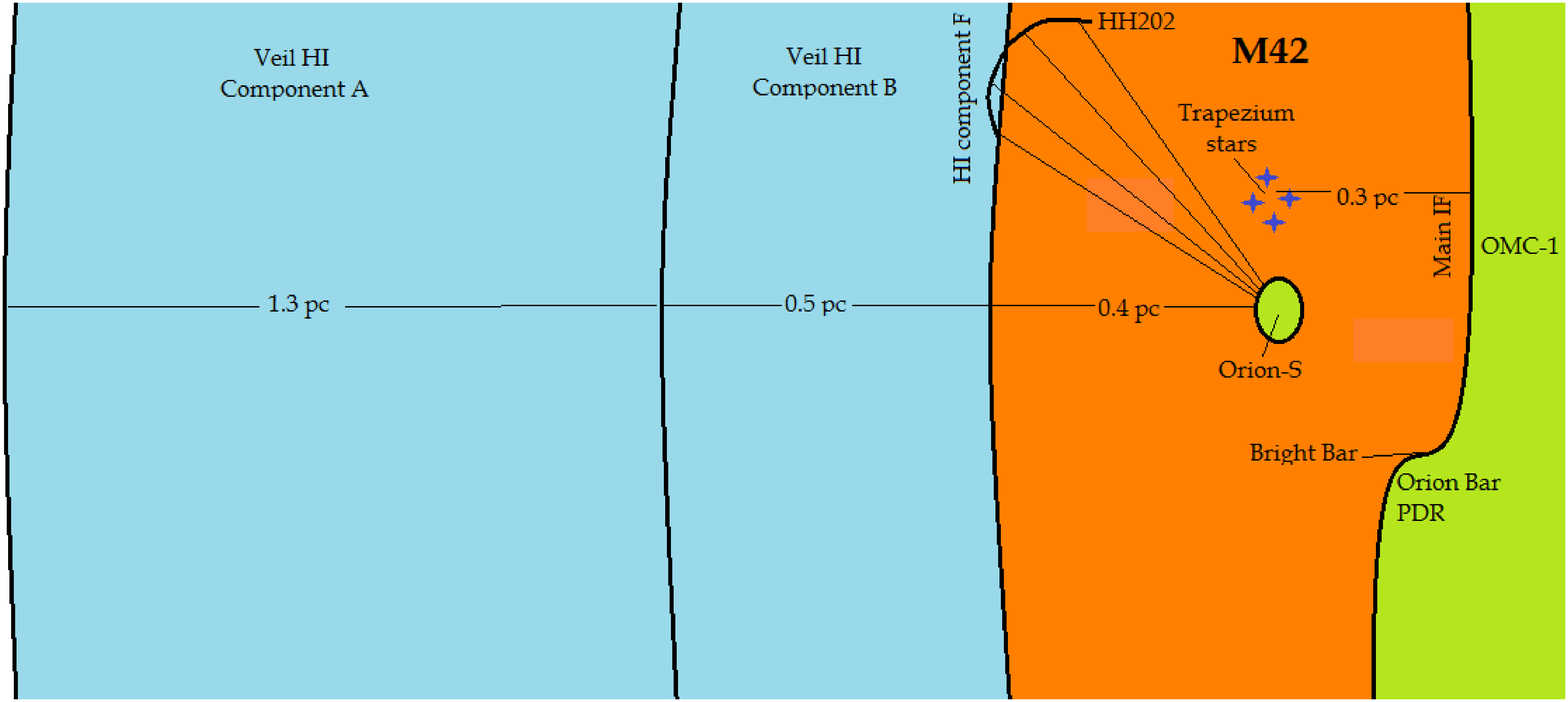}
\caption{Schematic geometry of M42 and the neutral Veil, illustrating
  the line-of-sight depths (approximately but not exactly to scale) of
  the various components. Orange indicates ionized gas, green
  indicates molecular gas and blue indicates atomic gas. The outflow
  originating in Orion-S, which powers the kinematics of $\HH{202}$
  and $\HI$ component~F is schematically indicated by a set of thin
  lines.}
\vspace{0.5cm}
\label{fig.Geometry}
\end{figure*}

\subsection{Geometry of the Orion Nebula and the Veil}

Our results reveal direct interaction of ionizing flows originating close to
Orion-S with the Veil. The evidence is particularly strong for the $\HH{202}$
($\HI$ component F, \secref{sec.DiscF}), which shows that only velocity
component~B of the Veil is affected. This result implies that component~B is
located closer to the ionized gas than component~A, and for the first time
reveals the relative locations of the two velocity components that constitute
the Veil.  As noted in \secref{sec.DiscVeil}, these two $\HI$ components are
large-scale features covering both M42 and M43, and exhibiting a common velocity
gradient, indicating that they are related. The fact that they both display a
high and similar line-of-sight magnetic field strength \citep{Troland.etal1989}
reinforces this conclusion. As shown by \citet{Abel.etal2004}, the magnetic
energy exceeds the energy in turbulent motions at least in component~A, where it
constitutes the dominant pressure term. \citet{Ferland2009} has suggested that
this situation is the result of compression by the expanding $\HII$ region and
radiation pressure from the ONC\null.

The $\HI$ column densities in components~A and B, based on $\Lya$
absorption measurements, are approximately 1.6 and
$3.2\cdot10^{21}\pcmsqu$ respectively \citep{Abel.etal2006}. With the
best density estimates of approximately $10^{2.5}$ and
$10^{3.4}\pcmcub$ for A and B \citep{Abel.etal2006}, their thicknesses
are approximately 1.3 (A) and $0.5\pc$ (B). The resulting geometry is
schematically shown in \figref{fig.Geometry}.  The lateral extent of
the Veil is at least $1.5\pc$ (its total extent covering M42 and M43),
but probably significantly larger, since it represents the atomic
envelope of the background molecular cloud (vdWG89), which extends
over at least $10\pc$, approximately the length of integral-shaped
filament of molecular gas and dust
\cite{Bally.etal1987,Johnstone.Bally1999}. The total depth of the
ionized cavity is about $0.7\pc$, which is small compared both to the
lateral extent of the Veil and to its depth of approximately
$1.8\pc$. These results therefore demonstrate the blister-type nature
of the M42 $\HII$ region, emphasized by main IF behind the Trapezium
stars, and also reveal its relation to the envelope of the background
molecular cloud, represented by the Veil.

\section{Conclusions}
\label{sec.Conclusions}

High resolution $\HI$ observations of the Orion Nebula reveal in
detail the feedback of the $\HII$ region and the ONC on their
neutral environment. This feedback takes place through both radiative
and mechanical interaction. Furthermore, our data provide new
information on the geometry of the complex.

Radiative feedback manifests itself principally by the non-ionizing UV radiation
from the ONC, which creates a prominent PDR outside the IF at the interface
between M42 and OMC-1. Prominent $\HI$ emission is detected from the Orion Bar
PDR, southeast of the Bright Bar IF\null. This result enables for the first time
a study of the abundance profile of atomic hydrogen and its relation to other
tracers in an edge-on PDR\null. We find that the $\HI$ emission arises from a
region with a temperature of $\sim 540\K$, with $\approx 5-10\%$ of the hydrogen
nuclei in the interclump region in the PDR in the form of $\HI$. Most of the
hydrogen in the region probed therefore remains molecular, in agreement with
theoretical PDR models.  This result shows that dense PDRs such as the Orion Bar
PDR have a low $\HI$ production, even if the impinging UV radiation field is
very strong.  The $\HI$ column density peak matches the peak of UV-excited
vibrational $\Ht$ line emission. This result confirms the two-step $\Ht$
photodissociation process through UV line absorption in the Lyman and Werner
bands.

Mechanical feedback results from the interaction of ionized and molecular
outflow systems with the neutral environment of the $\HII$ region. The $\HI$
data provides direct evidence for interaction of the HH object $\HH{202}$ and
the flow system responsible for the optical Southwest Shock of
\citet{Henney.etal2007} with the Veil.  The interaction of these flow systems
with the Veil causes velocity shifts in the absorbing $\HI$.  The evidence for
direct interaction between $\HH{202}$ and the neutral Veil implies that the Veil
must be located quite close to the Orion-S source of outflows. Using the known
space trajectory of $\HH{202}$, we find that the surface of Veil facing the
Orion Nebula is located at most $0.4\pc$ in front of Orion-S\null. Of the two
velocity components~A and B that constitute the Veil, we find that component~B
lies closer to the ionized nebula. 

The total depth of the ionized cavity is approximately $0.7\pc$. This depth is
small compared to both the lateral extent and the depth of Veil. These
dimensions confirm the nature of the M42 $\HII$ region as a thin blister, and
simultaneously reveal the relation of the ionized region to the background
molecular cloud and its neutral envelope.

\section*{Acknowledgments}

\acknowledgeNRAO We thank Nick Abel for extensively discussing the
distance between the Veil and the Trapezium stars with us. We are
grateful to John Bally for making available the $\Ht$ $v=1{-}0$ S(1)
image shown in \twofigsref{fig.BN}{fig.H2G}. We also thank Tom
Troland, Will Henney, and Gary Ferland for useful comments.

{\it Facilities:} \facility{VLA}, \facility{EVLA}

\appendix

\section{Relation between $\HI$ column density and 21\,cm line optical depth
  and brightness temperature}

The $\HI$ column density is related to the velocity-integrated optical depth of
the $21\cm$ line by
\begin{equation}
N(\HI)={32\pi k \Ts \over 3 \qu{A}{ul} h c \qu{\lambda}{ul}^2}\int\tau(v)\,dv,
\label{eq.NHI}
\end{equation}
where $\qu{\lambda}{ul}=21.10612\cm$ is the wavelength of the transition,
$\qu{A}{ul}=2.88426\cdot10^{-15}\ps$ \citep{Gould1994} is its Einstein $A$
value, $\Ts$ is the spin temperature (i.e., the excitation temperature of the
two spin levels), $h$ is Planck's constant, $k$ is the Boltzmann constant, and
$c$ is the speed of light. This expression makes use of the approximation $\exp
(-\qu{E}{ul}/k\Ts)\approx 1-\qu{E}{ul}/k\Ts$, where $\qu{E}{ul}$ is the energy
difference of the two spin levels. This approximation is valid to high accuracy,
since $\qu{E}{ul}/k=0.06816\K$ and $\Ts\gg\qu{E}{ul}/k$ under all practical
conditions. Inserting numbers, \eqref{eq.NHI} can be written as
\begin{equation}
{N(\HI)\over {\rm cm}^{-2}}=1.81267\cdot 10^{18}\,{\Ts \over {\rm
      K}}\int\tau(v)\,d{v\over {\rm km}\ps}.
\label{eq.NHInumbers}
\end{equation}

For optically thin ($\tau(v)\ll1$) $\HI$ emission, $\exp{[-\tau(v)]}$
can be expanded to first order and the equation of transfer
\eqref{eq.Transfer} reduces to
\begin{equation}
\Tb(v)=\tau(v)[\Ts-\Tc-\qu{T}{back}(v)].
\label{eq.TransferThin}
\end{equation}
If $\Ts\gg\Tc+\qu{T}{back}(v)$ in \eqref{eq.TransferThin}, the $\HI$ 
column density as given by \eqref{eq.NHI} can be expressed as
\begin{equation}
N(\HI)={32\pi k \Ts \over 3 \qu{A}{ul} h c \qu{\lambda}{ul}^2}\int\qu{T}{b}(v)\,dv,
\label{eq.NHIThin}
\end{equation}
which results in
\begin{equation}
{N(\HI)\over {\rm cm}^{-2}}=1.81267\cdot 10^{18}\,{\int\qu{T}{b}(v)\,dv\over
  {\rm K}\kms}.
\label{eq.NHIThinNumbers}
\end{equation}

\bibliographystyle{apj}
\bibliography{%
AtomicData,%
DataProcessing,%
HHObjects,%
H2,%
MolecularClouds,%
OrionA,%
PDRs,%
Photodissociation,%
Photoionization,%
Starformation,%
StellarData,%
XDRs%
}

\end{document}